\documentclass[12pt]{article}

\usepackage[margin=1in]{geometry}

\usepackage{array}
\usepackage{amssymb}%
\usepackage{mathrsfs}%
\usepackage[title]{appendix}%
\usepackage{textcomp}%
\usepackage{manyfoot}%
\usepackage{algorithm}%
\usepackage{algorithmicx}%
\usepackage{algpseudocode}%
\usepackage{listings}%
\usepackage{amsthm,amsmath}
\usepackage{natbib}
\usepackage{graphicx}

\usepackage[utf8]{inputenc}	
\usepackage[table]{xcolor}
\usepackage[T1]{fontenc}
\usepackage{url}%vmargin
\usepackage{booktabs}
\usepackage{tcolorbox}
\usepackage{indentfirst}	
\usepackage{longtable}
\usepackage{enumerate}
\usepackage{hyperref}
\usepackage{verbatim}
\usepackage{amsfonts}
\usepackage{makecell}
\usepackage{multirow}
\usepackage{lipsum}
\usepackage{helvet}
\usepackage{subfig}
\usepackage{natbib}
\usepackage{tikz}
\usepackage{rotating}

\usepackage{enumitem}

\DeclareMathOperator*{\tr}{tr}
\DeclareMathOperator*{\diag}{diag}

\DeclareMathOperator*{\GCV}{GCV}

\newcommand{\bb}{b}

\newcommand{\normal}{\mathrm{N}}
\newcommand{\prob}{\mathrm{P}}
\newcommand{\unif}{\mathrm{U}}
\newcommand{\ber}{\mathrm{Ber}}
\newcommand{\dexp}{\mathrm{Exp}}

\newcommand{\dBeta}{\mathrm{Beta}}
\newcommand{\GI}{\mathrm{IG}}
\newcommand{\E}{\mathrm{E}}

\renewcommand{\vec}[1]{\boldsymbol{#1}}

\usetikzlibrary{arrows}
\usetikzlibrary{matrix}
\usetikzlibrary{positioning}

\title{Bayesian Variable Selection for Function-on-Scalar Regression Models: a comparative analysis}
\author{Pedro Henrique Toledo de Oliveira Sousa$^{1*}$ \and \ Camila Pedroso\ Estevam\ de Souza$^2$ \and \ Ronaldo\ Dias$\dag$ \\  
\normalsize$^1$Department of Statistics, Federal University of Paraná, Brazil \\ 
\normalsize $^2$Department of Statistical and Actuarial Sciences, University of Western Ontario, Canada\\
\normalsize $\dag$Department of Statistics, University of Campinas, Brazil\\
\normalsize$^*$Corresponding author: pedro.henrique3@ufpr.br}
\date{}

\begin{document}
\maketitle

\begin{abstract}

In this work, we developed a new Bayesian method for variable selection in function-on-scalar regression (FOSR). 
Our method uses a hierarchical Bayesian structure and latent variables to enable an adaptive covariate selection process for FOSR.
Extensive simulation studies show the proposed method's main properties, such as its accuracy in estimating the coefficients and high capacity to select variables correctly. Furthermore, we conducted a substantial comparative analysis with the main competing methods: the BGLSS (Bayesian Group Lasso with Spike and Slab prior) method, the group LASSO (Least Absolute Shrinkage and Selection Operator), the group MCP (Minimax Concave Penalty), and the group SCAD (Smoothly Clipped Absolute Deviation). Our results demonstrate that the proposed methodology is superior in correctly selecting covariates compared with the existing competing methods while maintaining an excellent level of goodness of fit. In contrast, the competing methods could not balance selection accuracy with goodness of fit. We also considered a COVID-19 dataset and some socioeconomic data from Brazil as an application and obtained sound results. In short, the proposed Bayesian variable selection model is highly competitive, showing significant predictive and selective quality.		   
\end{abstract}


\section{Introduction}\label{sec1}

%\cite{sousa2023bayesian}

Linear regression models are well-established and thoroughly studied in the statistical literature \citep{kutner2004applied,rencher2008linear,weisberg2013applied}, but most have been developed for cross-sectional data rather than functional data. When the data are functions, standard linear regression models are generally unsuitable, as they require constructing a design matrix that incorporates not only covariate information but also additional elements, such as basis functions derived from functional data expansions. This approach has been explored in studies such as \cite{microarray}, \cite{Yakuan}, and \cite{Reimherr}. To address these limitations, functional regression models \citep[Chapters 12 to 17]{ReS} were introduced, becoming a key area of functional data analysis that has attracted considerable attention in both applied and methodological research \citep{ReS,infFDA,Mark,Yakuan,Philip,Xin_Qi}.

Among the various functional regression models described in \cite{ReS}, this paper focuses on function-on-scalar regression (FOSR), where the response is functional, covariates are scalar, and coefficients are functional. We propose a Bayesian model for parameter inference and covariate selection within the FOSR framework. Under a linear structure, the FOSR model is expressed as:

\begin{equation}
	y_{i}(t)=\beta_{0}(t)+\sum_{l=1}^{p}x_{li}\beta_{l}(t)+\epsilon_{i}(t).
	\label{resposta_funcional}
\end{equation}

\noindent where, $y_i(t)$ is the value of the $i$th response functional evaluated at point $t$, $x_{li}$ is the value of the $l$th scalar predictor corresponding to the $i$th curve. Notice that the intercept and variable coefficients are functional. The component $\epsilon_{i}(t)$ represents the error curve. It is often assumed that the error curves are independent and come from the same Gaussian process with a mean of zero and a given covariance structure. In many cases, in addition to the goal of prediction, the aim is to estimate the functional coefficients in (\ref{resposta_funcional}) while imposing some regularization to avoid overfitting. This can be done by assessing whether $\beta_{l}(t)=0$ only for some values of $t$ or whether $\beta_{l}(t)=0$ for all $t$. Here, we focus on the latter to perform variable selection.    

The literature on variable selection in functional regression analysis is still not dominant. However, there is strong evidence that this area has grown rapidly in recent years, especially in the last decade. Some examples of variable selection in functional regression can be found in \cite{matsui2011variable}, which studied the group SCAD regularization for the selection of functional regressors, \cite{mingotti2013lasso} and \cite{HL}, which generalized LASSO to the case of scalar covariates and functional response, among others \citep{FL,Ferraty2cap3,gertheiss2013variable,MSW,Julian}.  A related but different problem in functional data analysis is the selection of basis functions for functional data representation (smoothing). In this context, one can use penalty-based techniques such as LASSO \citep{lasso} and Bayesian LASSO \citep{art:bl-Casella} or latent variable approaches as in \cite{Kuo}, \cite{anselmo2005adaptive} and \cite{sousa2023bayesian}. 

There is growing interest in research problems focusing on FOSR. Some studies over the last decade have explored its potential from a Bayesian perspective \citep{Montagna, DFOSR, Bourgeois}, with the latter also addressing variable selection. However, the frequentist approach remains predominant for variable selection in FOSR, with applications spanning genetic studies \citep{microarray,Zhaohu,Reimherr,Parodi}, Alzheimer’s research \citep{jiguo}, and other domains \citep{Yakuan}.

This work proposes a new Bayesian approach to select covariates in an FOSR model by associating a Bernoulli latent variable with each functional coefficient. This makes it possible to assign zero to specific model functional coefficients with positive probability. From a hierarchical structure, considering a linear combination of basis functions to represent each functional coefficient, we obtain the full conditional distributions to build a Gibbs sampler to sample from the posterior distribution (R code implementation is available upon request). Unlike other Bayesian approaches that commonly use non-informative priors on the last layer, we use an exponential prior on the coefficient variance component to prevent problems of collinearity among variables. We also conduct an extensive comparative analysis between our proposed methodology and each of the following competing methods: \textit{group LASSO} \citep{groupLASSO}, \textit{group SCAD} \citep{microarray}, \textit{group MCP} \citep{Breheny12} and \textit{BGLSS} \citep{grBLASSO}.

%at %\url{https://github.com/phtosEST/Bayesian_Variable_Selection_for_FOSR}). 

This paper is organized as follows. Section \ref{mp_parte2} presents the proposed Bayesian model for variable selection in FOSR, while Section \ref{cap5} shows the design and results of several numerical experiments involving the proposed methodology. Also, in Section \ref{cap5}, we present a comparative study between the proposed model and the competing methods. In Section \ref{sec:4}, we conduct a study to evaluate the performance of the proposed model in a functional regression problem involving some socioeconomic data and COVID-19 data from the Federal District and Brazilian states. Finally, Section \ref{cap_conclusao} provides some general conclusions about this work.

\section{Proposed Bayesian model for variable selection in FOSR}
\label{mp_parte2}

For the introduction of the proposed Bayesian model, let us consider $p$ covariates, $m$ curves with $n_{i}$ observations at the points $t_{ij}\in A\subseteq \mathbb{R}$, in which $i \in \{1,\;2,\dots,\;m\}$ and $j \in \{1,\;2,\dots,\;n_{i}\}$. So, we assume that:
\begin{equation}
	y_{ij} = \beta_{0}(t_{ij})+\sum_{l=1}^{p}x_{li}Z_{l}\beta_{l}(t_{ij})+\epsilon_{ij}\text{,}
	\label{eq:FOSR_model}
\end{equation} where $Z_{l}$ is a Bernoulli latent variable that selects or not the $l$th covariate. While $y_{ij}$ represents the value of the $i$th curve at point $t_{ij}$ (meaning $y_{ij}=y_{i}(t_{ij})$), the partial functional coefficients associated with the covariates are represented by $\beta_{l}(.)$’s, just as $\beta_{0}(.)$ represents the intercept. Finally, $\epsilon_{ij}=\epsilon_{i}(t_{ij})$ is the random component that represents the error measurements with a normal distribution with zero mean and variance equal to $\sigma^2$.

For a given fixed curve $i$, the vector with the final functional coefficients of this functional linear regression is given by $\vec{\nu(.)}=(\nu_{1}(.),\;\nu_{2}(.),\dots,\;\nu_{p}(.))^{'}=(Z_{1}\beta_{1}(.),\;Z_{2}\beta_{2}(.),\dots,$ $\;Z_{p}\beta_{p}(.))^{'}$. For the sake of brevity and to avoid confusion in the notation, the components $\nu_{l}(.)$'s are called functional coefficients, while the $\beta_{l}(.)$'s are called partial functional coefficients.

We carry out a basis function expansion for all $\beta_{l}(.)$’s and $\beta_{0}(.)$ in (\ref{eq:FOSR_model}) with the same set of known basis functions, $B_{1}(.),\ldots,B_{K}(.)$, as follows:
\begin{gather}
	y_{ij}=\sum_{k=1}^{K}b_{k0}B_{k}(t_{ij})+\sum_{l=1}^{p}x_{li}Z_{l}\left[\sum_{k=1}^{K}b_{kl}B_{k}(t_{ij})\right]+\epsilon_{ij}\nonumber\\
	=\sum_{k=1}^{K}b_{k0}B_{k}(t_{ij})+\sum_{l=1}^{p}\sum_{k=1}^{K}x_{li}Z_{l}b_{kl}B_{k}(t_{ij})+\epsilon_{ij}\text{,}\nonumber%sem_citar
\end{gather}where the $b_{kl}$’s represent the coefficients of the basis expansion for the $l$th partial functional coefficient, and the $b_{k0}$’s are the coefficients of the basis expansion applied to the functional intercept. 

Hence, it is possible to develop a Bayesian model for adaptive variable selection with a hierarchical structure. First, however, some definitions should be established to facilitate understanding. Let
$\vec{\theta}=(\theta_{1},\;\theta_{2},\dots,\;\theta_{p})^{'}$ and
$\vec{Z}=(Z_{1},\;Z_{2},\dots,\;Z_{p})^{'}$. While $\vec{\theta}$ and $\vec{Z}$ are vectors of dimension $(p\times1)$, there are two other vectors of higher dimension $(Kp\times1)$ that are: $\vec{\tau^{2}}=(\vec{\tau^{2}_{.1}}^{'},\;\vec{\tau^{2}_{.2}}^{'},\dots,\;\vec{\tau^{2}_{.p}}^{'})^{'}$, where $\vec{\tau^{2}_{.l}}=(\tau^{2}_{1l},\tau^{2}_{2l},\dots,\tau^{2}_{Kl})^{'}$ and $\vec{\bb}=(\vec{\bb_{.1}}^{'},\;\vec{\bb_{.2}}^{'},\dots,\;\vec{\bb_{.p}}^{'})^{'}$, with $\vec{\bb_{.l}}=(\bb_{1l},\bb_{2l},\dots,\bb_{Kl})^{'}$. 
%Separately from $\vec{\bb}$, 
Note that there is also a vector that contains all $\bb_{k0}$'s, say, $\vec{\bb_{0}}=(\bb_{10},\;\bb_{20},\dots, \;\bb_{K0})^{'}$. It is worth mentioning that the intercept is estimated separately; therefore, a prior is not assigned to the $\bb_{k0}$'s. Finally, consider $\vec{y}=(\vec{y_{1.}}^{'},\;\vec{y_{2.}}^{'},\dots,\;\vec{y_{m.}}^{'})^{'}$, where $\vec{y_{i.}}=(y_{i1},y_{i2},\dots,y_{in_{i}})^{'}$. Then, we have:
\begin{gather}	y_{ij}|\vec{Z},\vec{\bb},\sigma^2\sim\normal\left(\sum_{k=1}^{K}b_{k0}B_{k}(t_{ij})+\sum_{l=1}^{p}\sum_{k=1}^{K}x_{li}Z_{l}b_{kl}B_{k}(t_{ij}),\sigma^{2}\right)\text{;}\nonumber\\
	\bb_{kl}|\sigma^{2},\vec{\tau^{2}}\sim\normal(0,\sigma^2\tau_{kl}^{2})\text{;}\nonumber\\
	Z_{l}|\vec{\theta}\sim\ber(\theta_{l})\text{;}\nonumber\\
	\theta_{l}|\vec{\mu}\sim\dBeta\left(\mu_{l},(1-\mu_{l})\right)\text{;}\nonumber\\
	\mu_{l}\sim \unif(0,\psi), \quad\text{in which}\quad \psi<1\text{;}\nonumber\\
	\tau_{kl}^{2}\sim\dexp\left(\frac{\lambda^{2}}{2}\right)\quad \text{and}\quad\sigma^{2}\sim\GI(\delta_{1},\delta_{2})\text{.}
	\label{modeloAutoparte2}
\end{gather}

One can also simplify the hierarchical model in \eqref{modeloAutoparte2} by considering $\mu_{l}$ as a hyperparameter, obtaining:
\begin{gather}
	y_{ij}|\vec{Z},\vec{\bb},\sigma^2\sim\normal\left(\sum_{k=1}^{K}b_{k0}B_{k}(t_{ij})+\sum_{l=1}^{p}\sum_{k=1}^{K}x_{li}Z_{l}b_{kl}B_{k}(t_{ij}),\sigma^{2}\right)\text{;}\nonumber\\
	\bb_{kl}|\sigma^{2},\vec{\tau^{2}}\sim\normal(0,\sigma^2\tau_{kl}^{2})\text{;}\nonumber\\
	Z_{l}|\vec{\theta}\sim\ber(\theta_{l})\text{;}\nonumber\\
	\theta_{l}|\vec{\mu}\sim\dBeta\left(\mu_{l},(1-\mu_{l})\right)\text{;}\nonumber\\
	\tau_{kl}^{2}\sim\dexp\left(\frac{\lambda^{2}}{2}\right)\quad \text{and}\quad\sigma^{2}\sim\GI(\delta_{1},\delta_{2})\text{.}
	\label{modeloparte2}
\end{gather}

A sample from the posterior distribution of $\vec{Z}$, $\vec{\bb}$ and $\sigma^2$ is obtained via the Gibbs sampler described in Subsection \ref{gibbs}.

In the model with $\mu_{l}$ as a parameter, it is recommended to define $\psi$ as a value not too close to $1$ because very high values of $\psi$ can lead to numerical issues in the implementation of the Gibbs sampler. When $\mu_{l}$ is a hyperparameter, it is known that the closer $\mu_{l}$ is to $0.5$, the less informative will be the prior of $\theta_{l}$ since in addition to having an average located precisely in the middle of the interval between 0 and 1, the respective density will be symmetric and with maximum variance.
Although the $\theta_{l}$ prior only explicitly presents one hyperparameter, one can rewrite $\mu_{l}$ so that, for all $l$, $\mu_{l}=\frac{ C}{p}$, where $C$ is a tuning parameter and can be interpreted as a prior guess about how many variables, on average, one wants to select. With this perspective, the $\theta_{l}$ prior, now, has two hyperparameters, $C$ and $p$. To visualize this, one can think of a prior mean of the binomial generated by the sum of the $Z_{l}$’s, that is:
\begin{equation}
	\E\left(\sum_{l=1}^{p} Z_{l}\right)=\sum_{l=1}^{p}\E\left(\E\left(Z_{l}|\theta_{l}\right)\right)=\sum_{l=1}^{p}\E(\theta_{l})=\sum_{l=1}^{p}\mu_{l}=\sum_{l=1}^{p}\frac{C}{p}=p\frac{C}{p}=C\text{.}\nonumber%sem_citar
\end{equation} 

%For both models, it is natural that it has some sensitivity to the hyperparameter $\lambda$ since this is the only hyperparameter that truly plays the role of regularizer.

The hyperparameter $\lambda$ controls both the mean and variance of $\tau_{kl}^2$, which in turn affects the prior variance of the coefficients.

\subsection{Gibbs sampler}
\label{gibbs}

The panel \ref{GIBBS} presents the standard procedure for implementing the Gibbs sampler in a clear and summarized way. Details on the derivation of each full conditional distribution are shown in Appendix \ref{Ap_full}. \\

\begin{tcolorbox}[colback=gray!10]
\label{GIBBS}

\begin{enumerate}[leftmargin=*, itemsep=2pt, topsep=2pt]
\item Define the hyperparameters.
\item Assign initial chain values to parameters.
\item For $\text{c}=2,\dots,\text{Nint}$:
  \begin{enumerate}[label=\arabic{enumi}.\arabic*, leftmargin=*, itemsep=2pt, topsep=2pt]
  \item Sample ${\sigma^{2}}^{(c)}\sim f(\sigma^2|\vec{\bb}^{(c-1)},\vec{\tau^2}^{(c-1)},\vec{Z}^{(c-1)},\vec{y})$, an inverse gamma as in \eqref{cmpos_sigma2_PARTE2}.
  \item For $k=1,\dots,K$:
    \begin{enumerate}[label=\arabic{enumi}.\arabic{enumii}.\arabic*, leftmargin=*, itemsep=2pt, topsep=2pt]
    \item For $l=1,\dots,p$:
      \begin{enumerate}[label=\arabic{enumi}.\arabic{enumii}.\arabic{enumiii}.\arabic*, leftmargin=*, itemsep=2pt, topsep=2pt]
      \item Sample ${\eta_{kl}^{2}}^{(c)}\sim f(\eta_{kl}^2|\bb_{kl}^{(c-1)},{\sigma^2}^{(c)})$, an inverse normal as in \eqref{cmpos_eta2_PARTE2}.
      \item Set $\tau_{kl}^{2}=1/\eta_{kl}^{2}$.
      \end{enumerate}
    \end{enumerate}

  \item For $l=1,\dots,p$:
    \begin{enumerate}[label=\arabic{enumi}.\arabic{enumii}.\arabic*, leftmargin=*, itemsep=2pt, topsep=2pt]
    \item Sample $\mu_{l}^{(c)}\sim f(\mu_{l}|\theta_{l}^{(c-1)})$, a continuous Bernoulli as in \eqref{cmpos_mu_PARTE2}.
    \item Sample $Z_{l}^{(c)}$ with $\prob(Z_{l}=1|\vec{\bb}^{(c-1)},\theta_{l}^{(c-1)},{\sigma^2}^{(c)},\vec{Z_{-[l]}}^{(*)},\vec{y})$ as in \eqref{cmpos_Z_PARTE2}.
    \item Sample $\theta_{l}^{(c)}\sim f(\theta_{l}|\mu_{l}^{(c)},Z_{l}^{(c)})$, a beta as in \eqref{cmpos_theta_PARTE2}.
    \end{enumerate}

  \item Sample $\vec{\bb}^{(c)}\sim f(\vec{\bb}|{\sigma^2}^{(c)},\vec{\tau^2}^{(c)},\vec{Z}^{(c)},\vec{y})$, a multivariate normal as in \eqref{cmpos_beta_PARTE2}.
  \end{enumerate}
\end{enumerate}

\end{tcolorbox}

The term $\vec{Z_{-[l]}}^{(*)}$ in the panel \ref{GIBBS} represents a mixed vector with components from the previous iteration that have not yet been updated and with components that have already been updated in the current iteration. As $l$ evolves in 3.3, the vector $\vec{Z_{-[l]}}^{(*)}$ is updated. Also, it is worth mentioning that the complete conditional distributions are presented in the panel in summary form, since the independent components are omitted from the notation. The definition of the continuous Bernoulli distribution can be found in \citep{NEURIPS2019_f82798ec}.

At the end of this process with the convergence of the chain, there is a sample of the joint posterior distribution of $\vec{Z}$, $\vec{\bb}$, and $\sigma^2$.

\section{Simulation study}
\label{cap5}

Numerical experiments were carried out with simulated data to assess the main properties of the proposed model. To ensure an in-depth simulation study while avoiding an expensive computational cost, we deliberately opt to generate each simulated dataset considering $m=10$ curves and $n_i=n=25$ evaluation points per curve. This setup allows us to thoroughly explore the properties and potentialities of the proposed model. We further apply the proposed model using a larger dataset in Section \ref{sec:4}. In addition, we simulated data considering two dispersion levels for the data ($\sigma=0.2$ and $\sigma=20$). We fitted each simulated dataset using our proposed Bayesian approach, considering $K=10$ B-spline basis functions for the expansion of the functional coefficients. We also investigated the sensitivity of our approach to the choice of $\mu$. Therefore, ten model fit configurations were tested under each dispersion scenario, nine of which considered $\mu$ as a hyperparameter and one that considered it as a parameter. For all model configurations, regardless of the dispersion scenario, a total of one hundred simulated datasets were used. 

Although multiple values of the regularization parameter $\lambda$ could be tested, as is typically done for competing methods, we fixed $\lambda = \sqrt{2}$ (yielding a prior mean of one) due to the substantial computational cost of fitting the proposed model across one hundred replications and multiple configurations. Still, even with this fixed value, our proposed model showed excellent performance, with great fit and superior covariate selection accuracy.

With regard to the other hyperparameters, we set $\delta_{1}=\delta_{2}=0$. Thus, when using a degenerated prior to $\sigma^2$ with $\delta_{1}=0$ and $\delta_{2}=0$, there is an equivalence with the use of a non-informative improper prior $\frac{1}{\sigma^2}$. In situations where $\vec{\mu}$ is considered as a parameter, $\psi=0.6$ was defined.

Subsection \ref{sec:data_generation} describes details about the simulated data generation and the Gibbs sampler implementation. Subsection \ref{sec:metrics} presents the performance metrics used for model evaluation. Then, in Subsection \ref{secao_simulacoes_PARTE2}, we present the results of our numerical experiments. 

\subsection{Simulated data generation and Gibbs implementation}\label{sec:data_generation}

Each simulated dataset consists of $m=10$ curves generated as follows:

\begin{enumerate}
	\item Six covariates are generated ($X_{li}$'s, in which $i \in\{1,\;2,\dots,\;m\}$ and $l \in\{1,\;2,\dots,\;p=6\}$) represented by six vectors of $m$ random variables such that $X_{1i}\sim\normal(200,100^2)$; $X_{2i}\sim\normal(100,100^2)$; $X_{3i}\sim\normal(20,50^2)$; $X_{4i}\sim\normal(50,50^2)$; $X_{5i}\sim\normal(2,5^2)$ and $X_{6i}\sim\normal(25,50^2)$.	
	\item Define the functional coefficients $\beta_{0}(t)=\exp(t^{2})$; $\beta_{3}(t)=\cos(2t)$ and $\beta_{5}(t)=t^{3}$, in which $t \in [0,2]$;	
	\item All other partial functional coefficients do not need to be defined since it is considered that only two covariates $X_{3i}$ and $X_{5i}$ should be used for data generation;
	
	\item For each curve, a vector is defined for the error term, in which each component is represented by $\epsilon_{ij}$ and is generated from a normal distribution with mean equal to zero and variance equal to $\sigma^{2}$ (two sets of data were generated, one with $\sigma=0.2$ and another with $\sigma=20$);
	
	\item Fixing $n_{i}=n=25$ for the number of evaluation points in each functional and defining a common set of points $t_{ij}$ for every functional $i$, which is defined by the function \verb|seq(0, 2, length=n)| of the \textit{software} R, the discrete representation of the $m$ functional responses is generated such that
	\begin{equation}
		y_{ij}=y_{i}(t_{ij})=\beta_{0}(t_{ij}) + \beta_{3}(t_{ij})X_{3i} + \beta_{5}(t_{ij})X_{5i} + \epsilon_{ij}\text{.}
		\label{rep_disc}
	\end{equation}
\end{enumerate}

The equality presented in \eqref{rep_disc} is a summary representation, since its full version is given by
\begin{gather}
	y_{ij}=y_{i}(t_{ij})=\beta_{0}(t_{ij}) + Z_{1i}\beta_{1}(t_{ij})X_{1i} + Z_{2i}\beta_{2}(t_{ij})X_{2i} + Z_{3i}\beta_{3}(t_{ij})X_{3i}+\nonumber\\ Z_{4i}\beta_{4}(t_{ij})X_{4i} + Z_{5i}\beta_{5}(t_{ij})X_{5i}+ Z_{6i}\beta_{6}(t_{ij})X_{6i} + \epsilon_{ij}\text{,}
	\label{rep_disc_full}
\end{gather}in which $Z_{3i}$ and $Z_{5i}$ are equal to one, while $Z_{1i}$, $Z_{2i}$, $Z_{4i}$ and $Z_{6i}$ are equal to zero. 

The covariates chosen to compose the discrete representation \eqref{rep_disc} were purposely positioned in the design matrix $\vec{X}$ so that both were not neighbors. Thus, we can verify that the model truly selects covariates, not just blocks of covariates.

Therefore, any method that provides a good fit to this dataset should be able to identify that only $X_{3i}$ and $X_{5i}$ are useful covariates, and yield estimates of the functional coefficients that are close to the original functional parameters. 
When building the synthetic data,  we establish that $\beta_{3}(t)$ and $\beta_{5}(t)$ are the so-called ``functional coefficients'' instead of ``partial functional coefficients'' since it is implied that the latent variable is equal to one in the generation of synthetic data in both cases. 

Since the goal of the model is to select covariates in a functional context, we decided to direct attention to the functional coefficients associated with the covariates. In view of this, the estimation of the functional intercept took place a step prior to the model fit. Thus, the covariates were standardized by their mean and standard deviation, and the functional intercept was estimated through the functional mean of the functional responses (see Supplementary Section 1 for more details). Therefore, the result of subtracting the functional intercept estimate from the initial response variable is used as the response variable to fit the proposed model. 

Bayesian inference was implemented through the Gibbs sampler described in Subsection \ref{gibbs} using two chains to enable the convergence diagnosis, which start at different points, and with the calculation of 10\,000 iterations. Considering a \textit{burn-in} period of 50\% of the size of each chain and spacing of 50 points between each sampled value, each chain contains 100 sampled points. Thus, at the end of the process, a posterior sample of size 200 is obtained for each parameter and latent variable.

The convergence diagnosis used in our analyses takes into account the convergence test of the chains proposed by \cite{Gelman}. In our numerical experiments carried out herein, the chains were initialized as follows:
\begin{itemize}
	\item $\vec{\bb}=\vec{-1}$, $\vec{\theta}=\vec{\frac{1}{5}}$, $\sigma^2=1$ and $\vec{\tau^2}=\vec{1}$ for the first chain and
	\item $\vec{\bb}=\vec{1}$, $\vec{\theta}=\vec{\frac{4}{5}}$, $\sigma^2=5$ and $\vec{\tau^2}=\vec{5}$ for the second chain.
\end{itemize} 

The latent random variables $Z_{l}$’s were initialized in the first chain via random draw, and the complementary vector of the drawn variables for the first chain was used for the second chain. In situations where $\vec{\mu}$ is considered as a parameter, the first chain was initialized with $\vec{\mu}=\vec{\frac{1} {5}}$, while the second was with $\vec{\mu}=\vec{\frac{4}{5}}$.

To summarize the results, we calculated, for each parameter, the MAP (Maximum a Posteriori) estimate.  Thus, regarding the latent variables $Z_l$’s, we obtained the most frequent value, meaning the posterior mode.

\subsection{Performance metrics}\label{sec:metrics}

Let $\tilde{\beta}_{0}(.)$ be the simple functional mean obtained from the functional responses of $m$ individuals so that $\tilde{\beta}_{0}(t_{ij})=\frac{1}{m}\sum_{i=1}^{m} y_{i}(t_{ij})$ is the value of this functional mean evaluated at point $t_ {ij}$. Now define $\hat{y}_{ij}=\hat{y}_{i}(t_{ij})=\hat{\beta}_{0}(t_{ij}) + \sum_{l=1}^{p}x_{li}\hat{Z}_{l}\hat{\beta}_{l}(t_{ij})$ (see Expression 2 in Section 1 of the Supplementary Material) as the proposed model prediction of the $i$th functional response at the evaluation point $t_{ij}$, so then we have 
\begin{equation}	
	1 -	\dfrac{\left(\sum_{i=1}^{m} n_{i} - 1\right)\sum_{i=1}^{m}\sum_{j=1}^{n_{i}} (y_{ij}-\hat{y}_{ij})^2 }{ \left(\sum_{i=1}^{m} n_{i} - \sum_{l=1}^{p} I_{\left\{\hat{Z}_{l}>0\right\}}K\right)\sum_{i=1}^{m}\sum_{j=1}^{n_{i}} (y_{ij}-\tilde{\beta}_{0}(t_{ij}))^2}
	\text{,}
	\label{eq:r2_PARTE2}
\end{equation}as a goodness-of-fit performance metric for the functional context, being similar to the adjusted $R^{2}$. Thus, the closer the metric is to one, the better the performance. The $\hat{Z}_{l}$’s represent the most frequent values in the chain for each of the respective $Z_{l}$’s.

Comparisons and performance evaluations of results obtained from our numerical experiments were performed using the metric presented in the Expression \eqref{eq:r2_PARTE2} and the Mean Squared Error (MSE) defined as follows. Let $g_{i}(t_{ij})=\sum_{l=1}^{p}x_{li}Z_{l}\beta_{l}(t_{ij})$, so that the model in (\ref{eq:FOSR_model}) can also be written as $y_{ij} = \beta_{0}(t_{ij}) + g_{i}(t_{ij}) + \epsilon_{ij}$. In particular, in our numerical experiments, $g_{i}(t_{ij}) = \beta_{3}(t_{ij})X_{3i} + \beta_{5}(t_{ij})X_{5i}$. Thus, the MSE is calculated as follows:
\begin{equation}
	\mbox{MSE} = \frac{1}{\sum_{i=1}^{m}n_{i}}\sum_{i=1}^{m}\sum_{j=1}^{n_{i}} \big[(\beta_{0}(t_{ij})+g_{i}(t_{ij})) - (\hat{\beta}_{0}(t_{ij})+\hat{g}_{i}(t_{ij}))\big]^2,
	\label{eq:MSE}
\end{equation}

\noindent where $\hat{g}_{i}(t_{ij}) = \sum_{l=1}^{p}x_{li}\hat{Z}_{l}\hat{\beta}_{l}(t_{ij})$. Thus, the MSE evaluates the proximity between the true mean (expected) functional response curve, $\beta_{0}(t_{ij})+g_{i}(t_{ij})$, and the estimated one, $\hat{\beta}_{0}(t_{ij})+\hat{g}_{i}(t_{ij})$. It is worth noticing that as the points $t_{ij}$’s are equidistant in all curves, the MSE is proportional to the IMSE (Integrated Mean Square Error).

\subsection{Results} \label{secao_simulacoes_PARTE2}

Supplementary Figures 1 through 4 summarize the boxplots of the metrics \eqref{eq:r2_PARTE2} and MSE across all tested model configurations in our simulations, including scenarios in which $\mu$ is treated as a hyperparameter or a parameter. The results consistently demonstrate that the proposed model exhibits low sensitivity to the choice of $\mu$, regardless of the dispersion level of the data or how $\mu$ is treated. This robustness ensures the proposed model can effectively smooth and select partial functional coefficients (and, consequently, covariates) regardless of the prior choice for $\mu$. Both metrics exhibit expected behaviour concerning $\sigma$, with improved performance when data are generated with low dispersion ($\sigma = 0.2$). Additionally, the symmetry and limited number of outliers in the boxplots confirm the well-behaved nature of the metrics, highlighting the model’s efficiency and reliability across configurations.

\begin{figure}%[!htb]
	\centering
	\begin{tabular}{ccc}
		\subfloat[Functional intercept $\beta_{0}(t)$ (Data with $\sigma=0.2$).]{\includegraphics[scale=0.18]{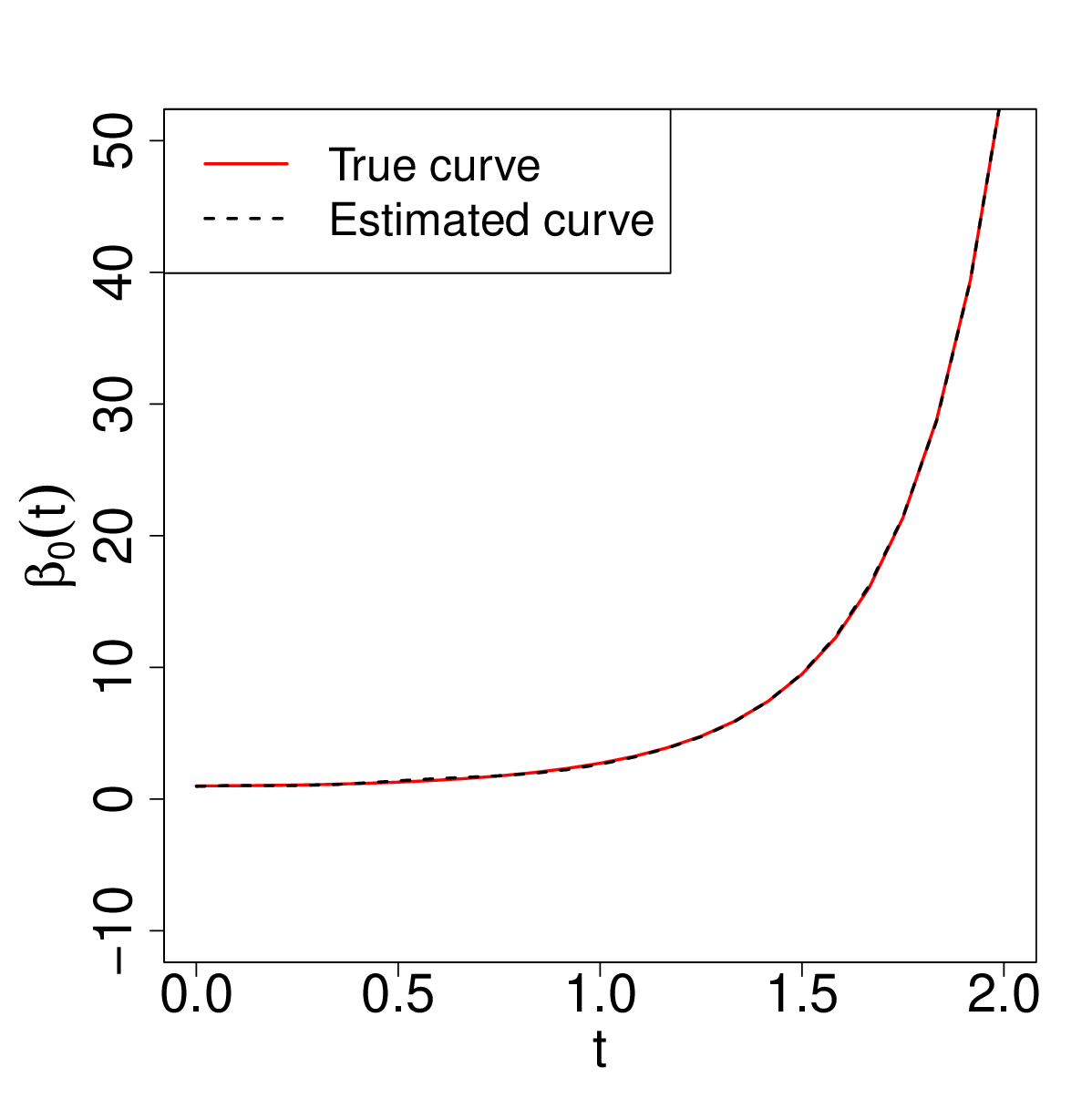}}&
		\subfloat[Partial functional coefficient $\beta_{3}(t)$ (Data with $\sigma=0.2$).]{\includegraphics[scale=0.18]{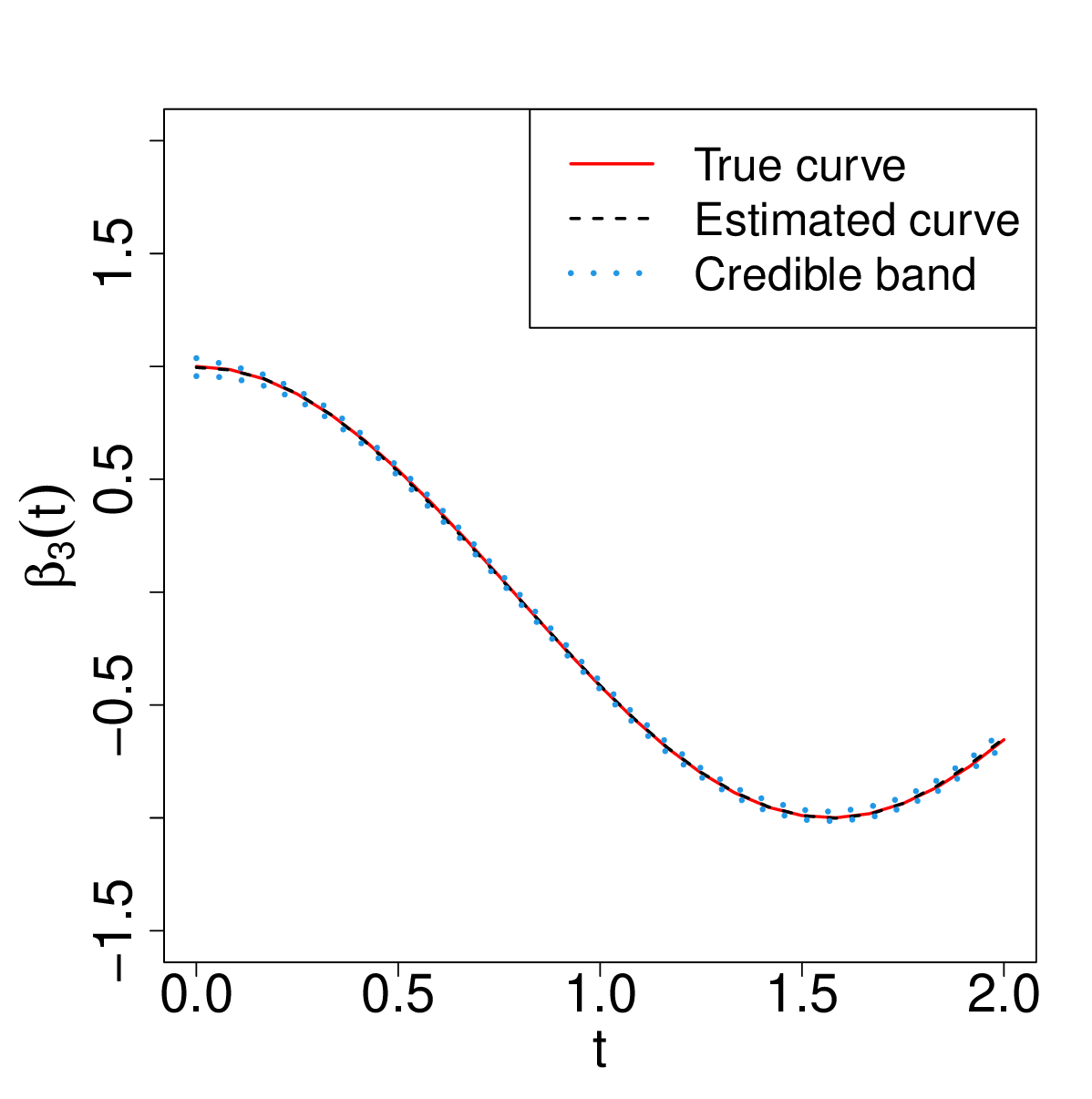}}&
		\subfloat[Partial functional coefficient $\beta_{5}(t)$ (Data with $\sigma=0.2$).]{\includegraphics[scale=0.18]{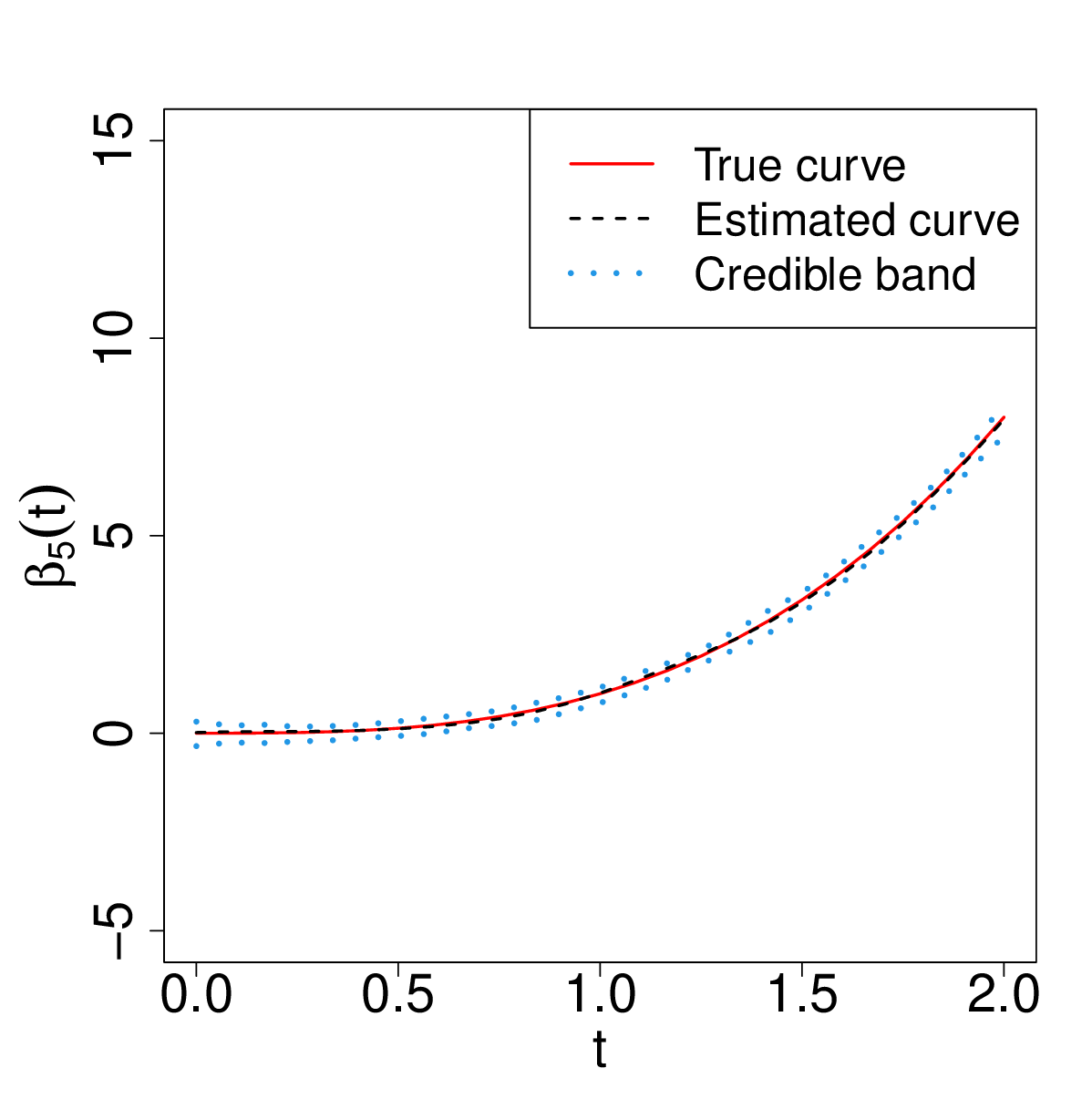}}\\
		\subfloat[Functional intercept $\beta_{0}(t)$ (Data with $\sigma=20$).]{\includegraphics[scale=0.18]{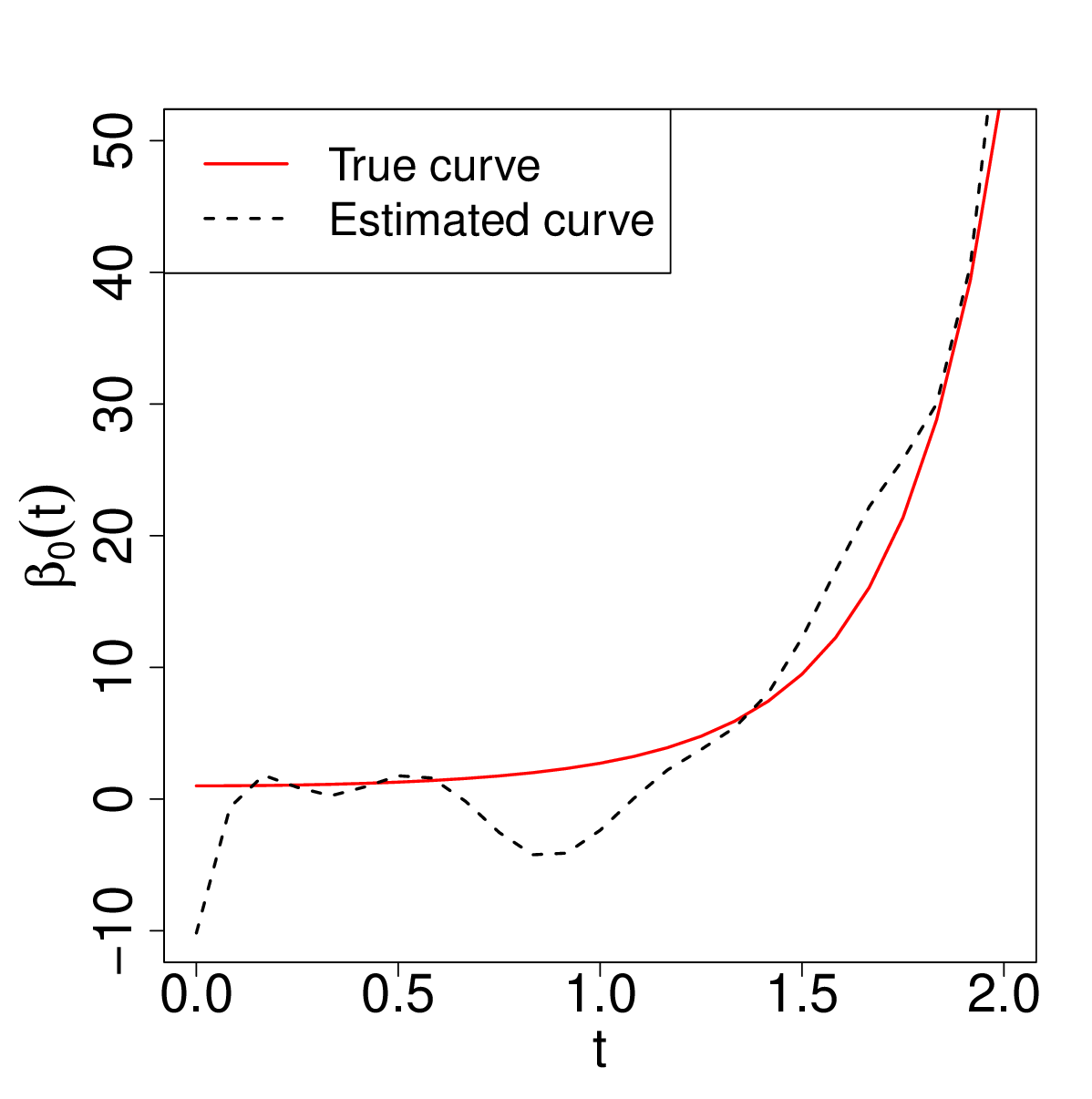}}&
		\subfloat[Partial functional coefficient $\beta_{3}(t)$ (Data with $\sigma=20$).]{\includegraphics[scale=0.18]{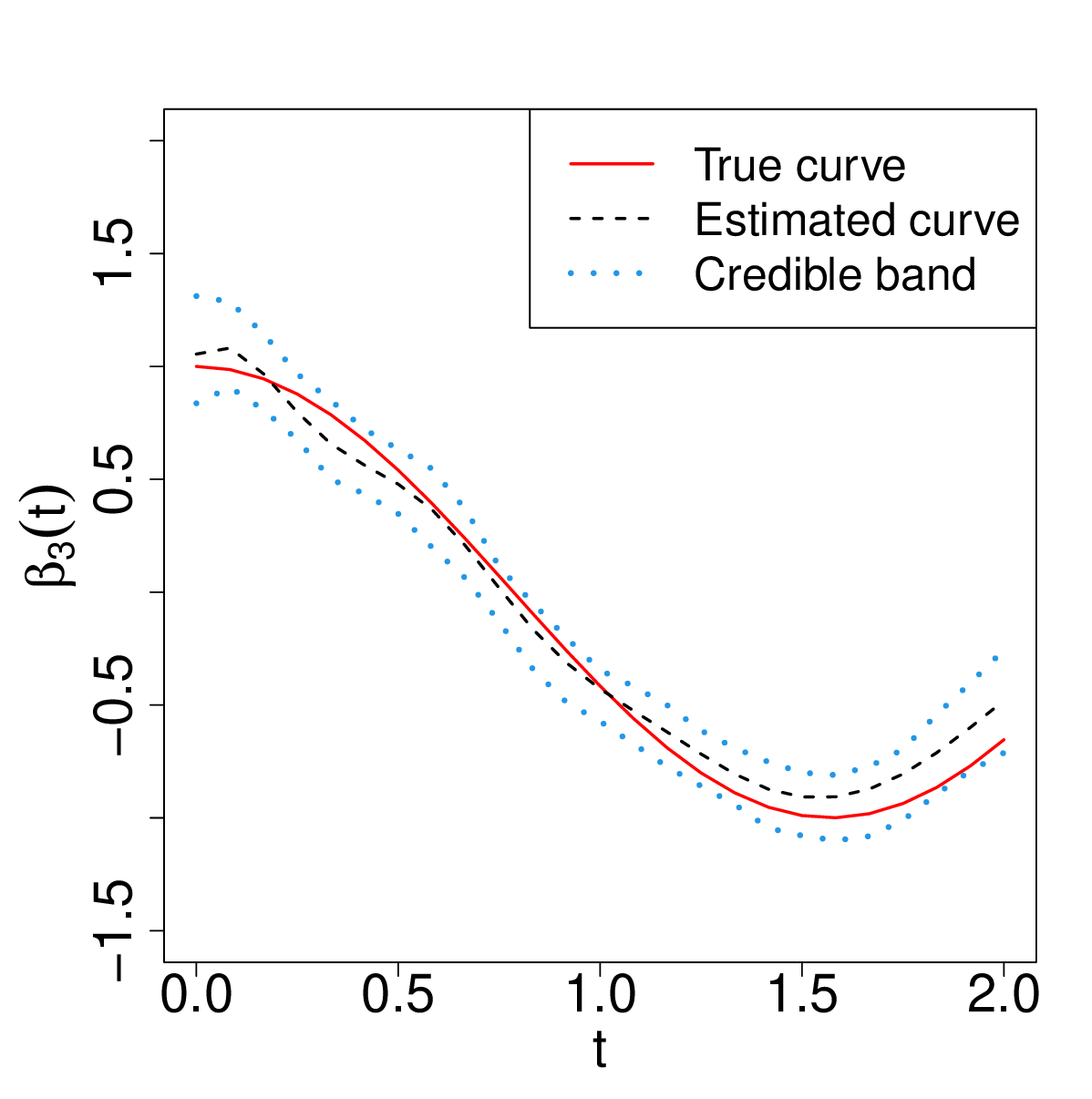}}&
		\subfloat[Partial functional coefficient $\beta_{5}(t)$ (Data with $\sigma=20$).]{\includegraphics[scale=0.18]{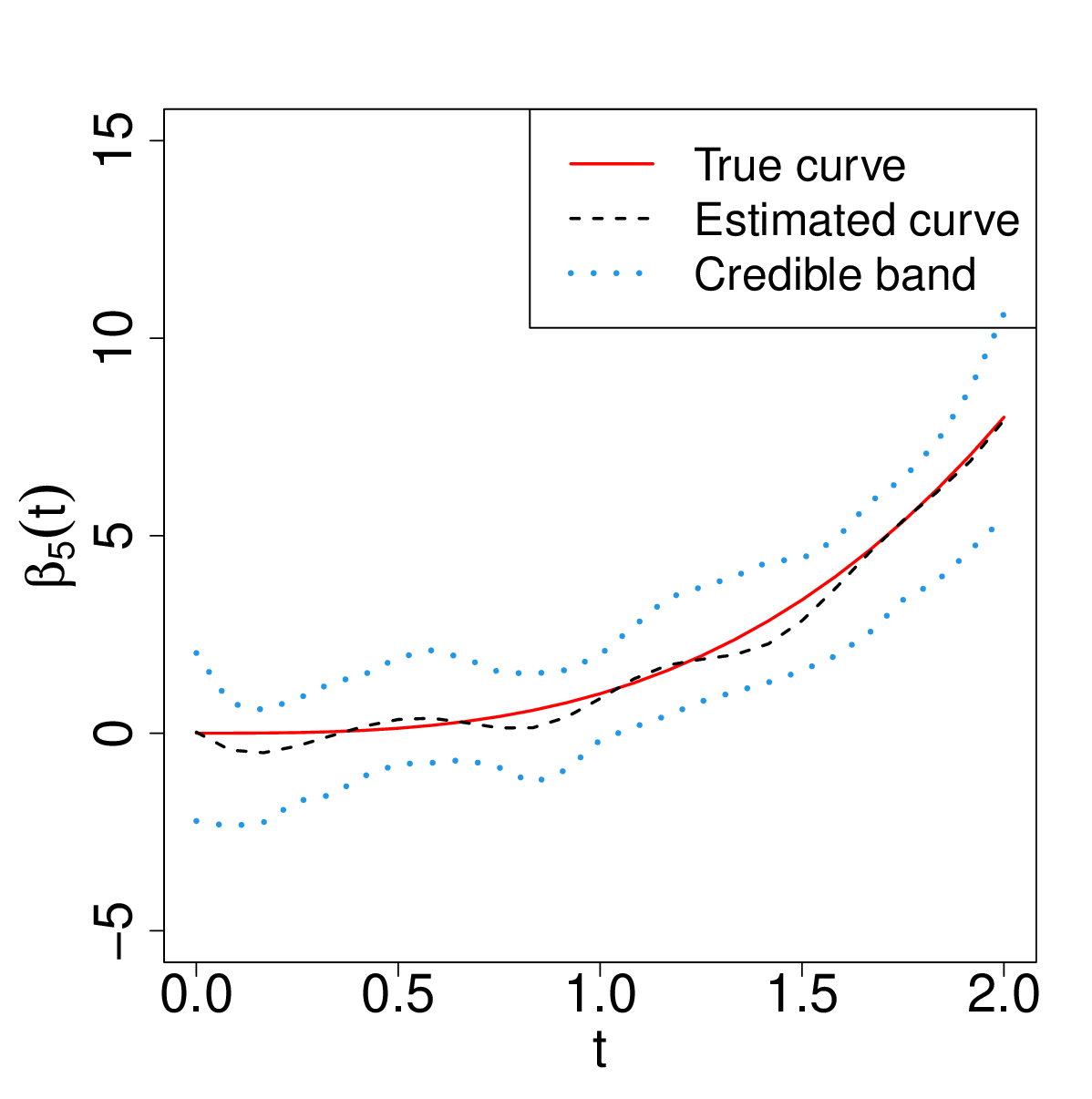}}  
	\end{tabular}
	\caption{Intercept and partial functional coefficients (true and estimated) that were selected by the model, according to the data dispersion degree (configurations with $K=10$ and $\mu=0.1$ for the two levels of dispersion). 95\% credible bands are also shown for the partial functional coefficients.}
	\label{coefsp_selecionados}
\end{figure}

As an example of results from one of the simulated datasets, Figure \ref{coefsp_selecionados} shows the estimates \(\hat{\beta}_{l}(t)\) of the partial functional coefficients selected by the model, along with 95\% credible bands for a configuration with \(\mu=0.1\) under two dispersion levels (\(\sigma=0.2\) and \(\sigma=20\)). The credible bands were computed using two hundred parameter vectors sampled by the Gibbs sampler, generating two hundred corresponding curves for each partial functional coefficient. Quantiles at \(2.5\%\) and \(97.5\%\) were then used to define the lower and upper bounds at each point \(t\). 

\begin{figure}%[!htb]
	\centering
	\begin{tabular}{ccc}
		\subfloat[$1^{\text{st}}$ functional response (Data with $\sigma=0.2$).]{\includegraphics[scale=0.18]{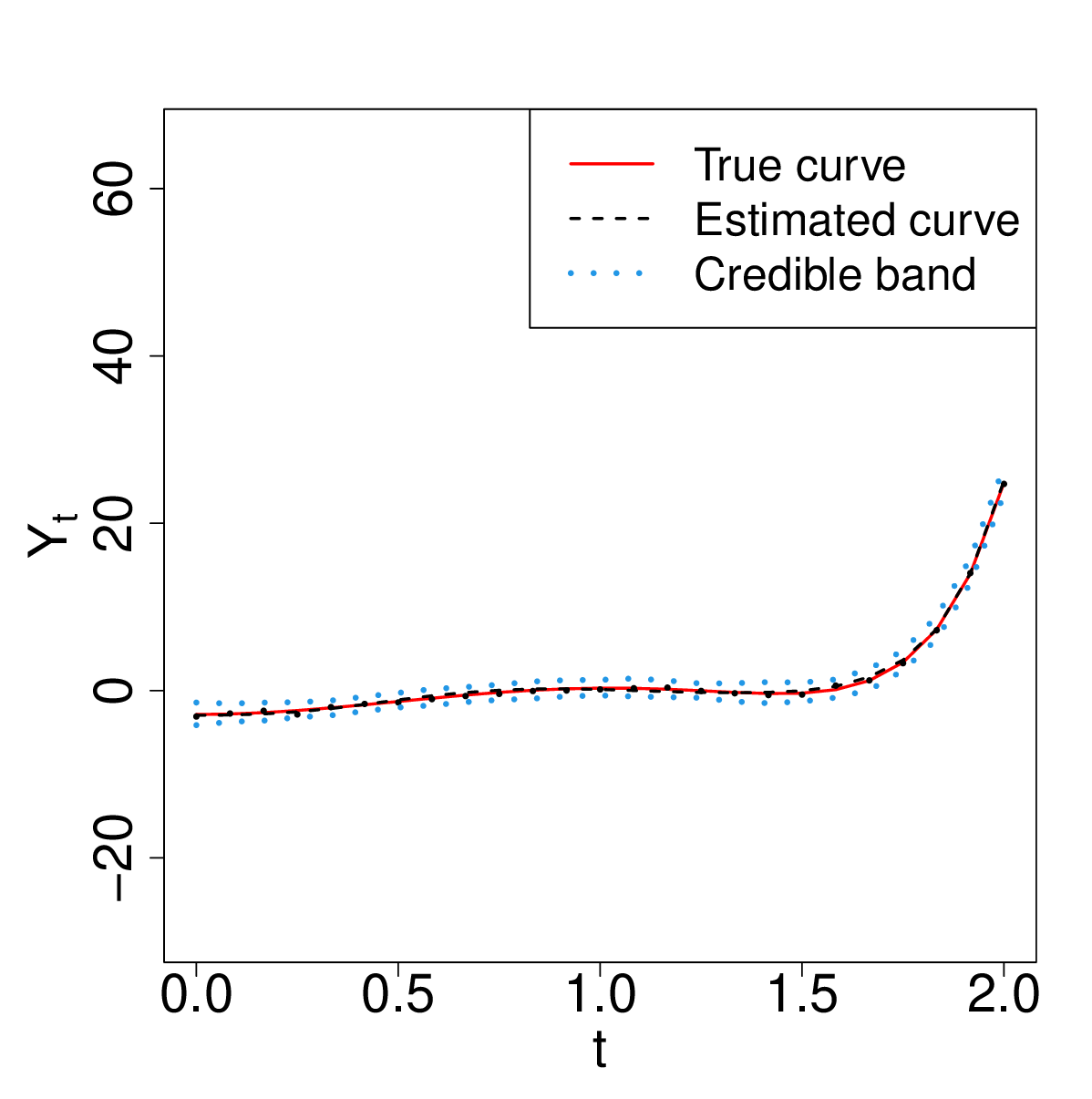}}&
		\subfloat[$2^{\text{nd}}$ functional response (Data with $\sigma=0.2$).]{\includegraphics[scale=0.18]{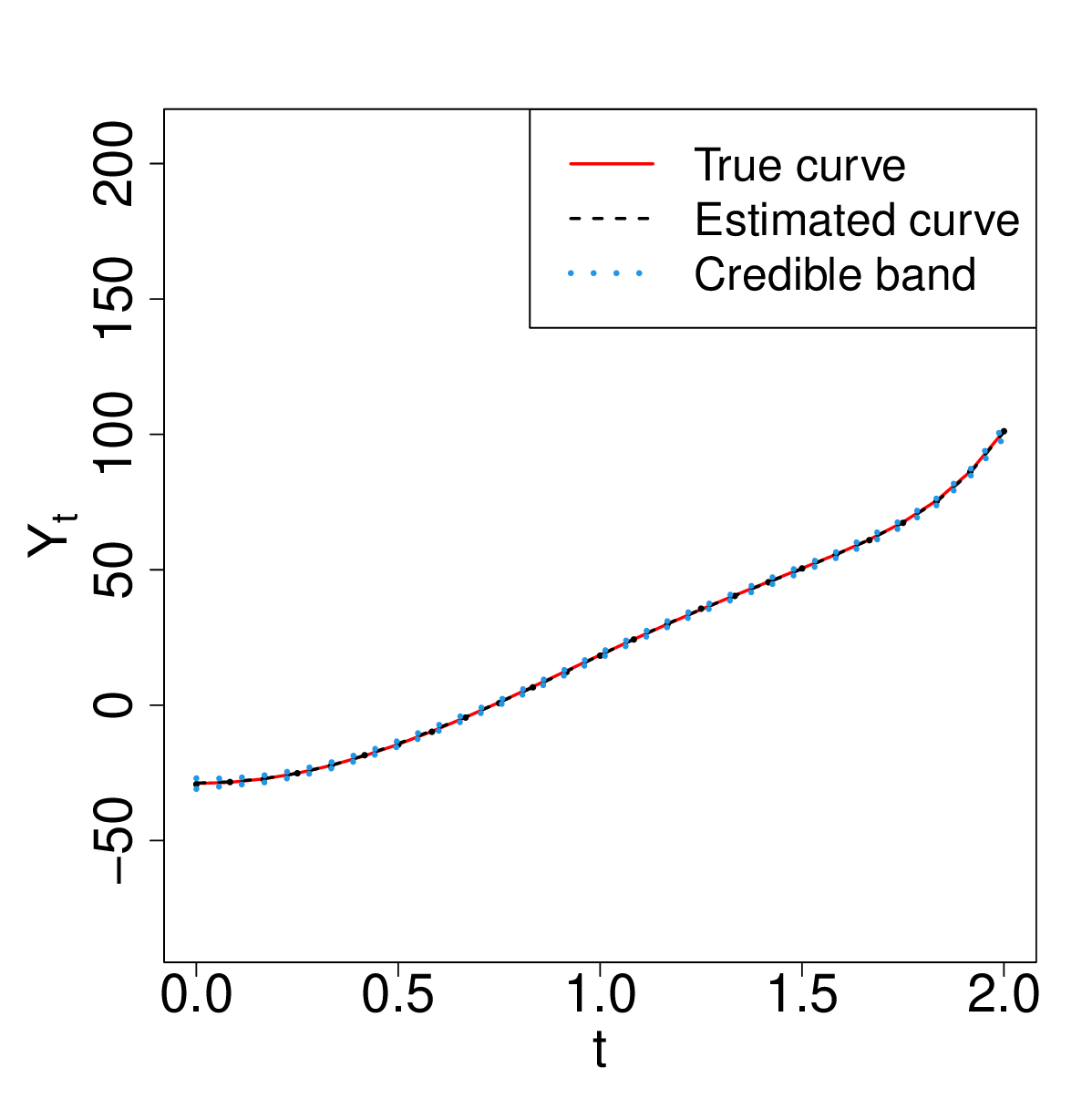}}&
		\subfloat[$3^{\text{rd}}$ functional response (Data with $\sigma=0.2$).]{\includegraphics[scale=0.18]{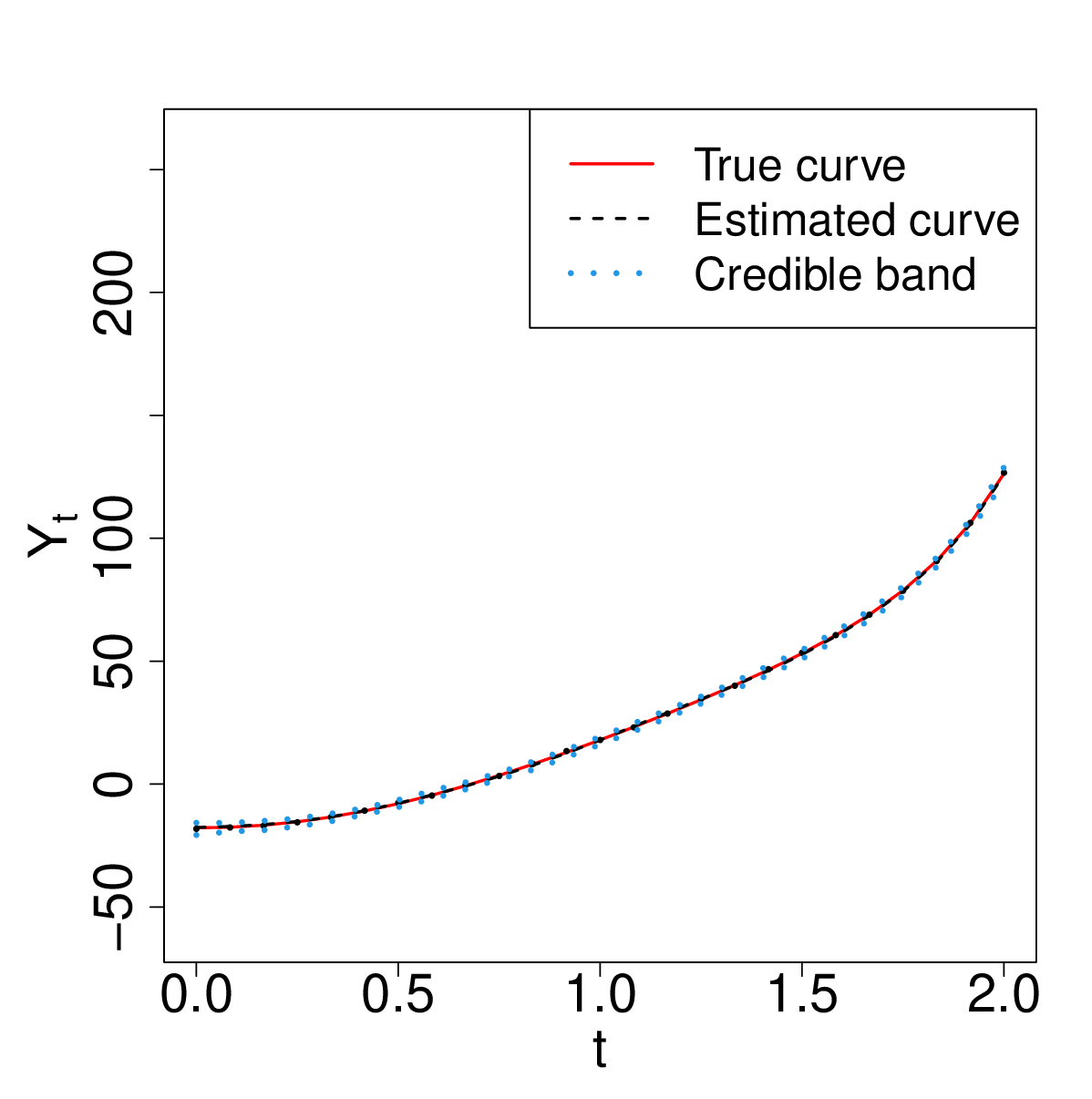}}\\
		\subfloat[$1^{\text{st}}$ functional response (Data with $\sigma=20$).]{\includegraphics[scale=0.18]{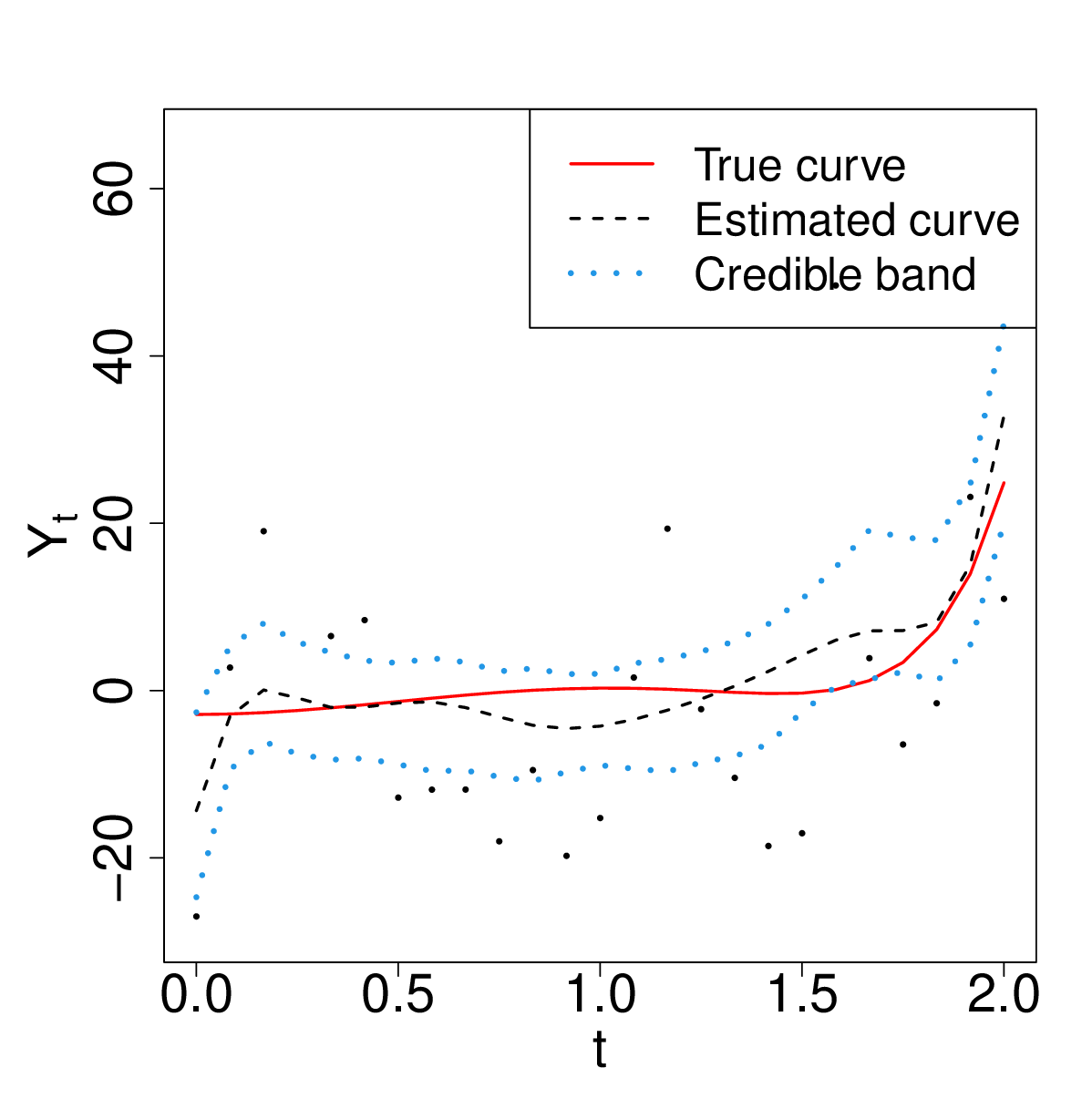}}&
		\subfloat[$2^{\text{nd}}$ functional response (Data with $\sigma=20$).]{\includegraphics[scale=0.18]{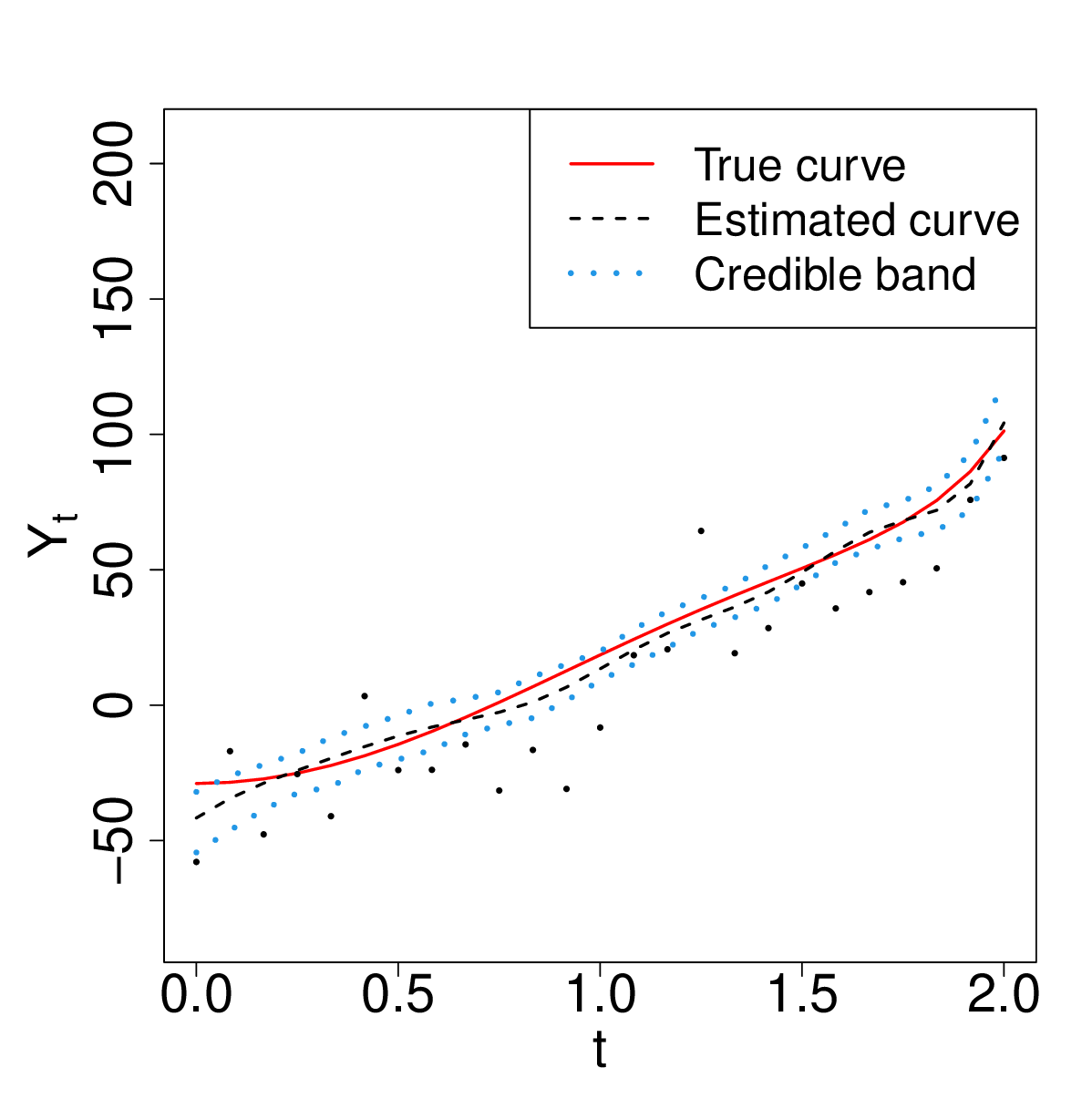}}&
		\subfloat[$3^{\text{rd}}$ functional response (Data with $\sigma=20$).]{\includegraphics[scale=0.18]{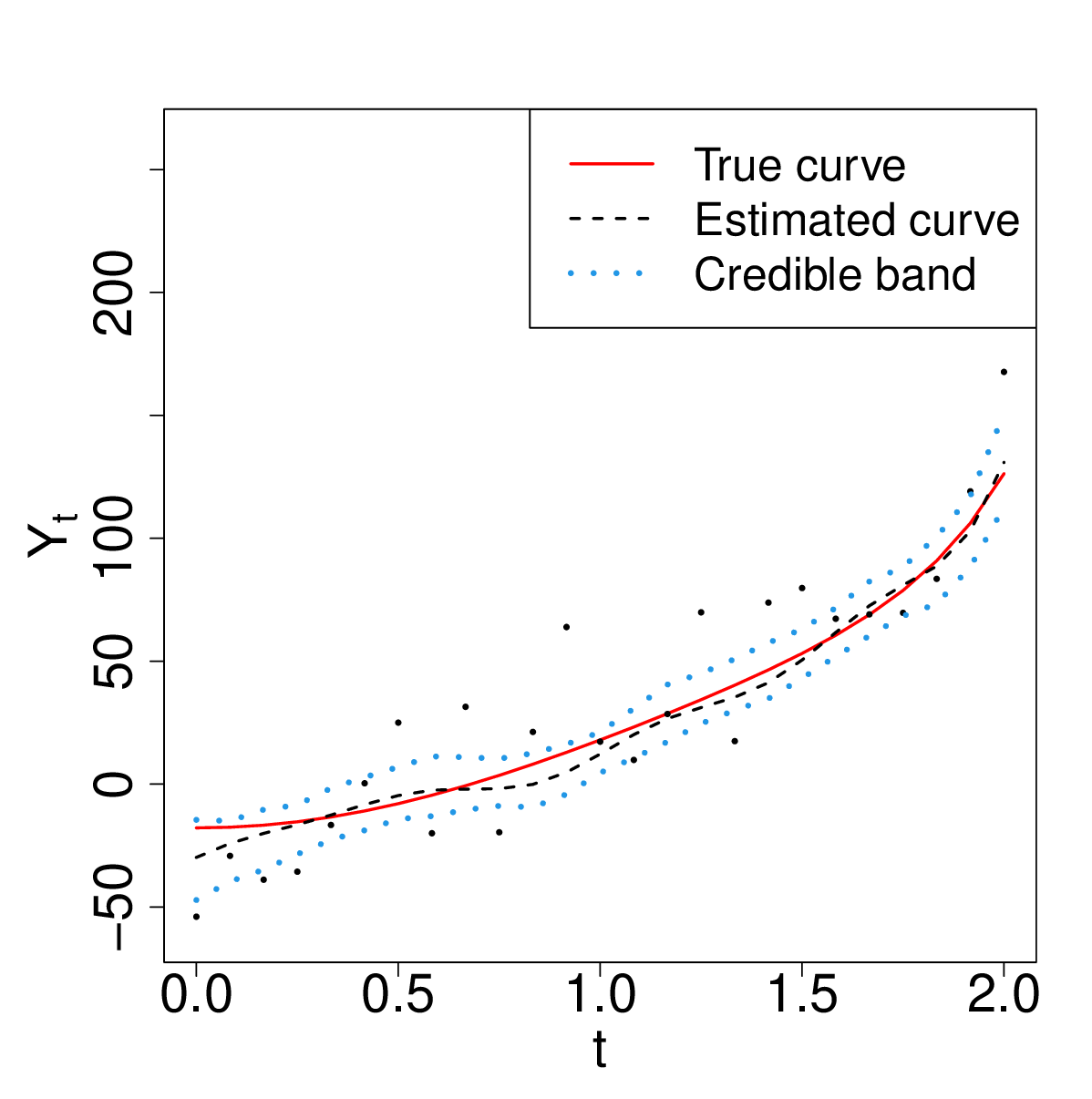}}  
	\end{tabular}
	\caption{Data and mean functional responses, $\beta_0(t_{ij})+g_i(t_{ij})$, for $i=1, 2, 3$ and $j=1\ldots,25$ (true and estimated along with credible bands), according to the data dispersion degree.}
	\label{obs_123}
\end{figure}

We can also observe in Figure \ref{coefsp_selecionados} that the proposed model demonstrates strong fitting performance, particularly with low dispersion data, where the true and estimated curves overlap almost perfectly. Although precision diminishes with higher noise levels, the estimates remain accurate. Notably, the model assigns near-zero estimates to the partial functional coefficients of excluded covariates, as shown in Supplementary Figure 5. This pattern reflects the dependence of the distributions of the basis expansion coefficients on the respective latent variables. Additionally, Figure \ref{obs_123} presents the first three estimated mean (or expected) functional response curves (among the total of \(m=10\)) alongside their credible bands, true curves, and observed data. The credible bands, constructed similarly to those in Figure \ref{coefsp_selecionados}, further demonstrate the model's ability to capture mean responses across different dispersion levels effectively.

In addition to goodness-of-fit, it is also interesting to assess the variable selection performance of each model configuration across the one hundred replications. Table \ref{tab:props} presents the proportion among the 100 replications that each functional coefficient estimate ($\hat{Z}_l\hat{\beta}_l(.)$, for $l =1,\ldots, 6$) was different from zero. For data with a low dispersion level ($\sigma=0.2$), we can observe in Table \ref{tab:props} that all tested model configurations correctly performed the selection in all replications. However, the selection of data with higher dispersion ($\sigma=20$) is not perfect in some replications, although it remains highly accurate.

\begin{table}%[!htb]
	\centering
    		\caption{Proportions of non-zero estimated functional coefficients among the 100 replications for the various model configurations tested ($\mu$ as a parameter or hyperparameter), according to the data dispersion degree ($\sigma=0.2$ and $\sigma=20$).}
	{\fontsize{7.5}{12}\selectfont 
		\begin{tabular}{cc|ccccccccc|c}\hline
			\multirow{2}{*}{$\sigma$} &
			\multicolumn{1}{c|}{\multirow{2}{*}{$\hat{Z}_{l}\hat{\beta}_{l}(.)$}} &
			\multicolumn{9}{c}{$\mu$ (Hyperparameter)} &
			\multicolumn{1}{|c}{\multirow{2}{*}{$\mu$ (Parameter)}}\\\cmidrule{3-11}
			& \multicolumn{1}{c|}{} & 0.1  & 0.2  & 0.3  & 0.4  & 0.5  & 0.6  & 0.7  & 0.8  & 0.9  & \multicolumn{1}{c}{} \\\hline
			\multirow{6}{*}{0.2} & $\hat{Z}_{1}\hat{\beta}_{1}(.)$          & 0    & 0    & 0    & 0    & 0    & 0    & 0    & 0    & 0    & 0                    \\
			&$\hat{Z}_{2}\hat{\beta}_{2}(.)$          & 0    & 0    & 0    & 0    & 0    & 0    & 0    & 0    & 0    & 0                    \\
			& $\hat{Z}_{3}\hat{\beta}_{3}(.)$          & 1    & 1    & 1    & 1    & 1    & 1    & 1    & 1    & 1    & 1                    \\
			& $\hat{Z}_{4}\hat{\beta}_{4}(.)$          & 0    & 0    & 0    & 0    & 0    & 0    & 0    & 0    & 0    & 0                    \\
			& $\hat{Z}_{5}\hat{\beta}_{5}(.)$          & 1    & 1    & 1    & 1    & 1    & 1    & 1    & 1    & 1    & 1                    \\
			& $\hat{Z}_{6}\hat{\beta}_{6}(.)$          & 0    & 0    & 0    & 0    & 0    & 0    & 0    & 0    & 0    & 0                    \\\hline
			\multirow{6}{*}{20}  & $\hat{Z}_{1}\hat{\beta}_{1}(.)$          & 0.01 & 0.03 & 0.03 & 0.04 & 0.04 & 0.04 & 0.05 & 0.05 & 0.13 & 0.01                 \\
			& $\hat{Z}_{2}\hat{\beta}_{2}(.)$          & 0.01 & 0.02 & 0.02 & 0.02 & 0.02 & 0.02 & 0.05 & 0.07 & 0.12 & 0.01                 \\
			& $\hat{Z}_{3}\hat{\beta}_{3}(.)$          & 1    & 1    & 1    & 1    & 1    & 1    & 1    & 1    & 1    & 1                    \\
			& $\hat{Z}_{4}\hat{\beta}_{4}(.)$          & 0    & 0    & 0.01 & 0.01 & 0.02 & 0.02 & 0.02 & 0.04 & 0.09 & 0                    \\
			& $\hat{Z}_{5}\hat{\beta}_{5}(.)$          & 0.99 & 1    & 1    & 1    & 1    & 1    & 1    & 1    & 1    & 0.99                 \\
			& $\hat{Z}_{6}\hat{\beta}_{6}(.)$          & 0.03 & 0.03 & 0.03 & 0.04 & 0.03 & 0.04 & 0.04 & 0.05 & 0.09 & 0.03         \\\hline       
		\end{tabular}
		\label{tab:props}
	}
\end{table}

Table \ref{tab:props} shows that except for configurations with $\mu$ as a parameter and as hyperparameter with $\mu=0.1$ when $\sigma=20$, all other configurations (for both $\sigma=0.2$ and $\sigma=20$) correctly selected the covariates associated with the partial functional coefficients $\hat{\beta}_{3}(.)$ and $\hat{\beta}_{5}(.)$ in all replications, allowing us to conclude that the model has no difficulty in selecting the partial functional coefficients that should remain in the functional regression. Furthermore, we can see that most of the selection errors committed in the scenario with $\sigma=20$ happened when the model failed to exclude a coefficient that should have been excluded. The proportions associated with the coefficients that should be excluded from the regression are generally below 7\%. However, it is worth noting that slightly higher proportions were observed for the configuration that considers $\mu=0.9$.
\begin{figure}[!htb]
	\centering
	\begin{tabular}{ccc}
		\subfloat[$\sigma=0.2$, $\mu=0.3$.]{\includegraphics[scale=0.18]{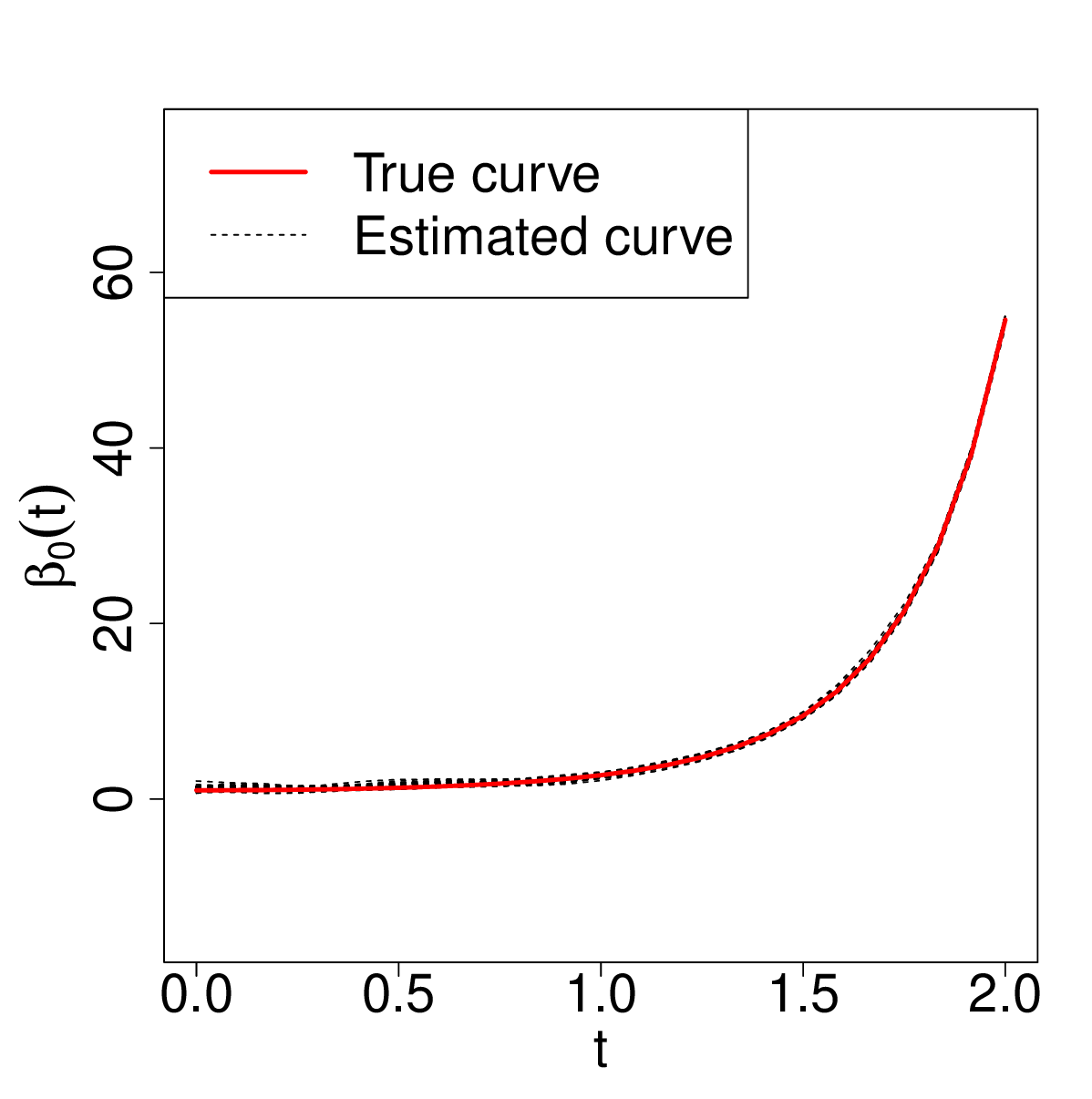}}&
		\subfloat[$\sigma=0.2$, $\mu=0.3$.]{\includegraphics[scale=0.18]{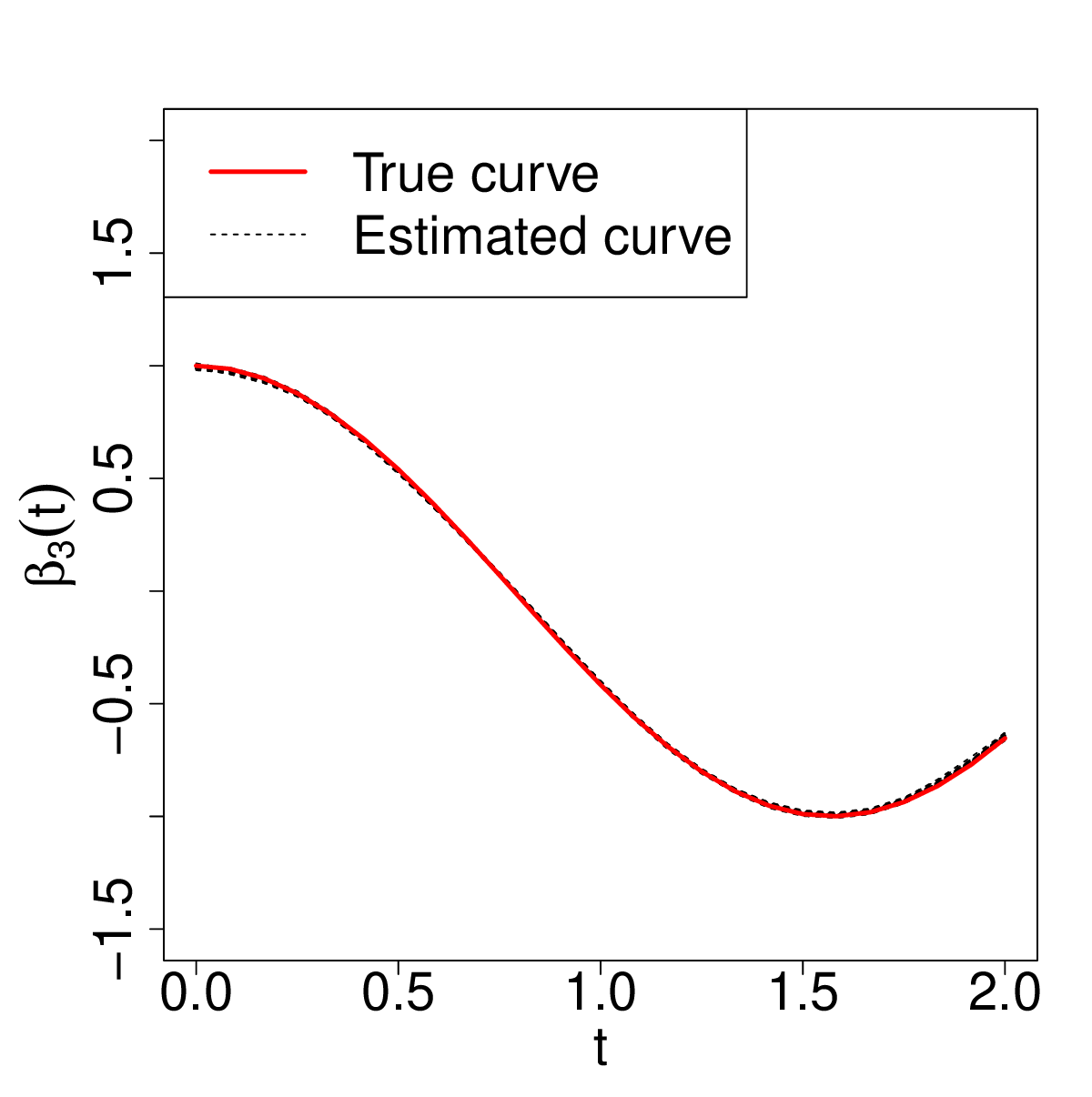}}&
		\subfloat[$\sigma=0.2$, $\mu=0.3$.]{\includegraphics[scale=0.18]{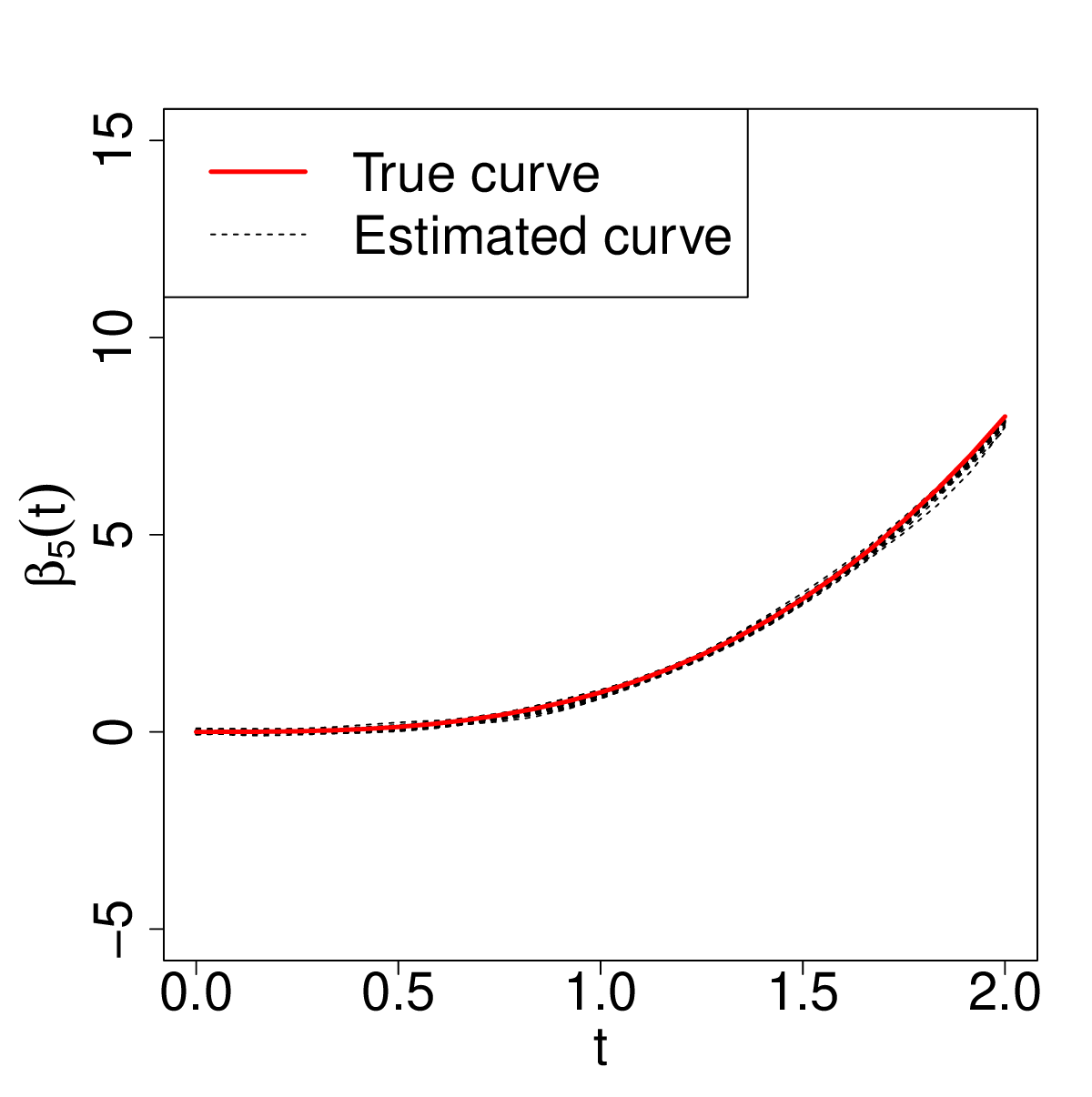}}\\
		\subfloat[$\sigma=20$, $\mu=0.5$.]{\includegraphics[scale=0.18]{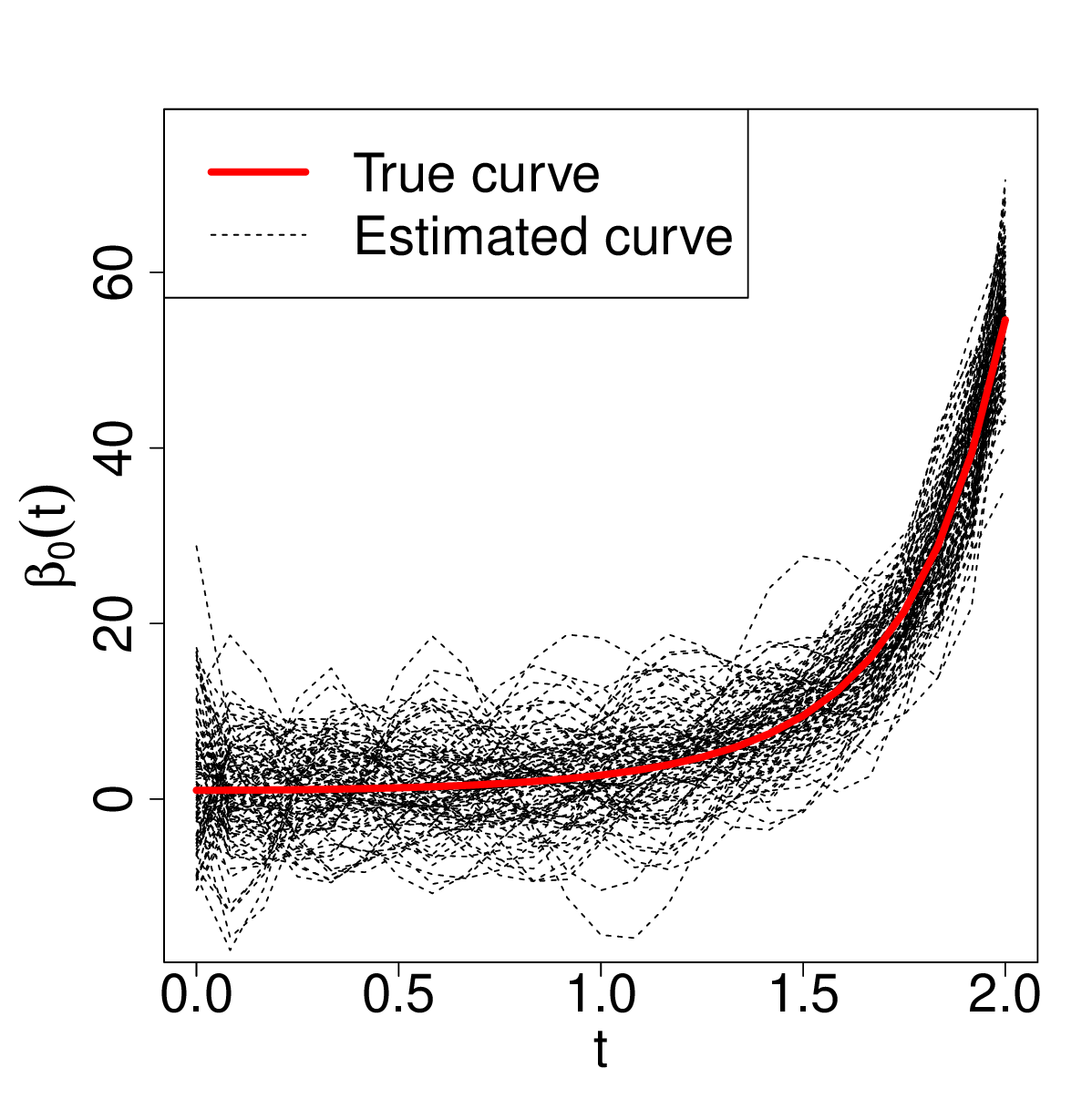}}&
		\subfloat[$\sigma=20$, $\mu=0.5$.]{\includegraphics[scale=0.18]{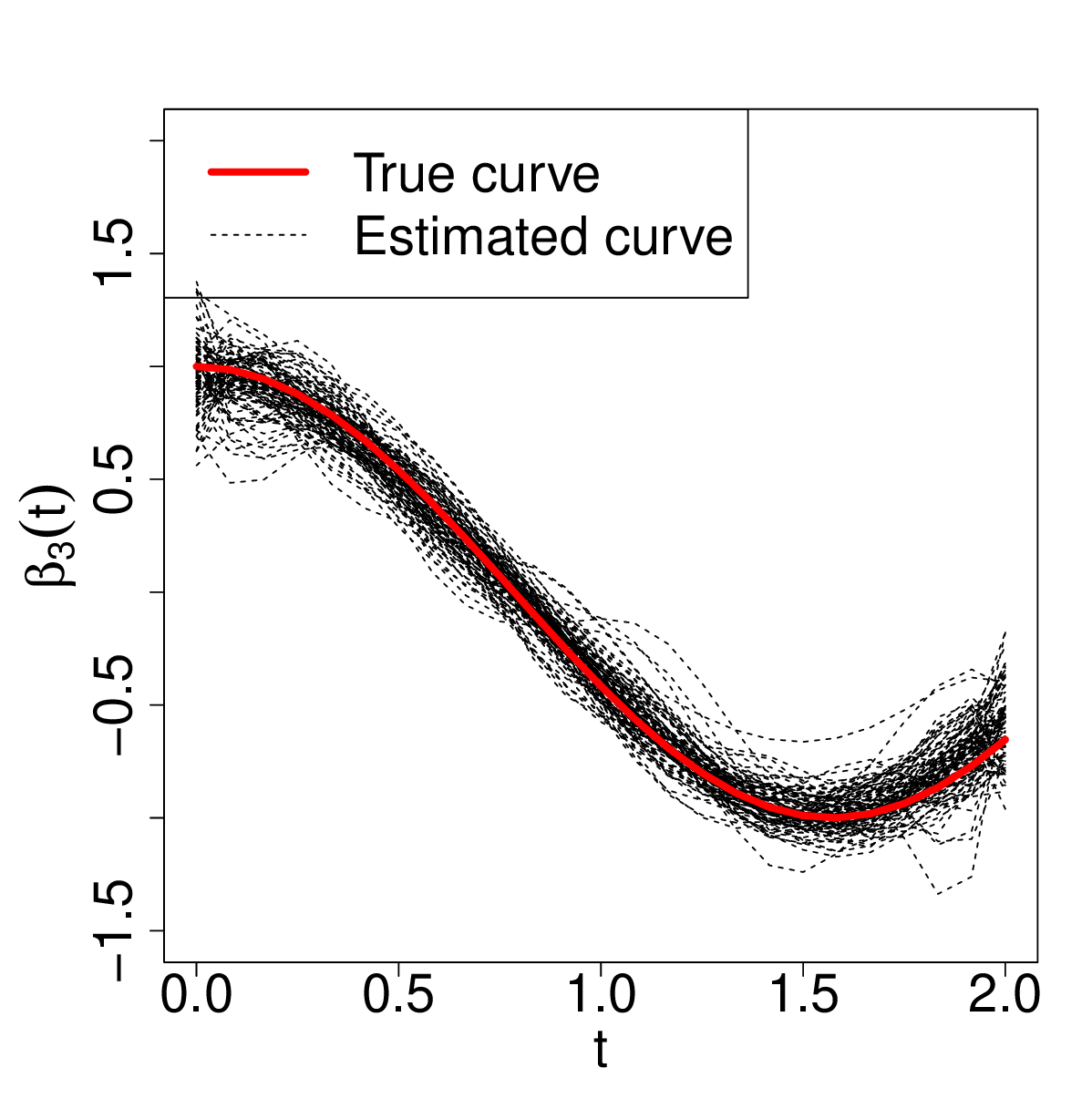}}&
		\subfloat[Data with $\sigma=20$, $\mu=0.5$.]{\includegraphics[scale=0.18]{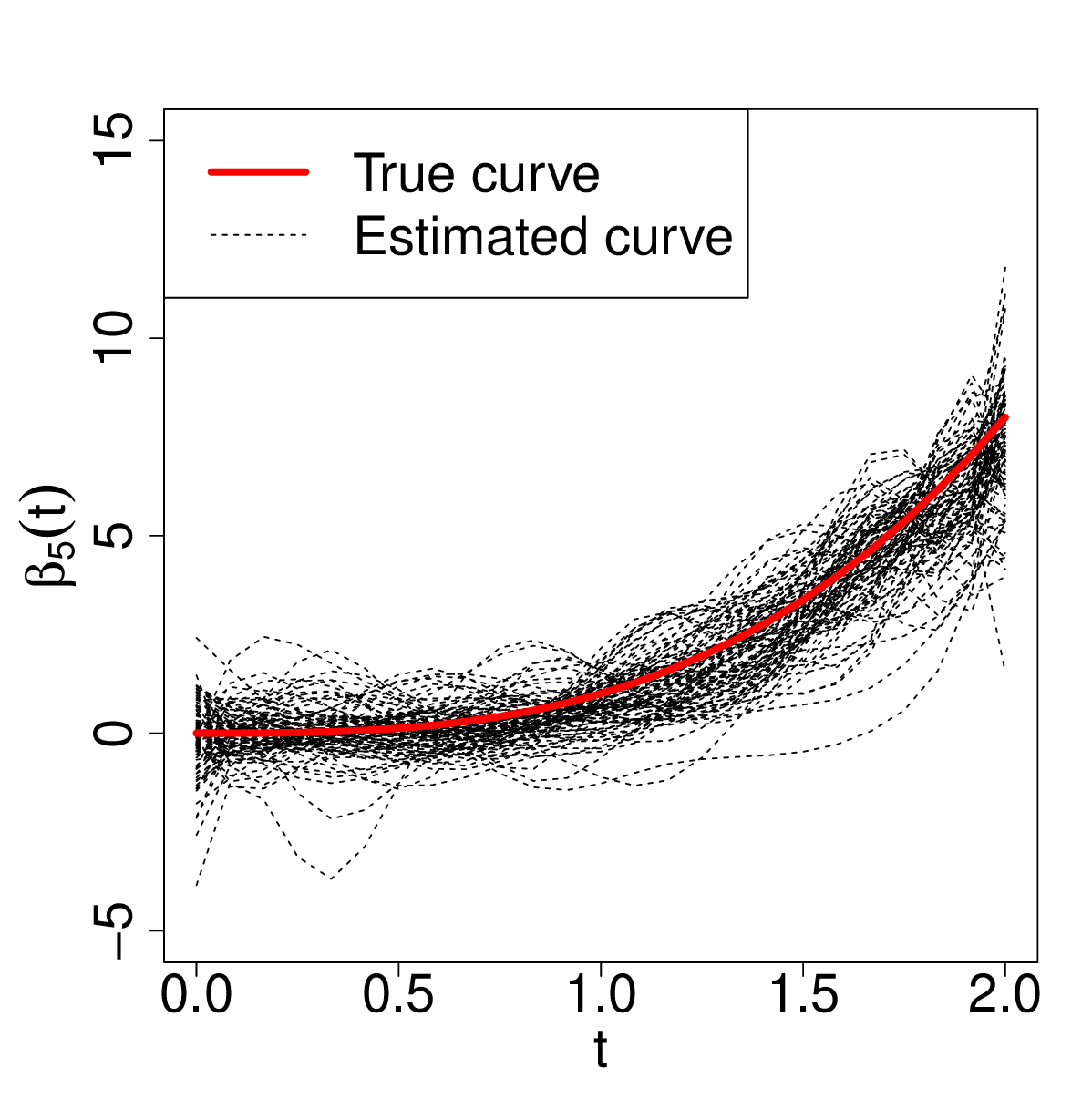}}		
	\end{tabular}
	\caption{Intercept and partial functional coefficients ($3^{\text{rd}}$ and $5^{\text{th}}$, true in red and estimated for the replications in which the coefficients were selected in black) from the model that considers $\mu$ as a hyperparameter (top row $\mu=0.3$, bottom row $\mu=0.5$), according to the data dispersion degree (top row $\sigma=0.2$, bottom row  $\sigma=20$).}
	\label{coefs_selecionados_simulacao}
\end{figure}

Figure \ref{coefs_selecionados_simulacao} shows the estimates of the partial functional coefficients that the proposed model predominantly selected ($\hat{\beta}_{3}(.)$ and $\hat{\beta}_{5}(.)$), as well as the estimates obtained in each replication for the functional intercept. The results in this figure correspond to the model that considers $\mu$ as a hyperparameter with a specific value of $\mu$ for each data dispersion level. We chose $\mu$ for each dispersion level as the value that resulted in the highest mean value of the performance metric in \eqref{eq:r2_PARTE2} over the one hundred replications. Thus, the model with $\mu=0.3$ was used for data with $\sigma=0.2$ and the model with $\mu=0.5$ for data with $\sigma=20$. When analyzing the intercept and partial functional coefficient results, we notice that the curves almost perfectly overlap in the scenario considering the data with low dispersion ($\sigma = 0.2$), even when plotting one hundred estimates. On the other hand, the estimated curves for the scenario with large dispersion ($\sigma=20$) are more dispersed but follow the same trend.

Similarly, Supplementary Figure 6 shows the estimated partial functional coefficients ($\hat{\beta}_{3}(.)$ and $\hat{\beta}_{5}(.)$) and functional intercept for the case when the model considers $\mu$ as a parameter. The conclusions are similar to those obtained from Figure \ref{coefs_selecionados_simulacao}, showing that the version with more layers in the hierarchical model does not differ significantly from the version that considers $\mu$ fixed.

To complement the analysis of the proposed methodology through simulations, Supplementary Figure 7 presents the estimates of all the other partial functional coefficients ($\hat{\beta}_{1}(.)$, $\hat{\beta}_{2}(.)$, $\hat{\beta}_{4}(.)$ and $\hat{\beta}_{6}(.)$) from the model that considers $\mu$ as a hyperparameter that are associated with the excluded variables, according to the data dispersion degree (Supplementary Figure 5 presents the corresponding plots obtained by the model that considers $\mu$ as a parameter). Regardless of the data dispersion degree and the approach to $\mu$, Supplementary Figures 5 and 7 show that all the functional estimates in question are around zero.

\subsubsection{Comparative analysis with grLASSO, grSCAD, grMCP and BGLSS}
\label{grLASSO_grSCAD_grMCP}

As emphasized in Section \ref{mp_parte2} , there are some methods that can compete with the method proposed in this paper for variable selection in FOSR. Through basis expansion of the functional coefficients, FOSR can be dealt with as a conventional regression problem, thus enabling the use of techniques developed for cross-sectional data. A relevant fact is that as the coefficients are expanded through basis functions, the characterization of each functional coefficient ($\beta_{l}(.)$) is determined by a distinct group of scalar coefficients ($\bb_{.l }$’s) from the basis expansion.

Among the frequentist methods that can be applied in this context of FOSR with functional coefficients expanded via basis functions, we can highlight the \textit{group Least Absolute Shrinkage and Selection Operator} (grLASSO) \citep{groupLASSO}, the \textit{group Smoothly Clipped Absolute Deviation} grSCAD) \citep{microarray} and the \textit{group Minimax Concave Penalty} (grMCP) \citep{Breheny12}. These methods are generalizations for group structures and derive from LASSO (Least Absolute Shrinkage and Selection Operator) \citep{lasso}, SCAD (Smoothly Clipped Absolute Deviation) penalized regression \citep{SCAD} and MCP (Minimax Concave Penalty) penalized regression \citep{MCP}, which are regularization and variable selection methods for regression analysis.

Thus, the regularization and selection methods in groups described above (grLASSO, grSCAD and grMCP) are considered in this section to carry out a comparative analysis with our proposed method. To complete the comparisons with other existing methods, we also consider the Bayesian Group LASSO with Spike and Slab prior (BGLSS) method \citep{grBLASSO}. In order to obtain the results of the grLASSO, grSCAD and grMCP models, we used the \texttt{grpreg} function from the \textit{grpreg} package \citep{grpreg} of the R statistical software, while the \texttt{BGLSS} function of the R package \textit{MBSGS} \citep{MBSGS} was used to implement BGLSS.

We used synthetic data replications generated as described in Subsection \ref{sec:data_generation} and fitted our model using $K=10$ B-spline bases for each functional coefficient. Finally, it is worth noting that the models are compared in two scenarios, one simulating data with low dispersion ($\sigma=0.2$) and another with higher dispersion ($\sigma=20$). For each dispersion scenario, 100 simulated datasets were used. Implementing the competing methods requires building appropriate response vectors and design matrices. For the frequentist methods, a grid of values for the regularization parameter $\lambda$ is also needed. In Section 2 of the Supplementary Material, we provide all the details for implementing the competing methods.

\begin{figure}[!htb]
	\centering
	\begin{tabular}{c}	\subfloat[Data with $\sigma=0.2$.]{\includegraphics[scale=0.3]{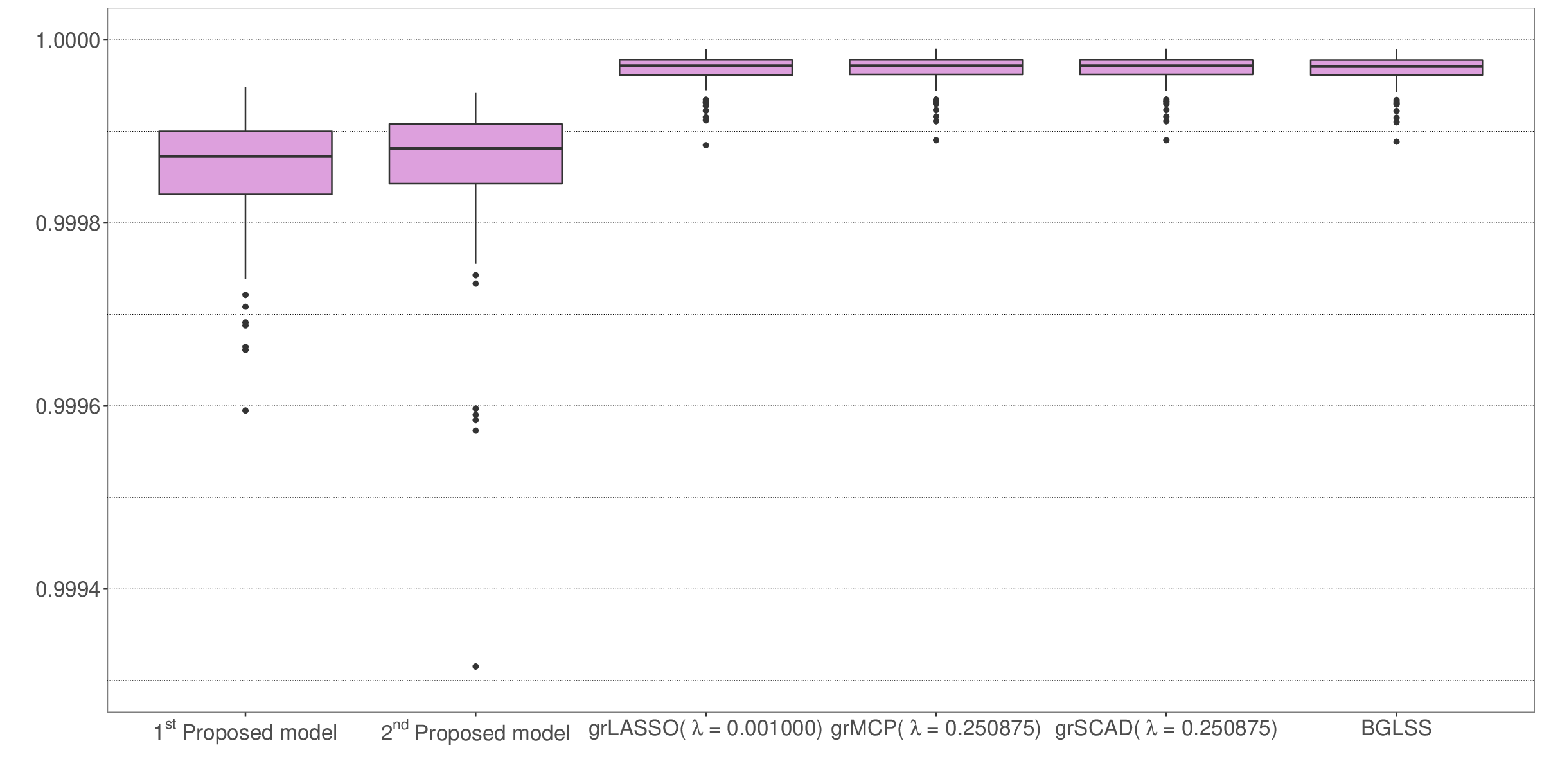}}\\
		\subfloat[Data with $\sigma=20$.]{\includegraphics[scale=0.3]{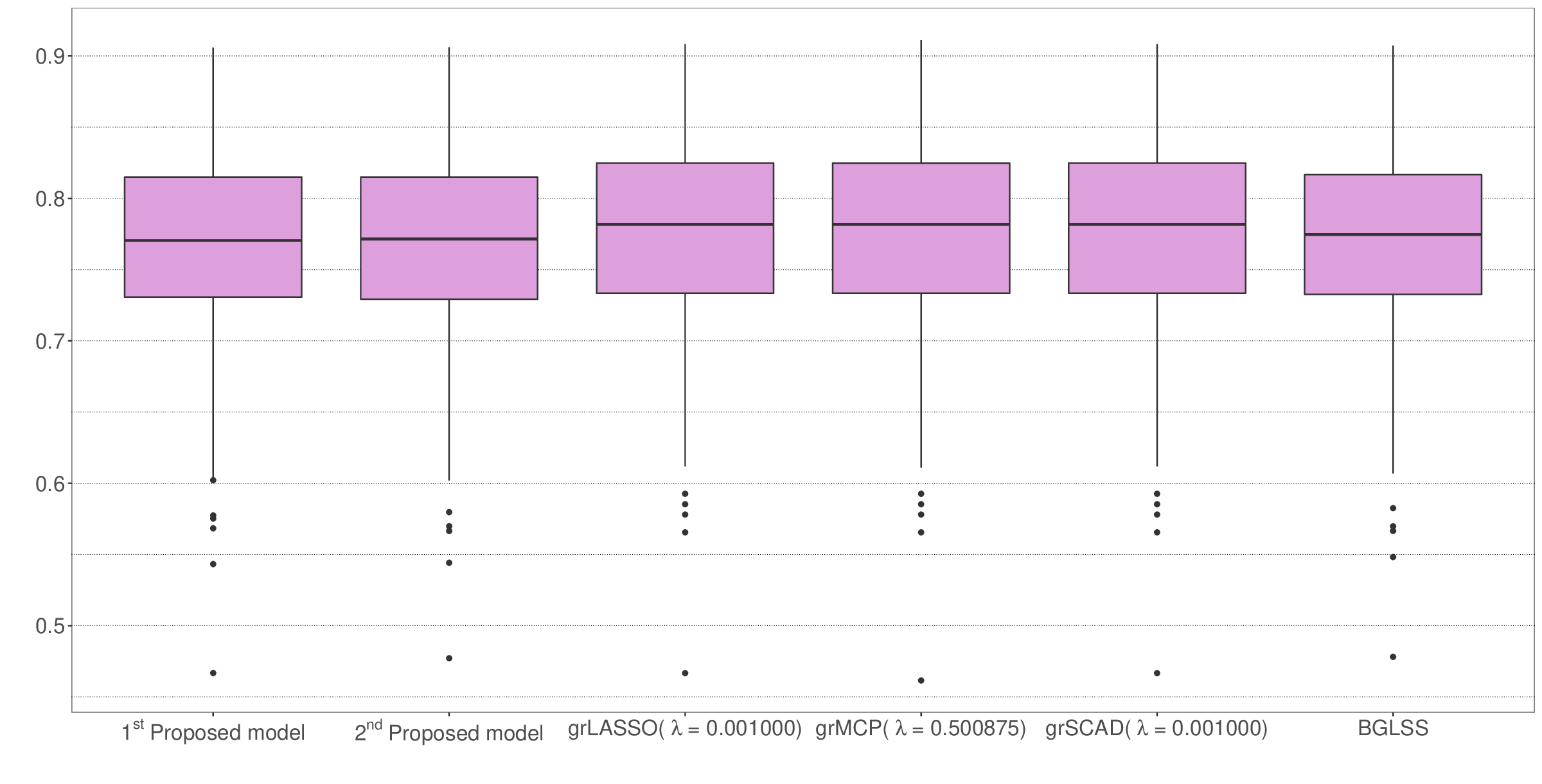}}
	\end{tabular}
	\caption{Boxplots of the metric \eqref{eq:r2_PARTE2}, according to the model used and the data dispersion degree. The ``$1^{\text{st}}$ Proposed Model'' is considered to be the version with  $\mu$ as a parameter, while the ``$2^{\text{nd}}$ Proposed Model'' is the version with $\mu$ as a hyperparameter ($\mu=0.3$ for $\sigma=0.2$ and $\mu=0.5$ for $\sigma=20$). For the competing methods grLASSO, grMCP, and grSCAD, the results correspond to the best $\lambda$ from the tested grid.}
	\label{grBLASSO}
\end{figure}

Supplementary Figures 8 and 9 show boxplots of metric \eqref{eq:r2_PARTE2} and MSE, respectively, for each competing model, grouped by data dispersion degree and tested $\lambda$ values. Regardless of the metric (\eqref{eq:r2_PARTE2} or MSE), the fit quality begins to decline beyond a certain $\lambda$, indicating that stronger regularization leads to suboptimal results when $\lambda$ becomes too large. Figure \ref{grBLASSO} and Supplementary Figure 10 compare our proposed model with grLASSO, grMCP, grSCAD, and BGLSS based on metric \eqref{eq:r2_PARTE2} and MSE, respectively, considering the best $\lambda$ from the tested grid for grLASSO, grMCP, and grSCAD.

In order to demonstrate the capacity of each frequentist competing model to select the correct covariates across different values of $\lambda$ and dispersion levels, Supplementary Tables 1 and 2 present the proportion of times that each functional coefficient estimate was different from zero among the total of 100 replications. We notice that none of the models performed variable selection at the lowest regularization level ($\lambda = 0.001$) across all three competing methods (all proportions were equal to 1). For the other $\lambda$ values tested, the models began selecting variables. Such selection for data with low dispersion is predominantly correct from $\lambda=0.250$, while for data with large dispersion from $\lambda = 2.005$. However, the non-zero proportions in Supplementary Table 2 associated with the fifth functional coefficient become lower for larger values of $\lambda$.

%lower than they should be

It is noted (Supplementary Table 2 and Supplementary Figure 9 (b)) that the competing models are not able to balance very well the accuracy in the selection in the scenario with greater data dispersion while preserving the quality of the fit according to the metrics under study, meaning that they need to lose fit quality and increase the regularization to gain better accuracy in the selection. The last finding from evaluating Supplementary Tables 1 and 2 is that the grSCAD and grMCP models stand out from the grLASSO model in the scenario with low data dispersion ($\sigma=0.2$), better selecting the variables for the different $\lambda$ values tested.

Tables \ref{tab:melhores02_parte2} and \ref{tab:melhores20_parte2} allow a brief comparison regarding the capacity to select the correct covariates among the best configurations tested for the competing models, as well as for our proposed model and for the BGLSS. The best model configuration (for the frequentist methods) was chosen according to the performance metric \eqref{eq:r2_PARTE2}. For this, the average metric value over the one hundred replications was calculated for each model configuration, and we then selected the model with the configuration that returned the highest of these averages. 

\begin{table}
	\centering
    		\caption{Proportion of non-zero estimates for each functional coefficient ($\hat{Z}_{l}\hat{\beta}_{l}(.)$ for our proposed model and $\hat{\beta}_{l}(.)$ for the competing methods) among the 100 replications, according to the best tested model configuration. Data with $\sigma=0.2$.}

		\begin{tabular}{>{\centering\arraybackslash}p{1.5cm}|
                >{\centering\arraybackslash}p{2.7cm}
                >{\centering\arraybackslash}p{2.8cm}
                >{\centering\arraybackslash}p{2cm}
                >{\centering\arraybackslash}p{1.8cm}
                >{\centering\arraybackslash}p{1.3cm}
                >{\centering\arraybackslash}p{1.3cm}}\hline
			\multirow{3}{*}{\begin{tabular}{c}
					$\hat{Z}_{l}\hat{\beta}_{l}(.)$  \\or\\ $\hat{\beta}_{l}(.)$ 
			\end{tabular}} &
			\multirow{3}{*}{\begin{tabular}{l}
					Proposed model\\($\mu=0.3$ and\\ $\lambda=\sqrt{2}$)
			\end{tabular}} &
			\multirow{3}{*}{\begin{tabular}{l}Proposed model\\ ($\mu$ as a parameter \\and $\lambda=\sqrt{2}$)\end{tabular}} &
			\multirow{3}{*}{\begin{tabular}{c}grLASSO\\ ($\lambda=$\\$0.001$)\end{tabular}} &
			\multirow{3}{*}{\begin{tabular}{c}grMCP\\ ($\lambda=$\\$0.250$)\end{tabular}} &
			\multirow{3}{*}{\begin{tabular}{c}grSCAD\\ ($\lambda=$\\$0.250$)\end{tabular}} &
			\multirow{3}{*}{\begin{tabular}{c}BGLSS\\\end{tabular}} \\
			&      &      &   &      &   \\
			&   &   &   &   &   \\\hline
			$l=1$ & 0 & 0 & 1 & 0 & 0 & 0 \\
			$l=2$ & 0 & 0 & 1 & 0 & 0 & 0.01\\
			$l=3$ & 1 & 1 & 1 & 1 & 1 & 1\\
			$l=4$& 0 & 0 & 1 & 0 & 0 & 0 \\
			$l=5$ & 1 & 1 & 1 & 1 & 1 & 0.99 \\
			$l=6$ & 0 & 0 & 1 & 0 & 0 & 0.01\\\hline
		\end{tabular}

		\label{tab:melhores02_parte2}
	
\end{table}

\begin{table}[htb]
	\centering
    		\caption{Proportion of non-zero estimates for each functional coefficient ($\hat{Z}_{l}\hat{\beta}_{l}(.)$ for our proposed model and $\hat{\beta}_{l}(.)$ for the competing methods) among the 100 replications, according to the best tested model configuration. Data with $\sigma=20$.}
	
	\begin{tabular}{>{\centering\arraybackslash}p{1.5cm}|
                >{\centering\arraybackslash}p{2.7cm}
                >{\centering\arraybackslash}p{2.8cm}
                >{\centering\arraybackslash}p{2cm}
                >{\centering\arraybackslash}p{1.8cm}
                >{\centering\arraybackslash}p{1.3cm}
                >{\centering\arraybackslash}p{1.3cm}}\hline
			\multirow{3}{*}{\begin{tabular}{c}
					$\hat{Z}_{l}\hat{\beta}_{l}(.)$ \\or\\ $\hat{\beta}_{l}(.)$
			\end{tabular}} &
			\multirow{3}{*}{\begin{tabular}{c}Proposed model\\ ($\mu=0.5$ and\\ $\lambda=\sqrt{2}$)\end{tabular}} &
			\multirow{3}{*}{\begin{tabular}{c}Proposed model\\ ($\mu$ as a parameter \\and $\lambda=\sqrt{2}$)\end{tabular}} &
			\multirow{3}{*}{\begin{tabular}{c}grLASSO\\ ($\lambda=$\\$0.001$)\end{tabular}} &
			\multirow{3}{*}{\begin{tabular}{c}grMCP\\ ($\lambda=$\\$0.500$)\end{tabular}} &
			\multirow{3}{*}{\begin{tabular}{c}grSCAD\\ ($\lambda=$\\$0.001$)\end{tabular}} &
			\multirow{3}{*}{\begin{tabular}{c}BGLSS\\\end{tabular}} \\
			&      &      &   &      &   \\
			&      &      &   &      &   \\\hline
			$l=1$ & 0.04 & 0.01 & 1 & 0.95 & 1 & 0.03\\
			$l=2$ & 0.02 & 0.01 & 1 & 0.96 & 1 & 0.01\\
			$l=3$ & 1    & 1    & 1 & 1    & 1 & 1\\
			$l=4$ & 0.02 & 0    & 1 & 0.94 & 1 & 0.01\\
			$l=5$ & 1    & 0.99 & 1 & 1    & 1 & 0.99\\
			$l=6$ & 0.03 & 0.03 & 1 & 1    & 1 & 0.02\\\hline
		\end{tabular}

		\label{tab:melhores20_parte2}
	
\end{table}

The results presented in Table \ref{tab:melhores02_parte2} show that most of the compared models correctly performed the covariate selection on low dispersion data ($\sigma=0.2$), except for grLASSO, which was unable to exclude any covariate in any of the hundred replications. When evaluating the selection performance for data with greater dispersion ($\sigma=20$), we can observe in Table \ref{tab:melhores20_parte2} that only the two versions of the proposed model ($\mu$ as a hyperparameter and as a parameter) and the BGLSS were able to identify the two true variables that were used in the construction of synthetic data ($3^{\text{rd}}$ and $5^{\text{th}}$) in almost all replications.

It is important to note that the value of \(\lambda = \sqrt{2}\) used in the various versions of the proposed model represents the lowest feasible setting for this parameter. Using a smaller value can lead to problems with the uniqueness of the stationary distribution due to two factors: low regularization and multicollinearity among the covariates. While this setting offers the minimum possible regularization, the proposed model versions still achieved a good fit to the data while accurately selecting the variables. In contrast, as mentioned earlier, the grLASSO, grSCAD, and grMCP models faced difficulties in balancing accurate variable selection with the quality of fit.

Finally, unlike the grLASSO, grSCAD and grMCP methods, the BGLSS method can select the variables well even in the case with large data dispersion ($\sigma = 20$) and is the only method among the ones considered here that can compete with our proposed methodology in terms of predictive quality (through the metric \eqref{eq:r2_PARTE2} and MSE) and variable selection capacity.

Figure \ref{grBLASSO} and Supplementary Figure 10 present the results regarding the performance metric in \eqref{eq:r2_PARTE2} and MSE, respectively, for the best model configurations. The metric \eqref{eq:r2_PARTE2}  was used as a reference for choosing the best model configurations to be compared in Figure \ref{grBLASSO}, as it was done for Tables \ref{tab:melhores02_parte2} and \ref{tab:melhores20_parte2}. Whereas, for Supplementary Figure 10, the MSE itself was the reference metric for defining the best model settings. Thus, the comparison via MSE becomes fair. Turning the analysis to the results from data with low dispersion ($\sigma = 0.2$), regardless of the metric being analyzed, slightly better performance of the competing methods can be observed in comparison to the two versions of the proposed model, especially when using the metric \eqref{eq:r2_PARTE2} as a reference. As for the data with greater dispersion, the models present similar goodness-of-fit and estimation accuracy. 

Note that for the BGLSS, in one replication with low dispersion ($\sigma=0.2$), the method had its worst performance, resulting in an outlier (for both evaluated metrics). Figure \ref{grBLASSO} and Supplementary Figure 10 omit this outlier since we adjusted the scale to focus on the interquartile range region to enable the comparison across the boxplots.

\begin{figure}
	\centering
	\begin{tabular}{cccc}
		\subfloat[{\scriptsize $\hat{\beta}_{3}(.)$ (proposed model, $\mu=0.3$).}]{\includegraphics[scale=0.15]{figures/beta3_02_ENG.eps}}&
		\subfloat[{\scriptsize $\hat{\beta}_{5}(.)$ (proposed model, $\mu=0.3$).}]{\includegraphics[scale=0.15]{figures/beta5_02_ENG.eps}}&
		\subfloat[{\scriptsize $\hat{\beta}_{3}(.)$ (proposed model, $\mu$ as a parameter).}]{\includegraphics[scale=0.15]{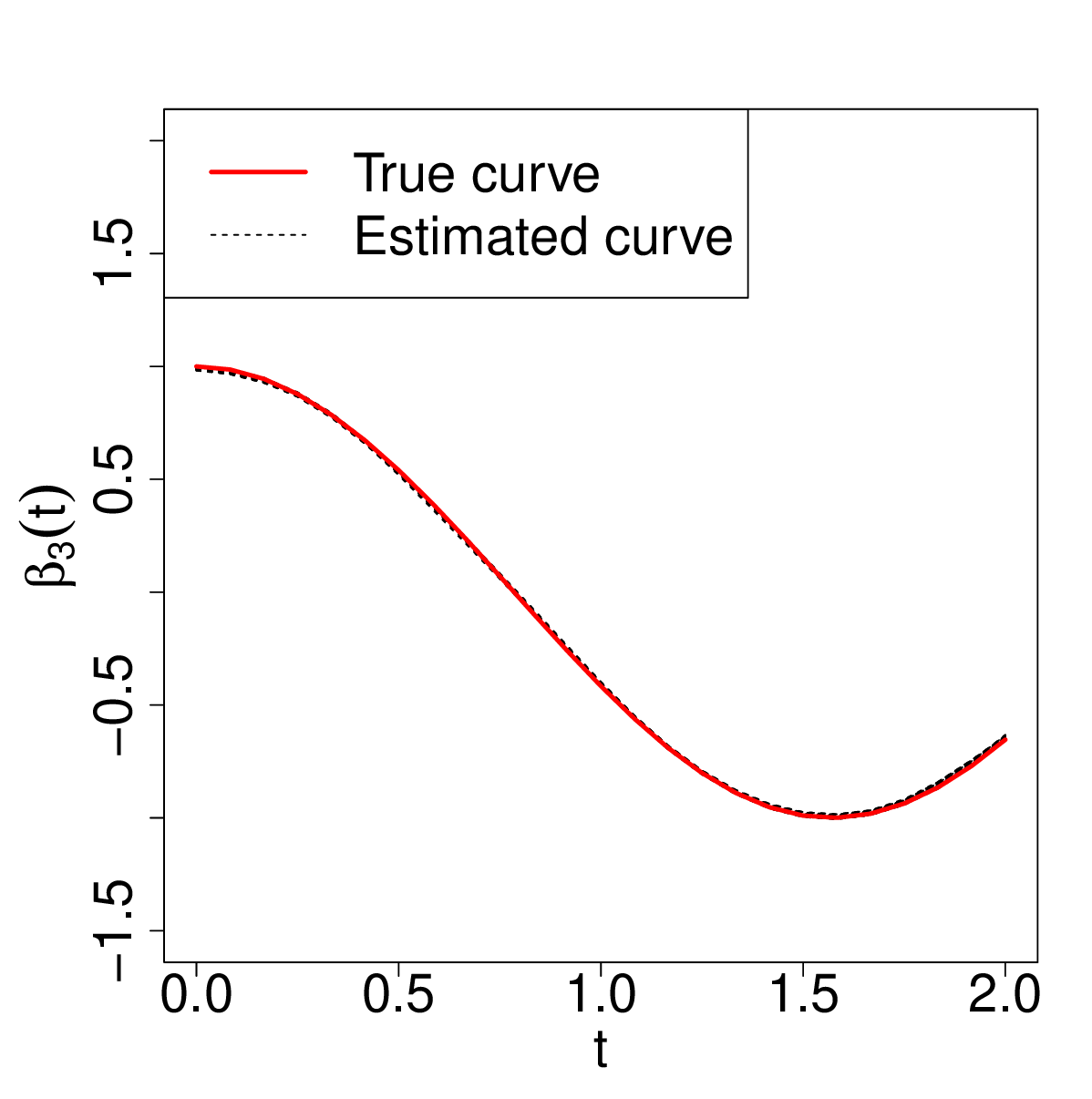}}&
		\subfloat[{\scriptsize $\hat{\beta}_{5}(.)$ (proposed model, $\mu$ as a parameter).}]{\includegraphics[scale=0.15]{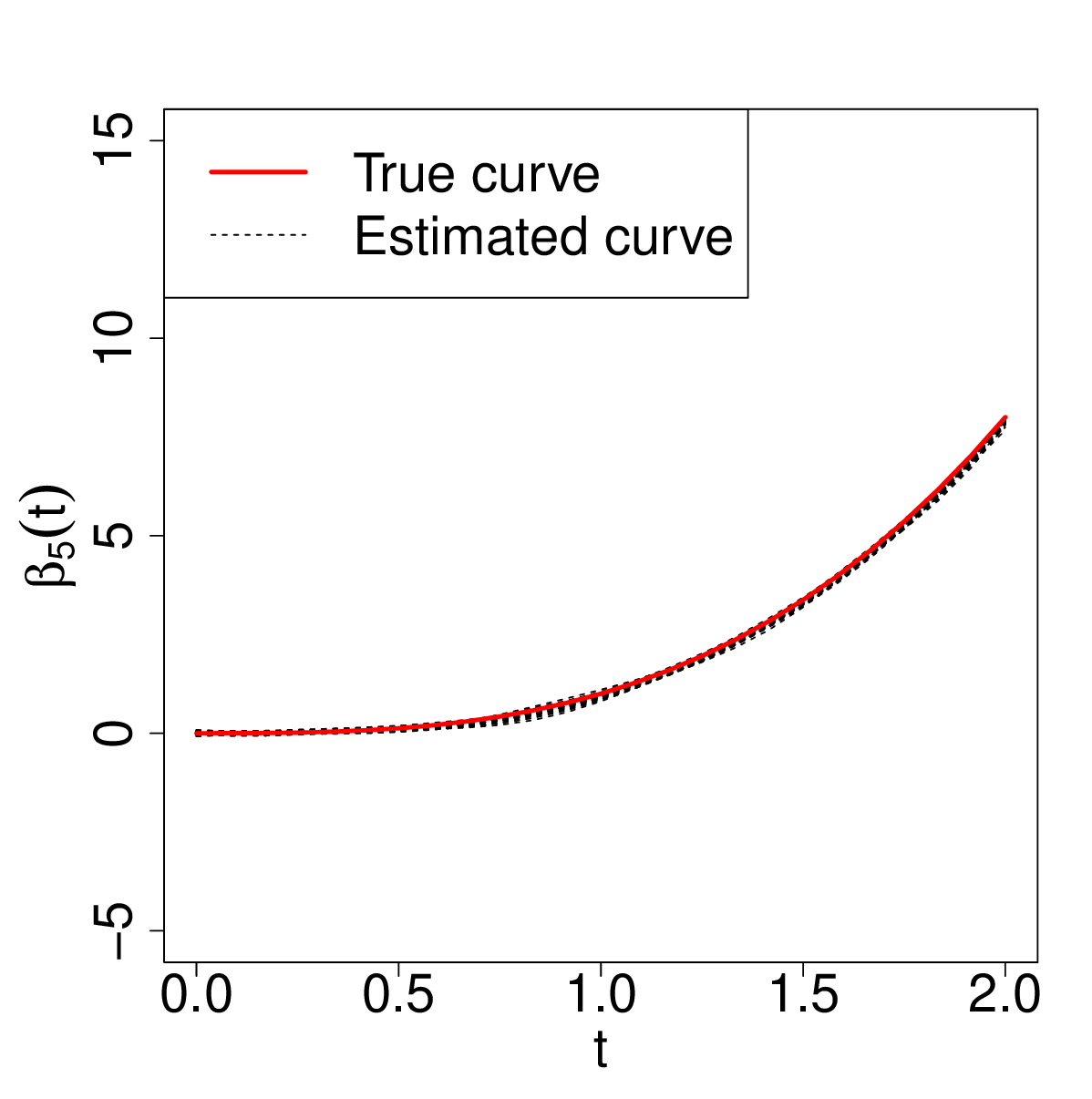}}\\
		\subfloat[{\scriptsize $\hat{\beta}_{3}(.)$ (grLASSO, $\lambda=0.001$).}]{\includegraphics[scale=0.15]{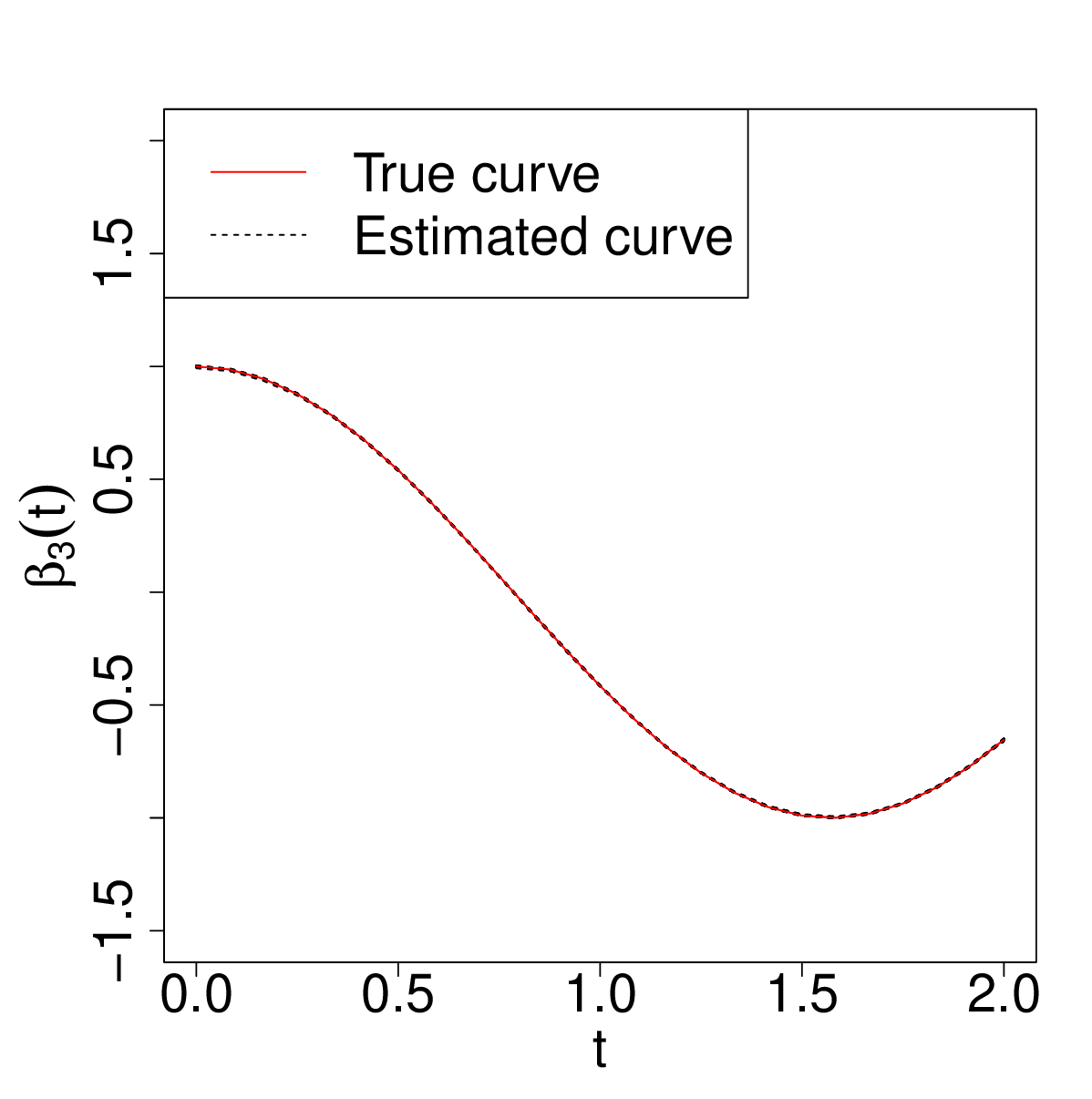}}&
		\subfloat[{\scriptsize $\hat{\beta}_{5}(.)$ (grLASSO, $\lambda=0.001$).}]{\includegraphics[scale=0.15]{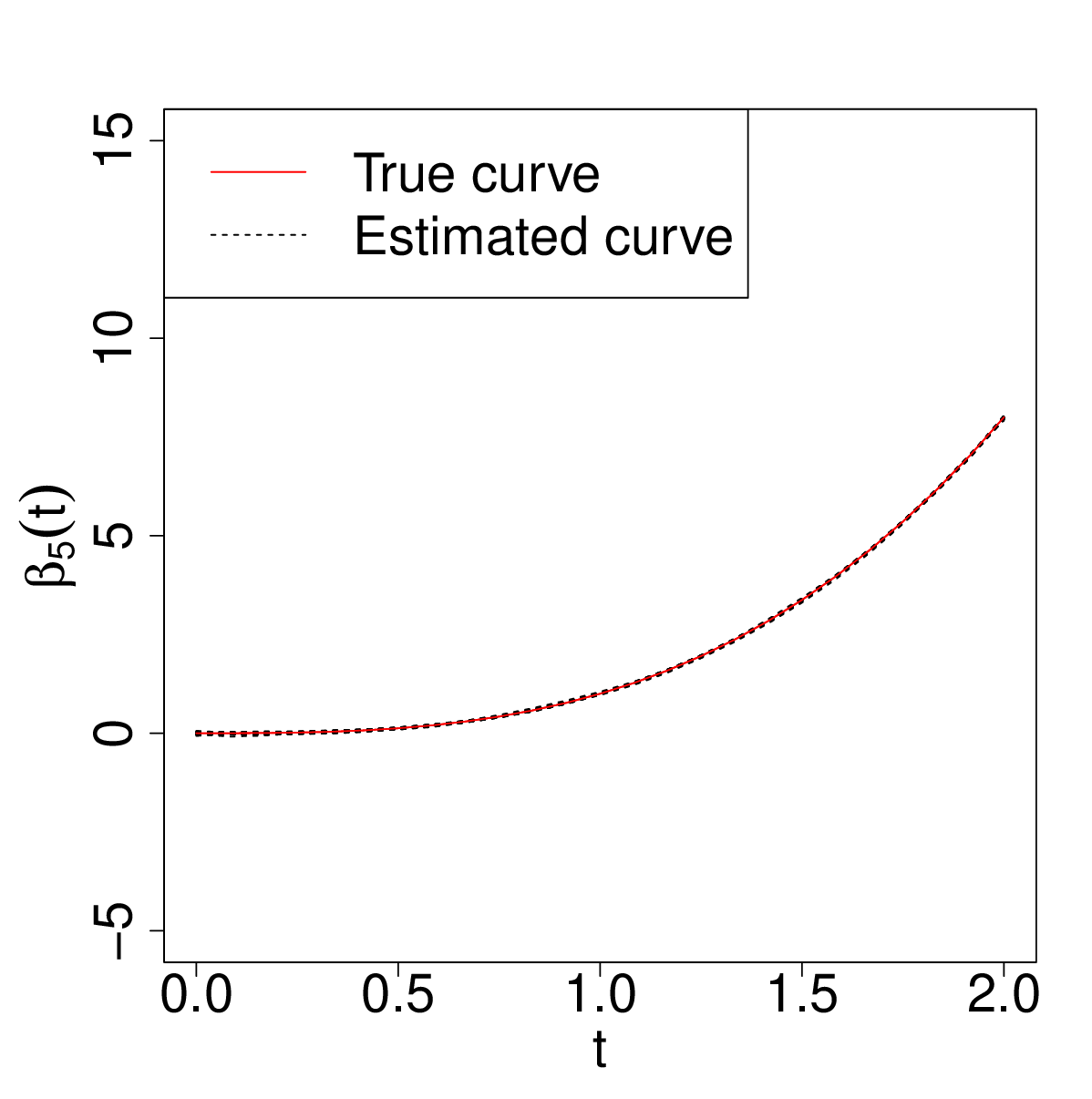}}&
		\subfloat[{\scriptsize $\hat{\beta}_{3}(.)$ (grMCP, $\lambda=0.250$).}]{\includegraphics[scale=0.15]{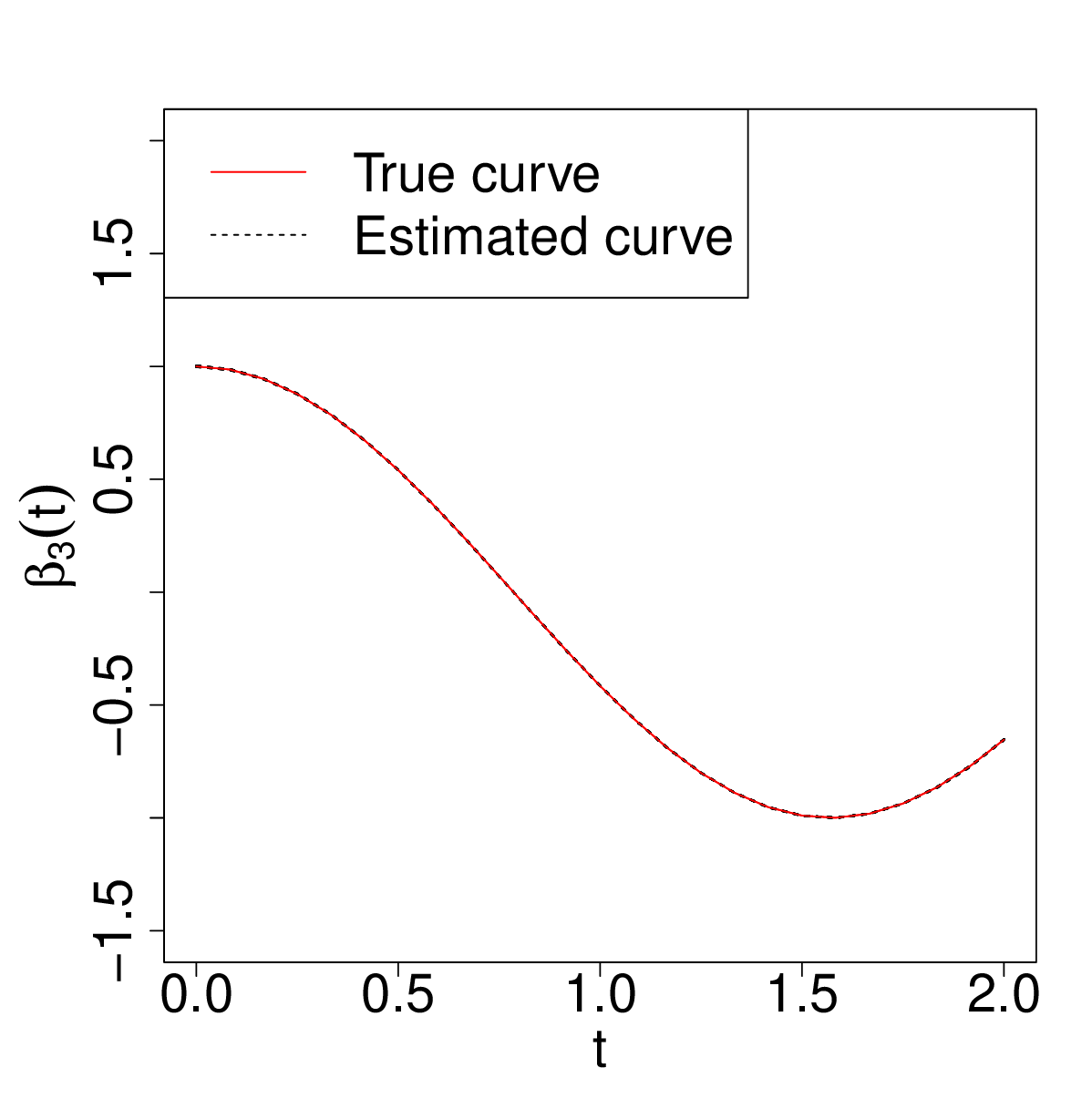}}&
		\subfloat[{\scriptsize $\hat{\beta}_{5}(.)$ (grMCP, $\lambda=0.250$).}]{\includegraphics[scale=0.15]{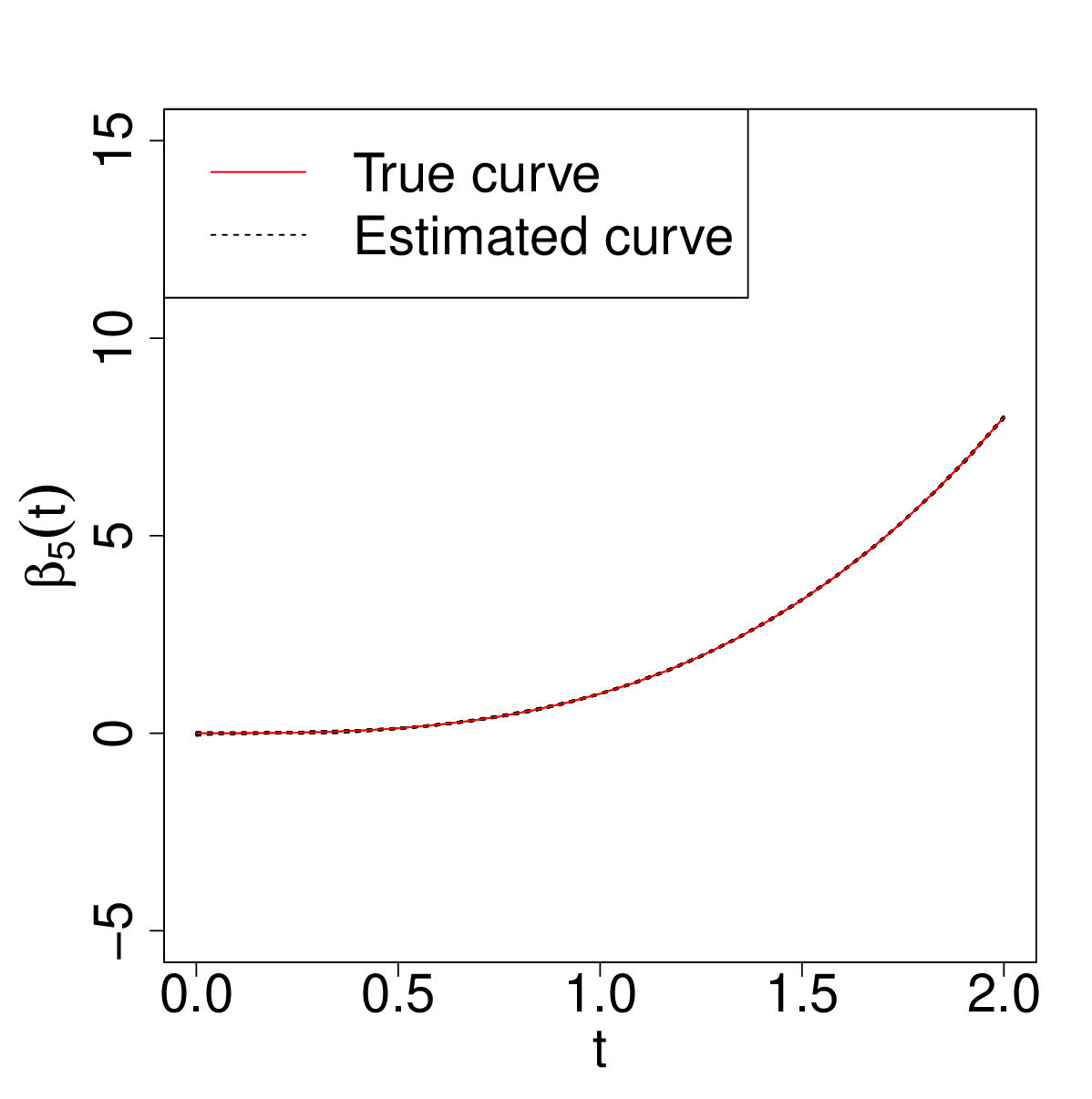}}\\
		\subfloat[{\scriptsize $\hat{\beta}_{3}(.)$ (grSCAD, $\lambda=0.250$).}]{\includegraphics[scale=0.15]{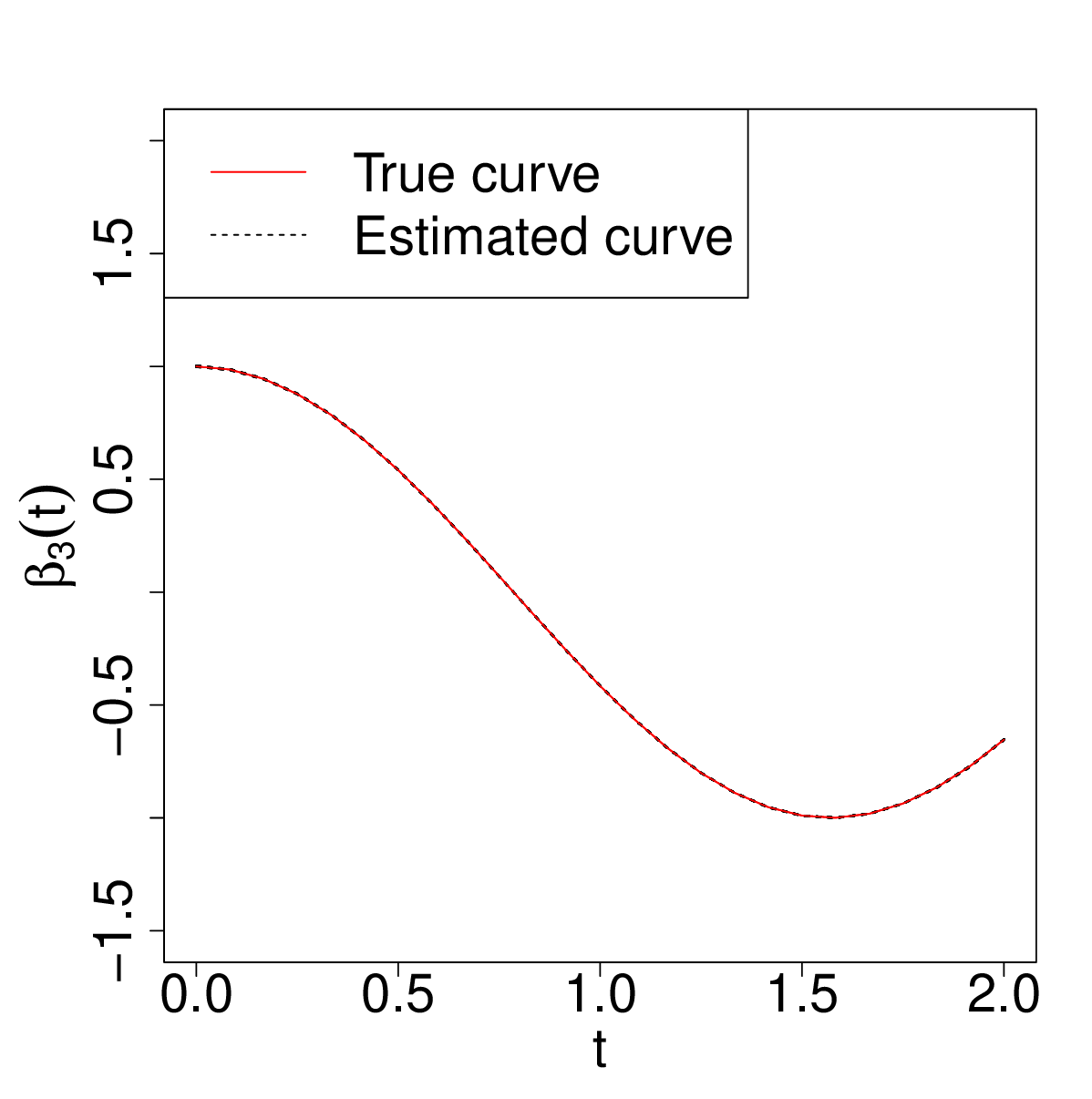}}&
		\subfloat[{\scriptsize $\hat{\beta}_{5}(.)$ (grSCAD, $\lambda=0.250$).}]{\includegraphics[scale=0.15]{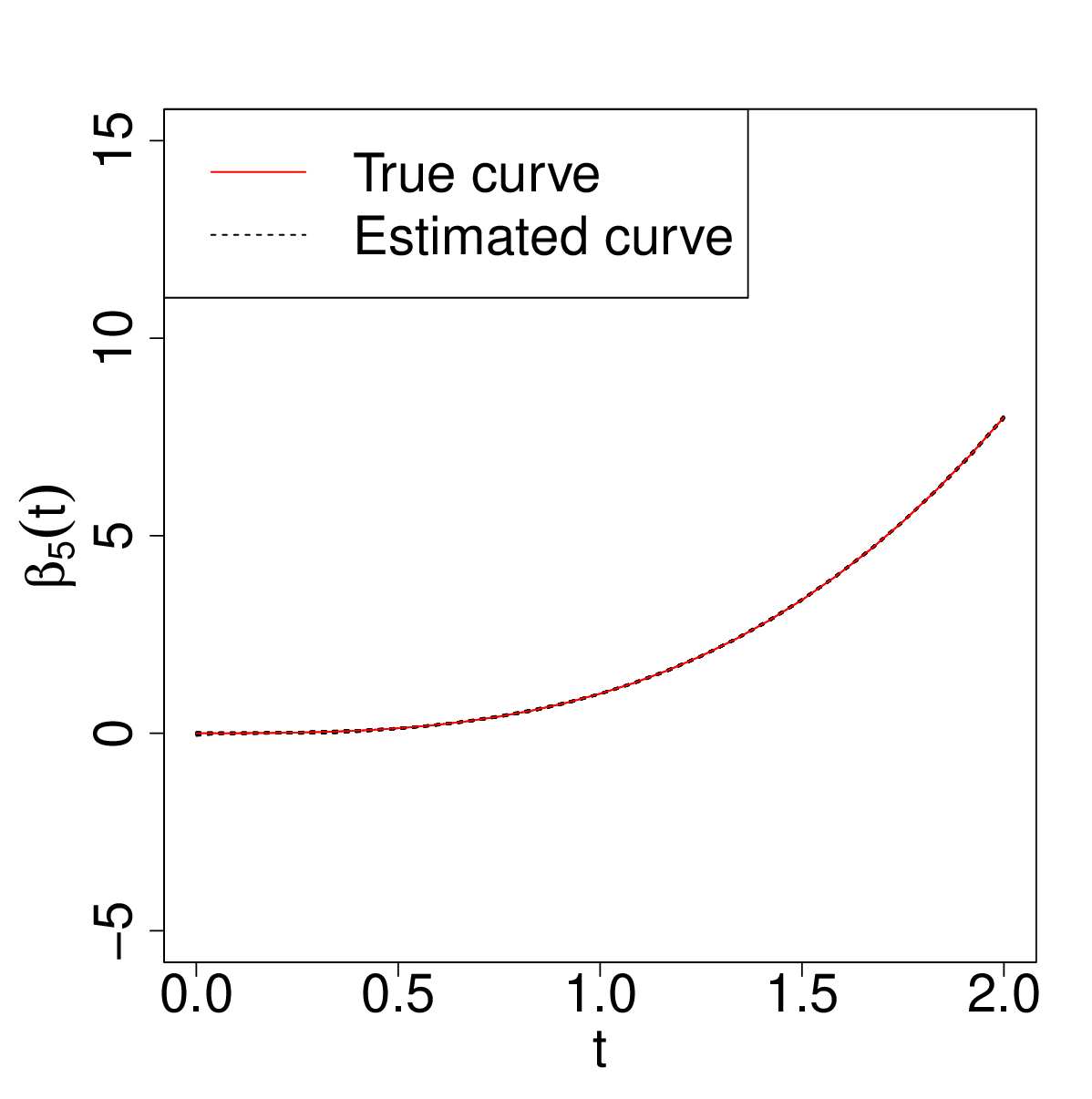}}&
        \subfloat[{\scriptsize $\hat{\beta}_{3}(.)$ (BGLSS).}]{\includegraphics[scale=0.15]{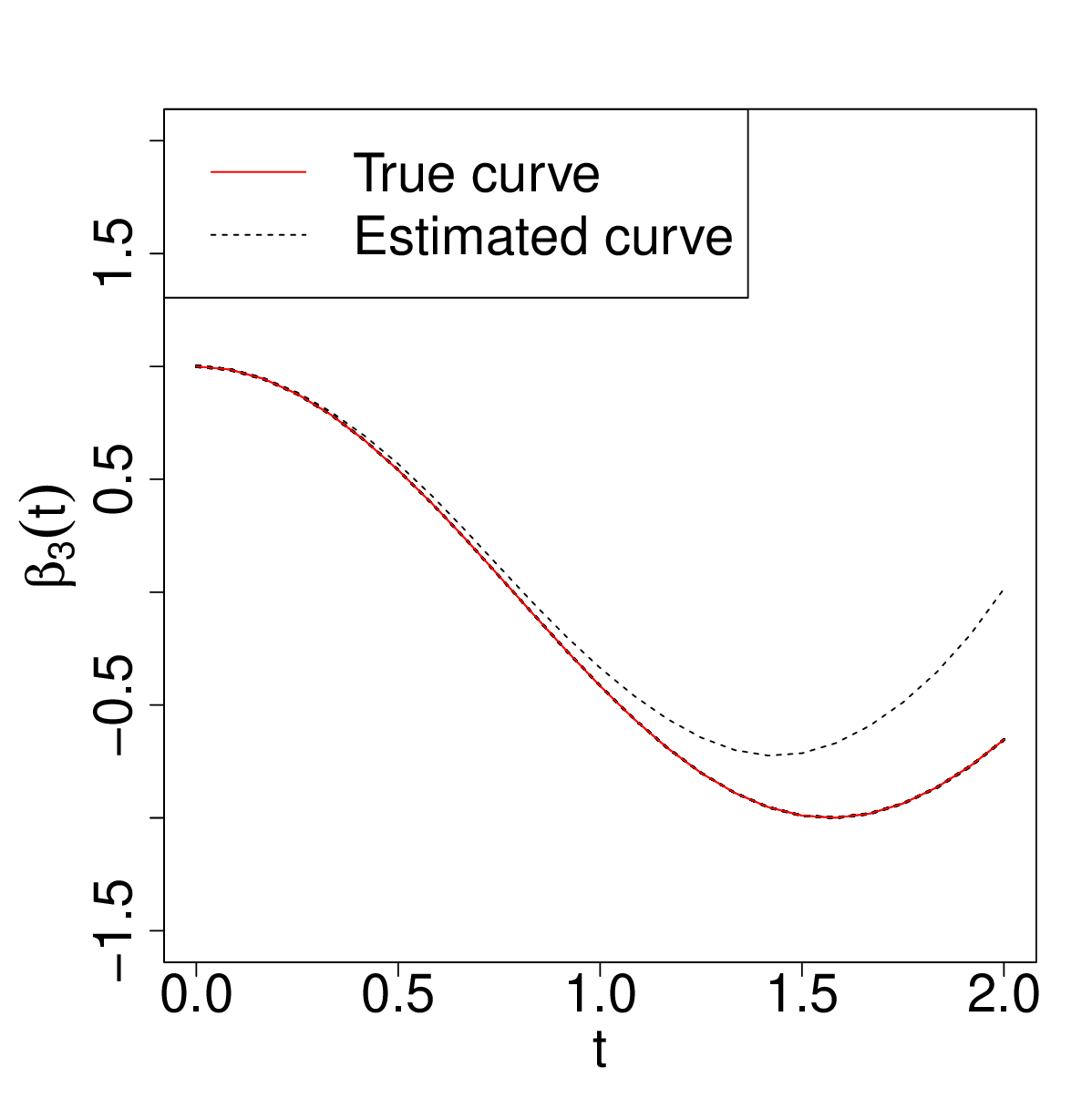}}&
		\subfloat[{\scriptsize $\hat{\beta}_{5}(.)$ (BGLSS).}]{\includegraphics[scale=0.15]{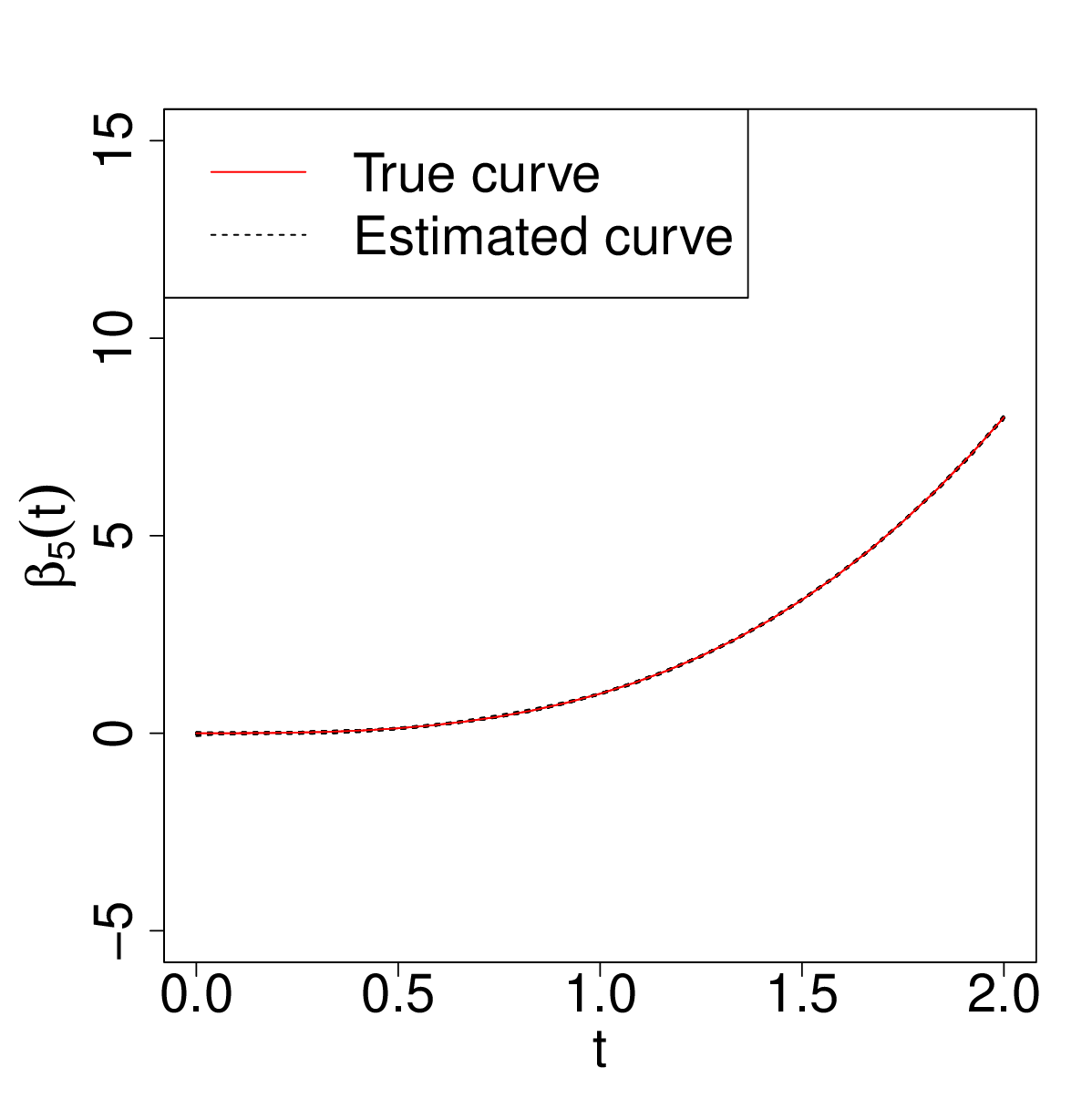}}\\
	\end{tabular}
	\caption{Partial functional coefficients ($3^{\text{rd}}$ and $5^{\text{th}}$, true in red and estimated for the replications in which the coefficients were selected in black), according to the model used, considering data with $\sigma=0.2$.}
	\label{coefs3e5_compara_modelos_sigma_02}
\end{figure}

\begin{figure}
	\centering
	\begin{tabular}{cccc}
		\subfloat[{\scriptsize $\hat{\beta}_{3}(.)$ (proposed model, $\mu=0.5$).}]{\includegraphics[scale=0.15]{figures/beta3_20_ENG.eps}}&
		\subfloat[{\scriptsize $\hat{\beta}_{5}(.)$ (proposed model, $\mu=0.5$).}]{\includegraphics[scale=0.15]{figures/beta5_20_ENG.eps}}&
		\subfloat[{\scriptsize $\hat{\beta}_{3}(.)$ (proposed model, $\mu$ as a parameter).}]{\includegraphics[scale=0.15]{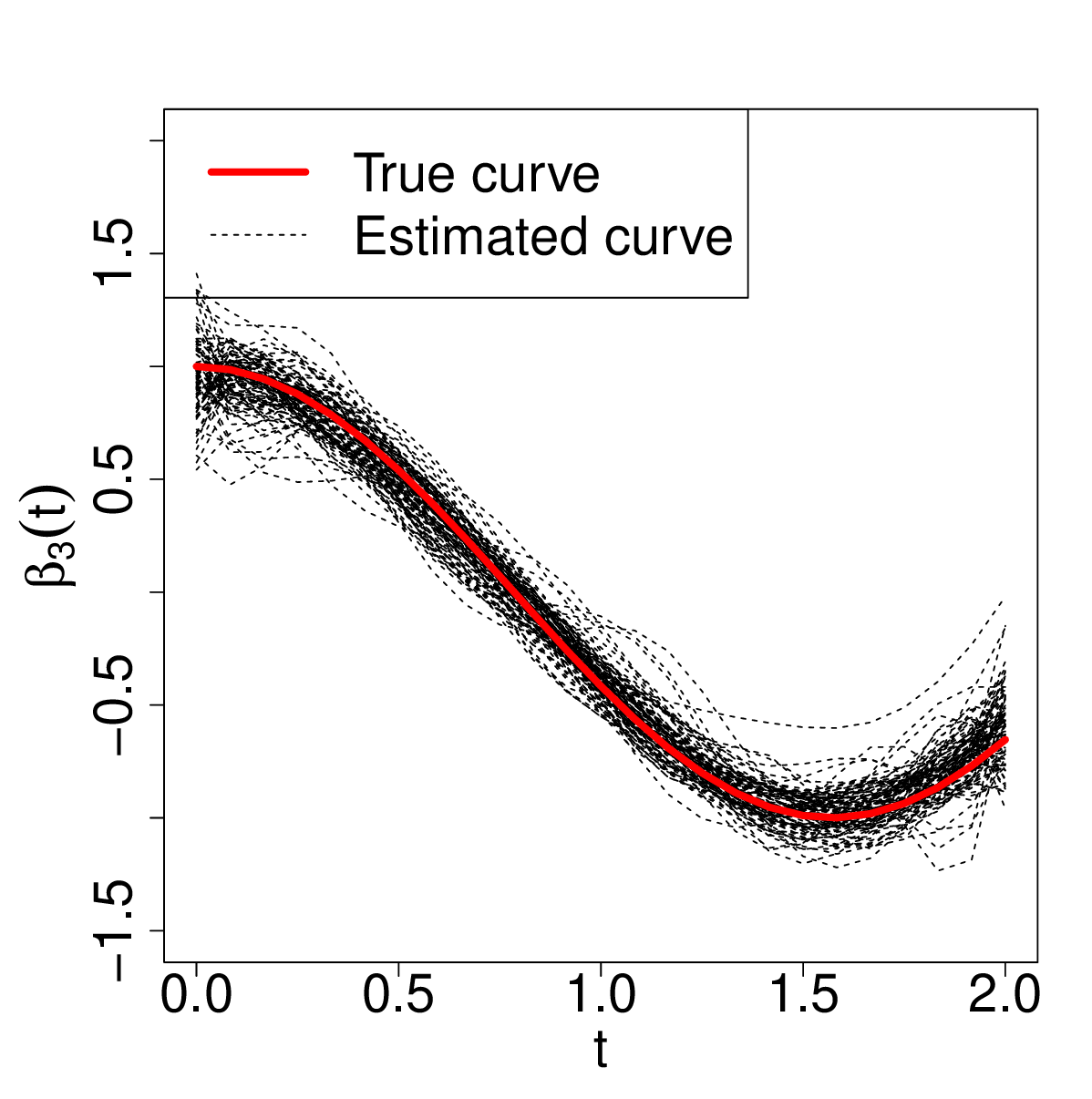}}&
		\subfloat[{\scriptsize $\hat{\beta}_{5}(.)$ (proposed model, $\mu$ as a parameter).}]{\includegraphics[scale=0.15]{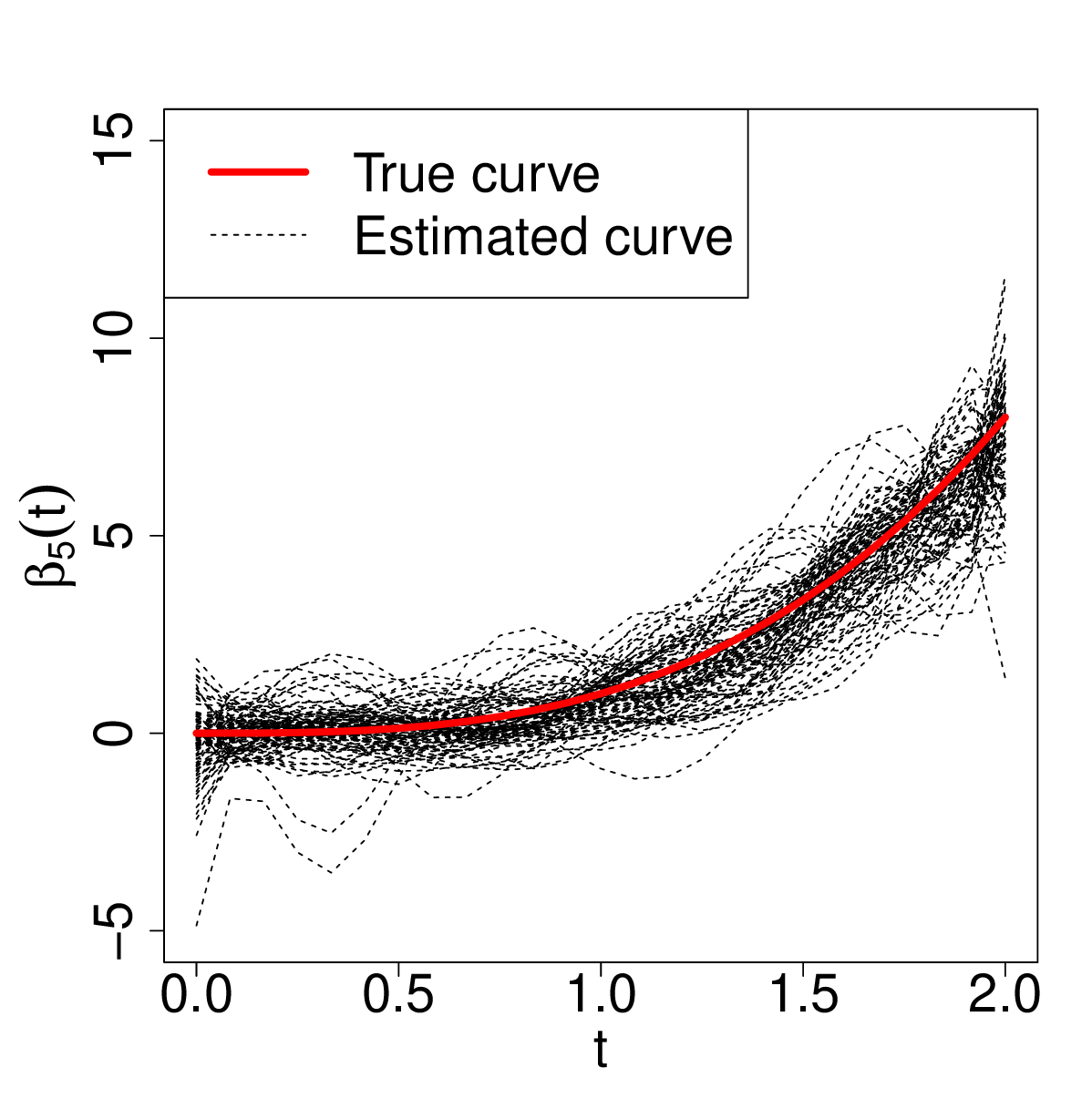}}\\
		\subfloat[{\scriptsize $\hat{\beta}_{3}(.)$ (grLASSO, $\lambda=0.001$).}]{\includegraphics[scale=0.15]{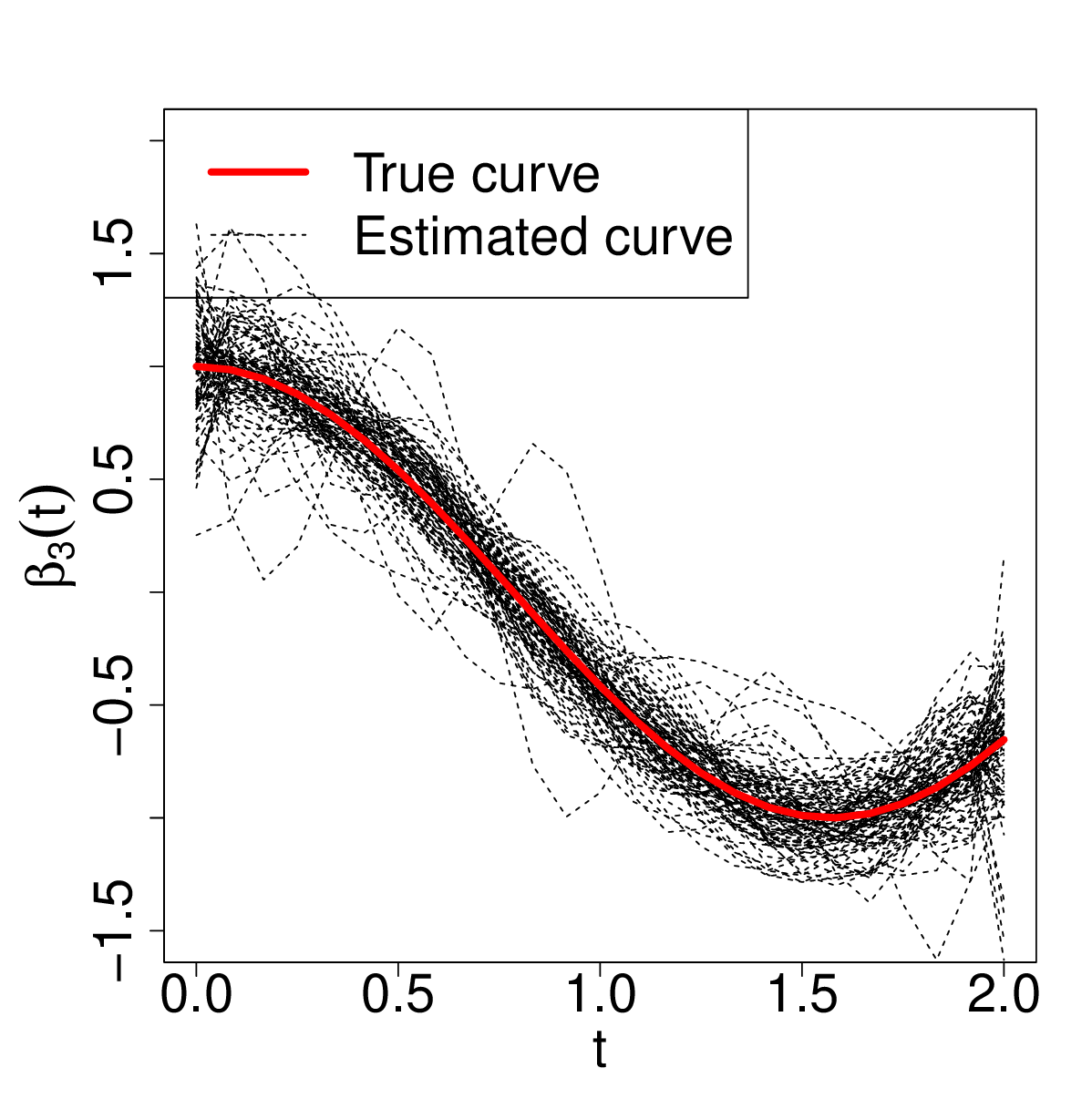}}&
		\subfloat[{\scriptsize $\hat{\beta}_{5}(.)$ (grLASSO, $\lambda=0.001$).}]{\includegraphics[scale=0.15]{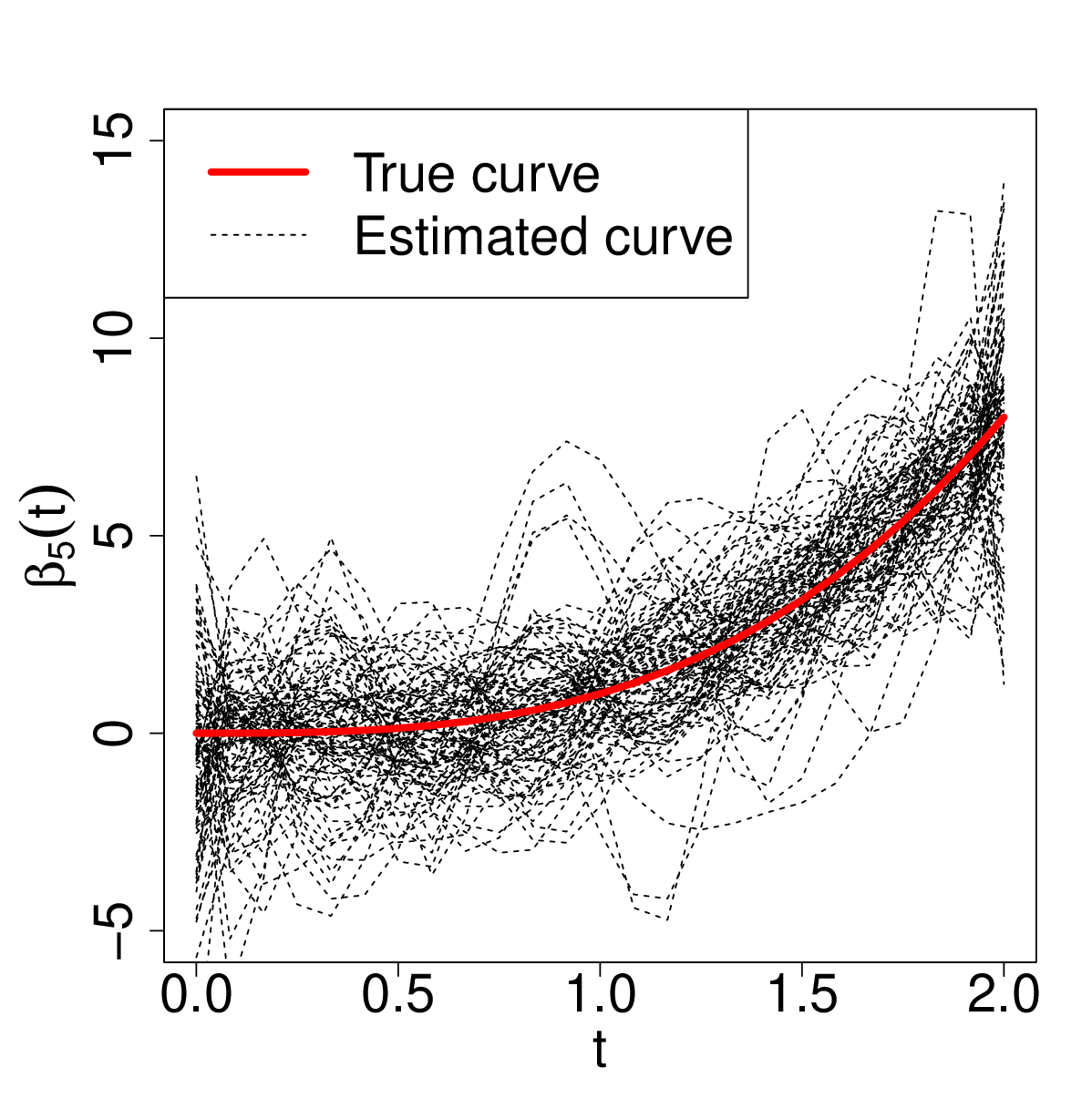}}&
		\subfloat[{\scriptsize $\hat{\beta}_{3}(.)$ (grMCP, $\lambda=0.500$).}]{\includegraphics[scale=0.15]{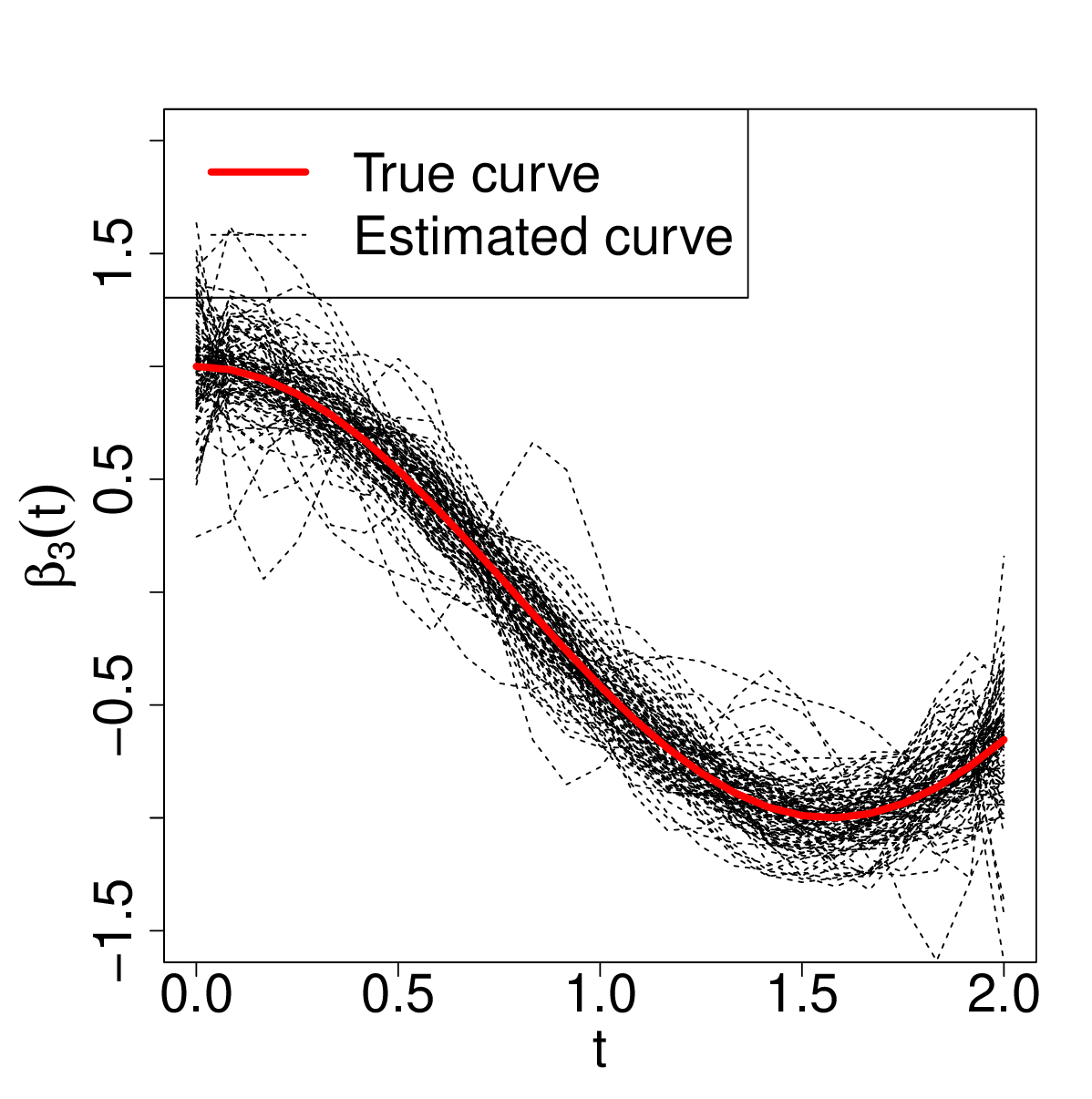}}&
		\subfloat[{\scriptsize $\hat{\beta}_{5}(.)$ (grMCP, $\lambda=0.500$).}]{\includegraphics[scale=0.15]{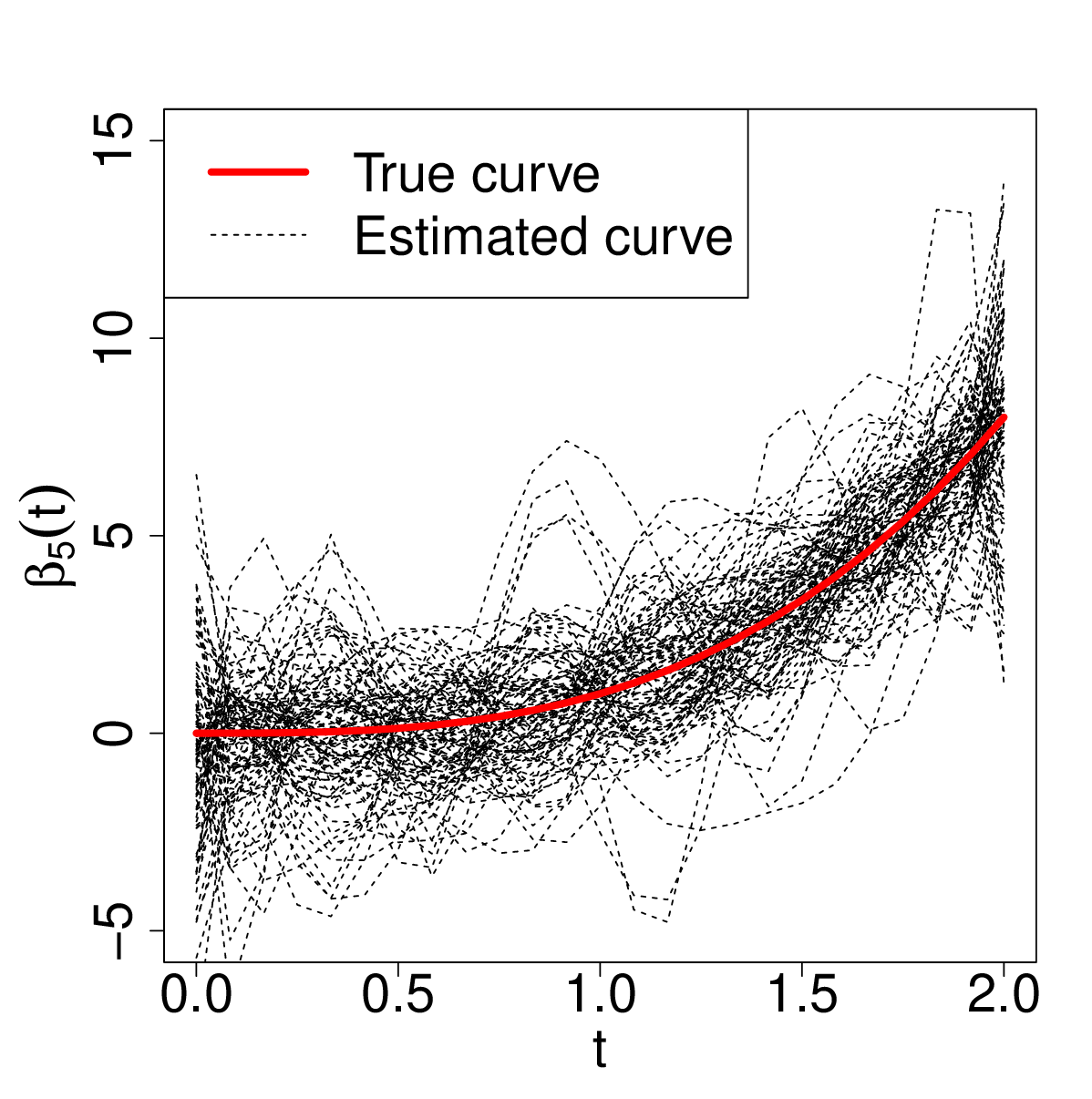}}\\
		\subfloat[{\scriptsize $\hat{\beta}_{3}(.)$ (grSCAD, $\lambda=0.001$).}]{\includegraphics[scale=0.15]{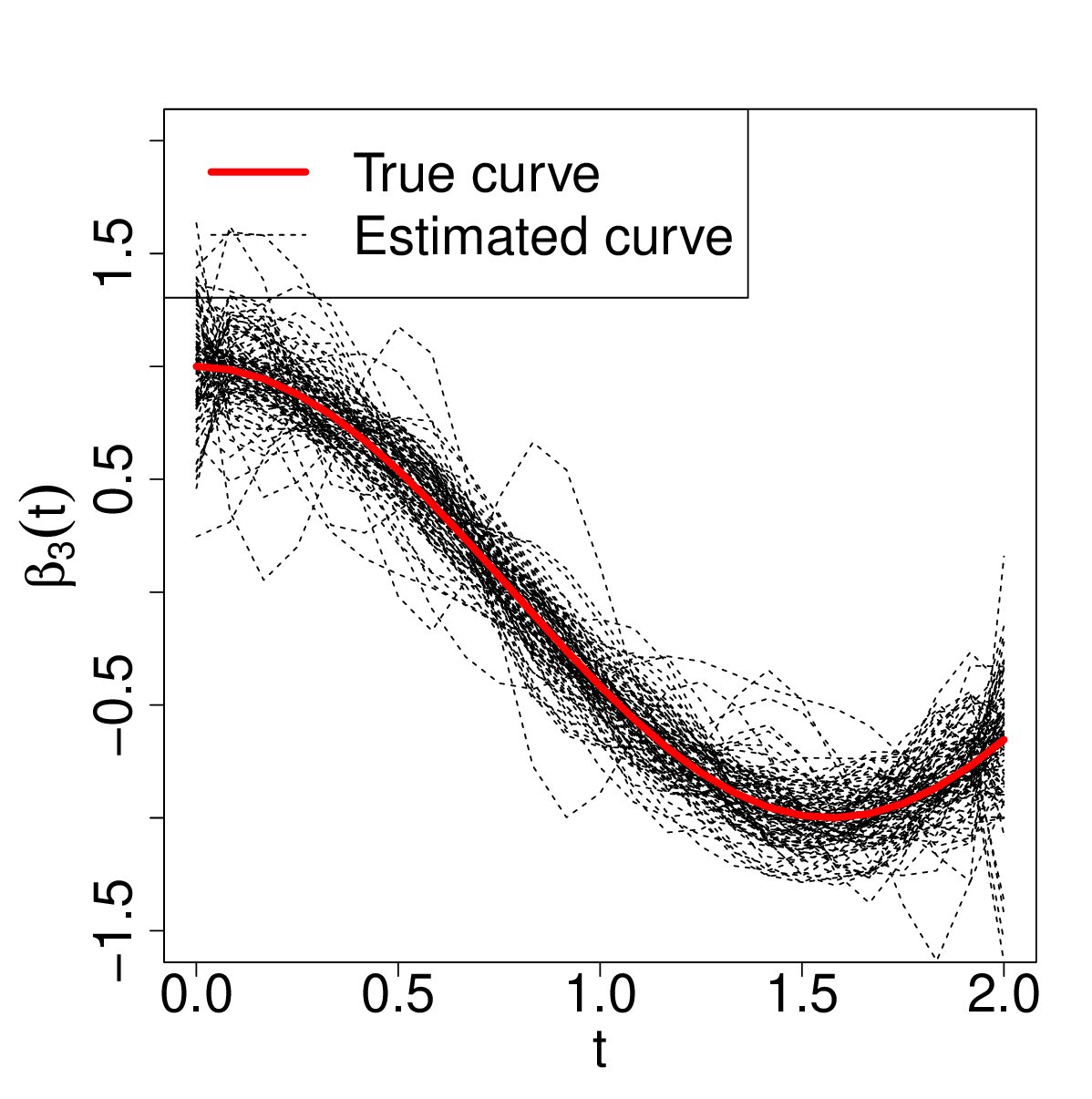}}&
		\subfloat[{\scriptsize $\hat{\beta}_{5}(.)$ (grSCAD, $\lambda=0.001$).}]{\includegraphics[scale=0.15]{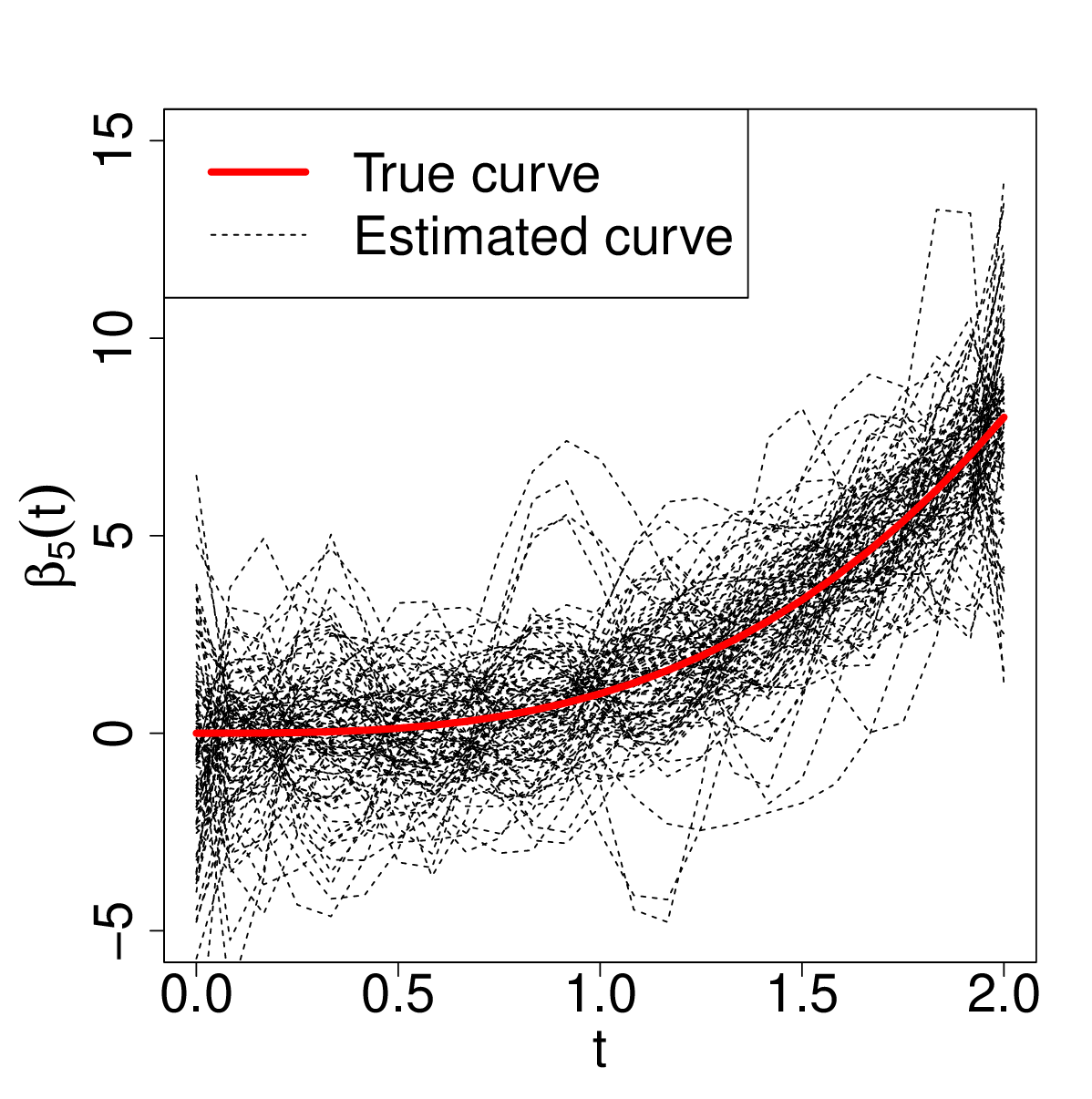}}&
        \subfloat[{\scriptsize $\hat{\beta}_{3}(.)$ (BGLSS).}]{\includegraphics[scale=0.15]{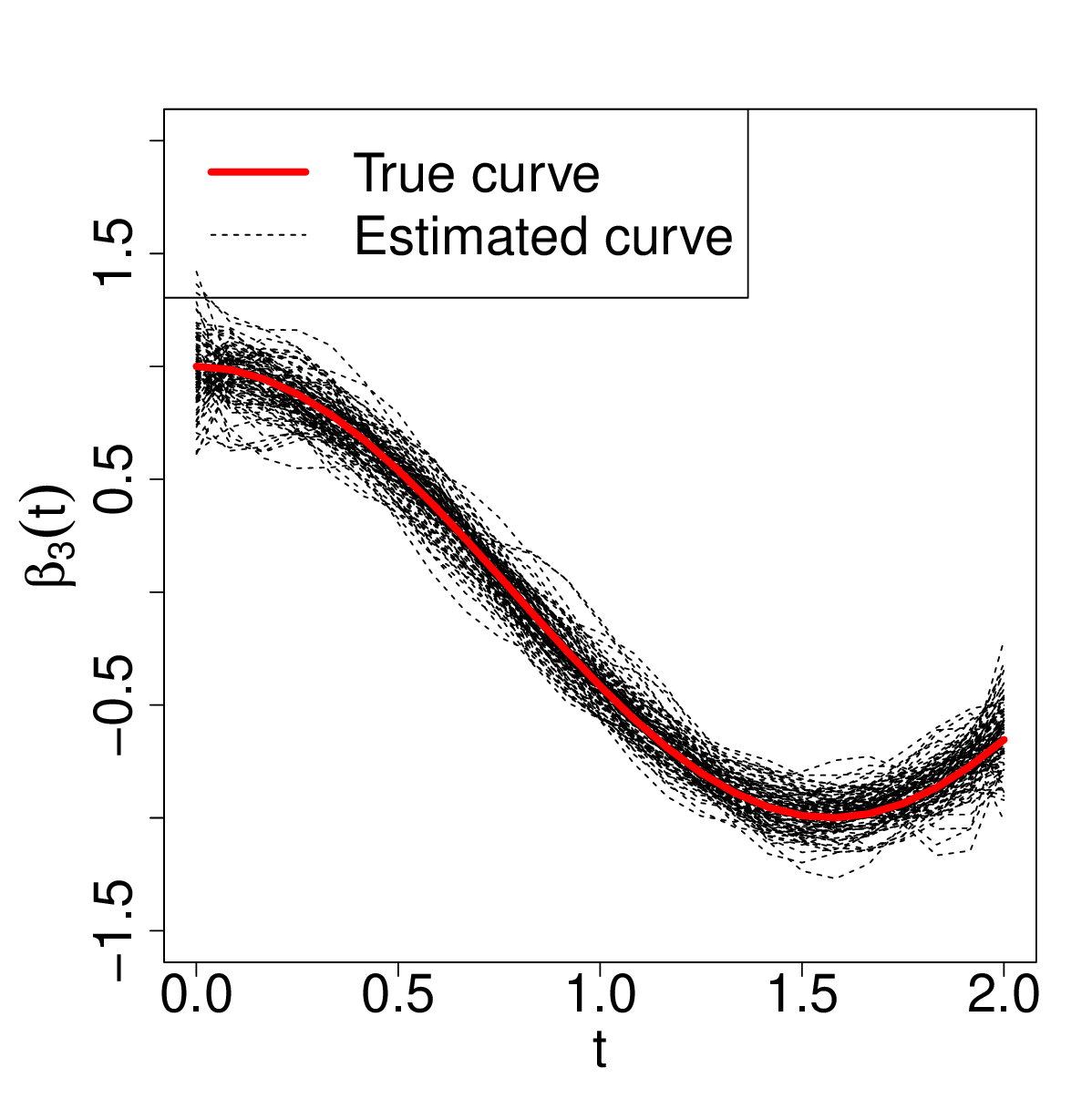}}&
		\subfloat[{\scriptsize $\hat{\beta}_{5}(.)$ (BGLSS).}]{\includegraphics[scale=0.15]{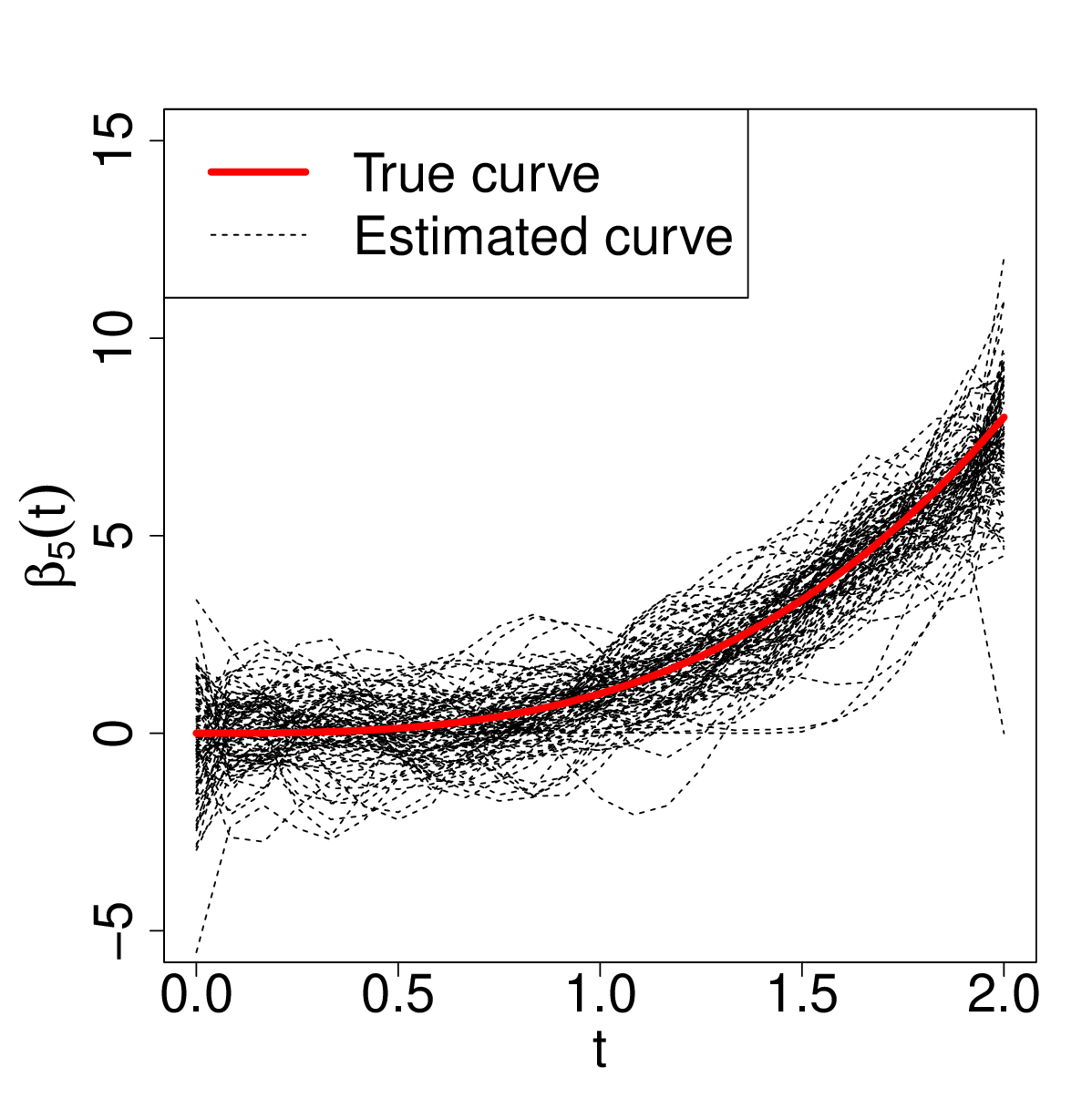}}\\
	\end{tabular}
	\caption{Partial functional coefficients ($3^{\text{rd}}$ and $5^{\text{th}}$, true in red and estimated for the replications in which the coefficients were selected in black), according to the model used, considering data with $\sigma=20$.}
	\label{coefs3e5_compara_modelos_sigma_20}
\end{figure}

Finally, Figures \ref{coefs3e5_compara_modelos_sigma_02} and \ref{coefs3e5_compara_modelos_sigma_20} show the estimates obtained in each replication for the third and fifth functional coefficients for each method and each level of data dispersion. In this case, the performance metric in \eqref{eq:r2_PARTE2} was used as a reference to choose the model configurations to compare. When $\sigma = 20$, it is noticeable that there is a smaller dispersion of the estimated functional coefficients obtained by our proposed model (first row of Figure \ref{coefs3e5_compara_modelos_sigma_20}) compared to the grLASSO, grMCP and grSCAD competing models. This observation is valid for both the third and the fifth functional coefficients.

\section{Application to real data and the GCV criterion}\label{sec:4}

We considered two databases to evaluate the performance of the proposed model on a set of real data: the first with COVID-19 data freely available through the Brasil.IO accessible open data portal \citep{BRASILio} for the definition of the response variable of the study; and the second with the socioeconomic data of the states and the Federal District made available by the IBGE (Brazilian Institute of Geography and Statistics) for the definition of the covariates for our analysis.

The response variable chosen for this study is the logarithm of the accumulated weekly number of deaths per 10\,000 inhabitants due to COVID-19 in the Brazilian states and the Federal District. Therefore, there are $m=27$ curves corresponding to the 26 states plus the Federal District. As the first cases of COVID-19 in each state appeared at different weeks, we decided to analyze the data from the 20th epidemiological week of 2020, the week from which new cases were already registered in all of the federation members (the 26 states and the Federal District), until the 32nd epidemiological week of the year 2021, totalling 66 weeks ($n_{i}=n=66$).

Next, nine covariates were defined ($p=9$) as explanatory variables, which are demographic density; HDI (Human Development Index); Gini index of the per capita household income distribution; unemployment rate of people aged 14 or over; QoLLI (Quality of Life Loss Index); percentage of population with health insurance; number of hospital beds per 10\,000 inhabitants; number of doctors per 10\,000 inhabitants; average real income from the main job usually earned per month by persons aged 14 years or over (income).

%\begin{itemize}
%	\item demographic density;
%	\item HDI (Human Development Index);
%	\item Gini index of the per capita household income distribution;
%	\item unemployment rate of people aged 14 or over;
%	\item QoLLI (Quality of Life Loss Index);
%	\item percentage of population with health insurance;
%	\item number of hospital beds per 10\,000 inhabitants;
%	\item number of doctors per 10\,000 inhabitants; 
%	\item average real income from the main job usually earned per month by persons aged 14 years or over (income).
%\end{itemize}

The design matrix was defined with the covariates positioned in the same order in which they were presented above, so we have $Z_{1}\beta_{1}(.)$ as the functional coefficient associated with the demographic density, $Z_{2}\beta_{2}(.)$ as a functional coefficient associated with the HDI, and so on. In addition, the covariates were standardized by their mean and standard deviation before applying the model (see Section 1 of the Supplementary Material). In addition, the intercept was estimated as described in Subsection \ref{sec:data_generation}.

The Gibbs sampler implemented the proposed Bayesian model using two chains to enable a convergence diagnosis. The initialization of the chains, the maximum number of iterations, the choice of the \textit{burn-in} period, the spacing of points between sampled values, as well as the choice of the hyperparameters and obtaining the Bayesian estimates were all established following what was described in Subsection \ref{sec:data_generation} for the studies with synthetic data. Furthermore, we considered the model with $\mu$ as a parameter in this study. 

Before presenting any results, it is worth mentioning that cubic B-splines were used for the basis expansion of all partial functional coefficients, considering three values for $K$ ($K=5$, $K=10$ and $K= 15$). The diagnostic analysis based on the method proposed by \cite{Gelman} indicated convergence of the chains of partial coefficients across all tested model configurations after the \textit{burn-in} period. 

Unfortunately, it is impossible to calculate the MSE between the points of the estimated curve and those of the true curve in a study with real data since the true curve is unknown. Since calculating the MSE from the distances between the observed values and the points on the estimated curve is not recommended due to the possibility of \textit{overfitting}, we consider an alternative measure based on a cross-validation procedure. Although the conventional cross-validation associated with the MSE is an effective procedure to overcome the problem of \textit{overfitting}, it often becomes inefficient due to the high computational cost. As an alternative to the cross-validation procedure, there is the Generalized Cross Validation (GCV) \citep{Wahba}. Adapting the GCV criterion presented by \cite{Wahba} to the reality of the proposed model, we obtain
\begin{gather}
	\GCV(K)=\frac{1}{\sum_{i=1}^{m}n_{i}}\frac{\sum_{i=1}^{m}\sum_{j=1}^{n_{i}}(\tilde{y}_{ij}-
		\sum_{l=1}^{p}x_{li}\hat{Z}_{l}\hat{\beta}_{l}(t_{ij}))^{'}(\tilde{y}_{ij}-
		\sum_{l=1}^{p}x_{li}\hat{Z}_{l}\hat{\beta}_{l}(t_{ij}))}{\left[1-\dfrac{1}{\sum_{i=1}^{m}n_{i}}\tr(\vec{S_{K}})\right]^2}\nonumber\\
	=\frac{1}{\sum_{i=1}^{m}n_{i}}\frac{(\vec{\tilde{y}_{..}}-\vec{\hat{O}_{...}}\vec{\hat{\bb}})^{'}(\vec{\tilde{y}_{..}}-\vec{\hat{O}_{...}}\vec{\hat{\bb}})}{\left[1-\dfrac{1}{\sum_{i=1}^{m}n_{i}}\tr(\vec{S_{K}})\right]^2}\text{,}\nonumber%sem_citar
	%\label{eq:GCV_parte2}
\end{gather}

\noindent where $\tilde{y}_{ij}=y_{ij}-\tilde{\beta}_{0}(t_{ij})$, $\vec{S_{K}}$ is the projection matrix of the proposed model with $K$ bases on the expansions of the functional coefficients, $\vec{\hat{O}_{...}}$ is the estimate for $\vec{O_{...}}$ (as defined in Subsection \ref{Ap_full}) and $\vec{\hat{\bb}}$ is the estimate for $\vec{\bb}$.

To understand how $\vec{S_{K}}$ is calculated, note that the mean of the complete conditional distribution of $\vec{\bb}$, which is equivalent to the mode since the normal distribution is symmetric,  is given as follows: 
\begin{equation}	\E(\vec{\bb}|\vec{\theta},\vec{\mu},\sigma^2,\vec{\tau^2},\vec{Z},\vec{y})=\vec{Q}^{-1}\vec{O_{...}}^{'}\vec{\tilde{y}_{..}}\text{,}
	\nonumber%sem_citar%\label{mediabeta_PARTE2}
\end{equation} as shown in the details of the full conditional distribution for $\vec{b}$ in Subsection \ref{Ap_full}. Finally, when considering $\vec{\hat{\bb}}=\vec{\hat{Q}}^{-1}\vec{\hat{O}_{...}}^{'}\vec{\tilde{y}_{..}}$, in which $\vec{\hat{Q}}$ and $\vec{\hat{O}_{...}}$ are the respective estimates of $\vec{Q}$ (also defined in Subsection \ref{Ap_full}) and $\vec{O_{...}}$, then the estimated curve is represented by
\begin{equation}
\vec{\hat{y}_{..}}=\vec{\hat{O}_{...}}\vec{\hat{\bb}}=\vec{\hat{O}_{...}}\vec{\hat{Q}}^{-1}\vec{\hat{O}_{...}}^{'}\vec{\tilde{y}_{..}}\text{,}\nonumber%sem_citar
\end{equation}so that the projection matrix can be represented by
\begin{equation}	\vec{S_{K}}=\vec{\hat{O}_{...}}\vec{\hat{Q}}^{-1}\vec{\hat{O}_{...}}^{'}\quad{.}\nonumber%sem_citar
\end{equation}

\noindent Therefore, it becomes completely possible to calculate the $\GCV(.)$ for all models tested as we vary $K$. The smaller the $\GCV$ value, the better the fit. 

Table \ref{tab:real_data_parte2} presents the value of the metric in \eqref{eq:r2_PARTE2} and the GCV criterion for each $K$ number of basis functions considered. We can observe that the fit quality is better with $K=5$, regardless of the metric used for analysis. Because the functional response data are observational, it is natural that the metric in \eqref{eq:r2_PARTE2} is not so close to one as it was in the simulation studies. Still, even with a relatively low number of curves ($m=27$), the model with $K=5$ can achieve almost 45\% on that metric.

\begin{table*}[!htb]
	\centering
    \caption{Values of the metric \eqref{eq:r2_PARTE2} and the GCV criterion, according to the value of $K$ used in the model.}
	\begin{tabular}{c|lll}\hline
		\multirow{2}{*}{Metric} & \multicolumn{3}{c}{$K$}                     \\\cmidrule{2-4}
		& \multicolumn{1}{c}{5} & \multicolumn{1}{c}{10} & \multicolumn{1}{c}{15} \\\hline
		Metric \eqref{eq:r2_PARTE2} &0.44907 &	0.11603 & 0.14010  \\
		GCV                          &  0.59545 & 0.94175 & 0.91838    \\\hline      
	\end{tabular}
\label{tab:real_data_parte2}
\end{table*}

\begin{figure}
	\centering
	\begin{tabular}{cccc}
		\subfloat[{\scriptsize Estimated $\beta_{0}(t)$ (Intercept).}]{\includegraphics[scale=0.15]{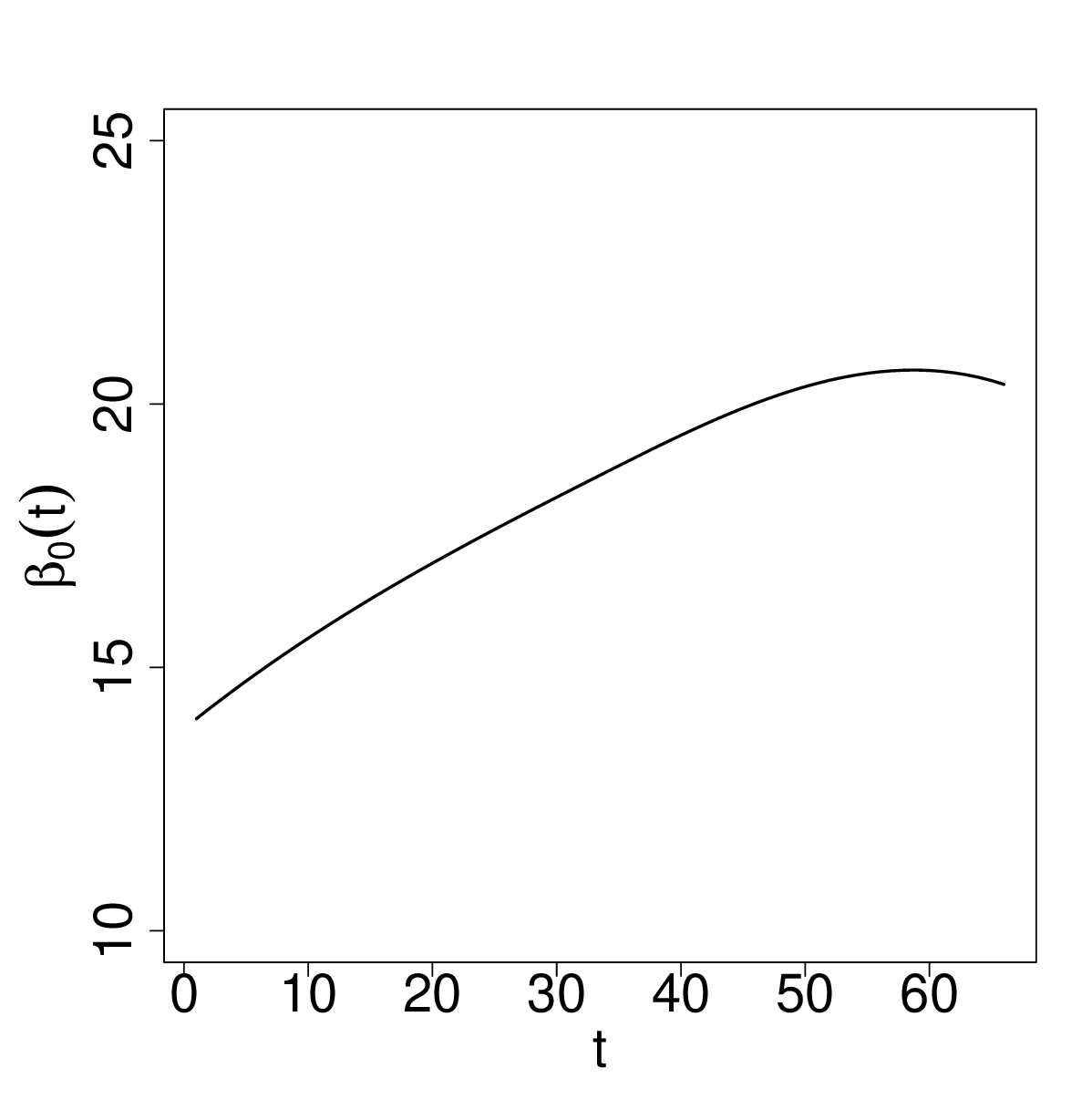}}&
		\subfloat[{\scriptsize Estimated $\beta_{1}(t)$ (Demographic density).}]{\includegraphics[scale=0.15]{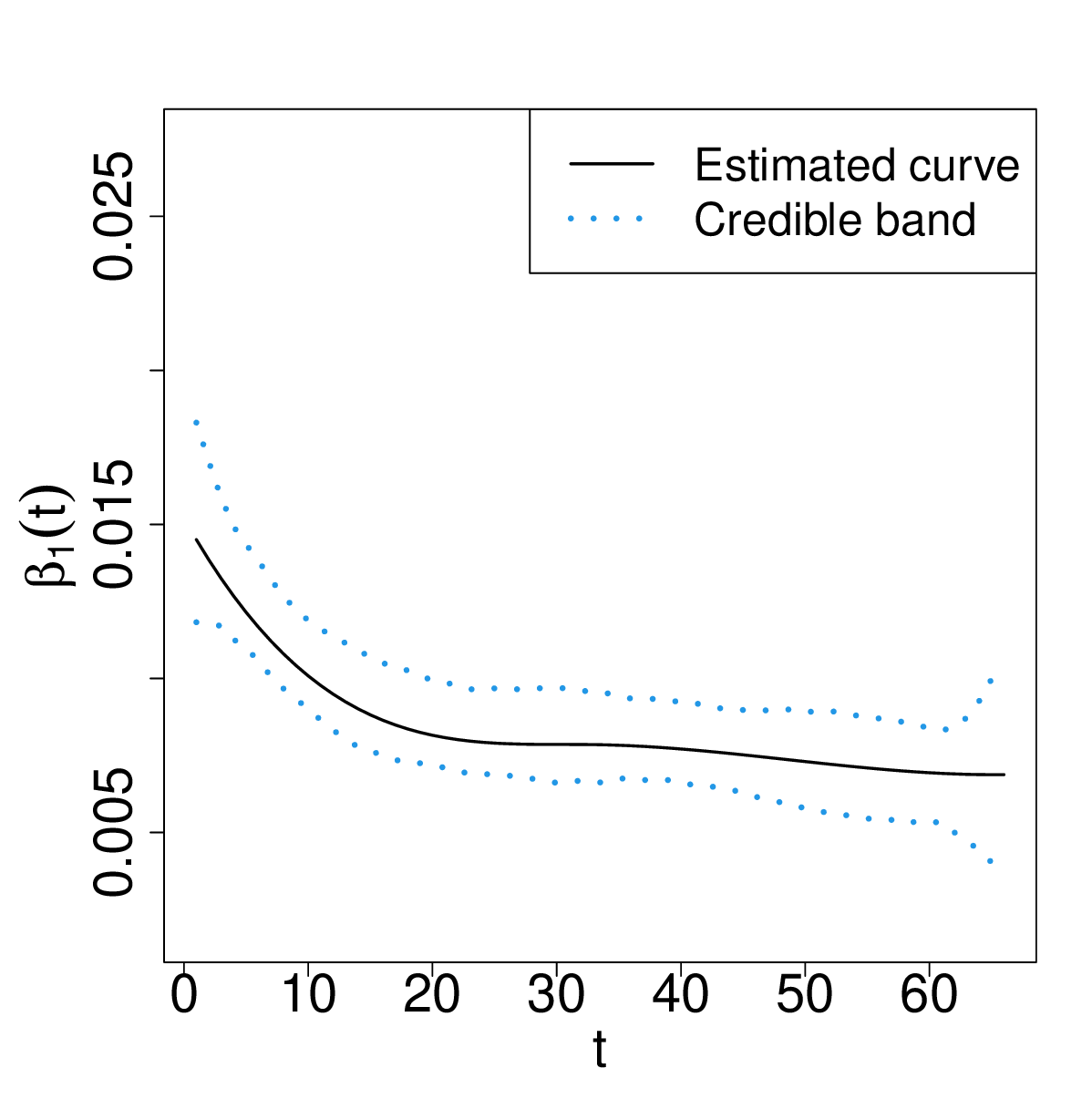}}&
		\subfloat[{\scriptsize Estimated $\beta_{3}(t)$ (Gini index).}]{\includegraphics[scale=0.15]{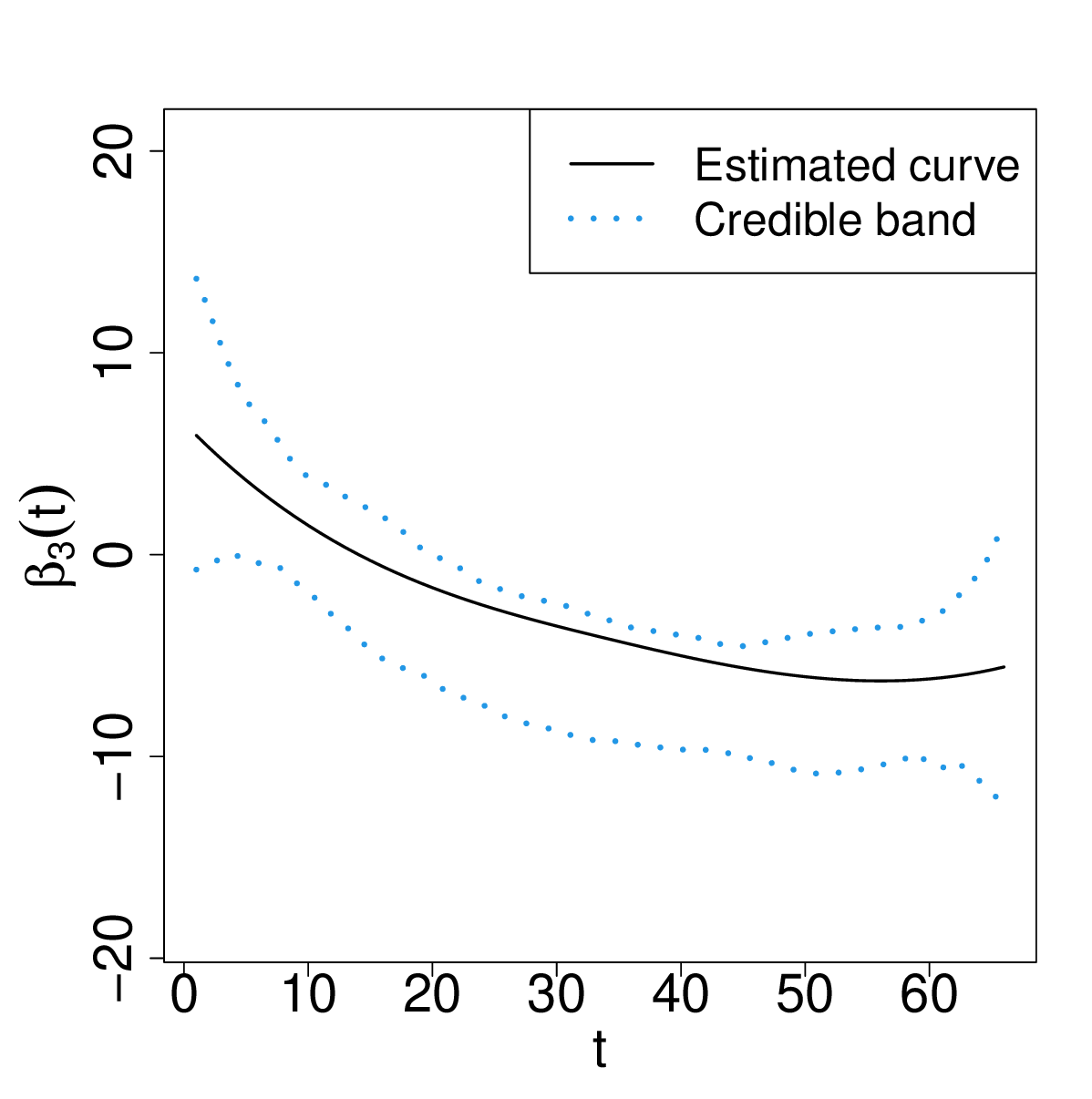}}&
		\subfloat[{\scriptsize Estimated $\beta_{4}(t)$ (unemployment rate).}]{\includegraphics[scale=0.15]{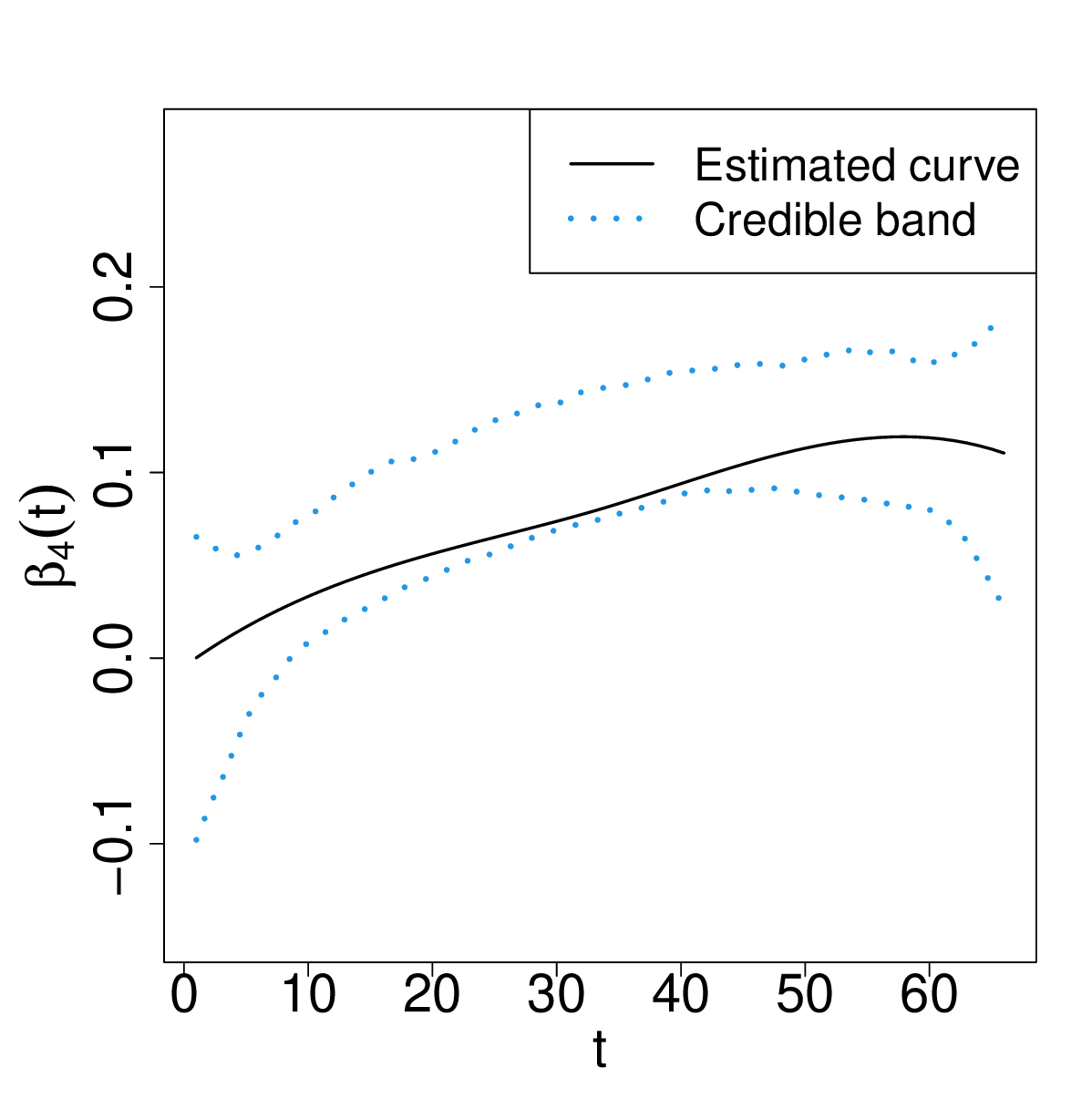}}\\
		\subfloat[{\scriptsize Estimated $\beta_{5}(t)$ (QoLLI).}]{\includegraphics[scale=0.15]{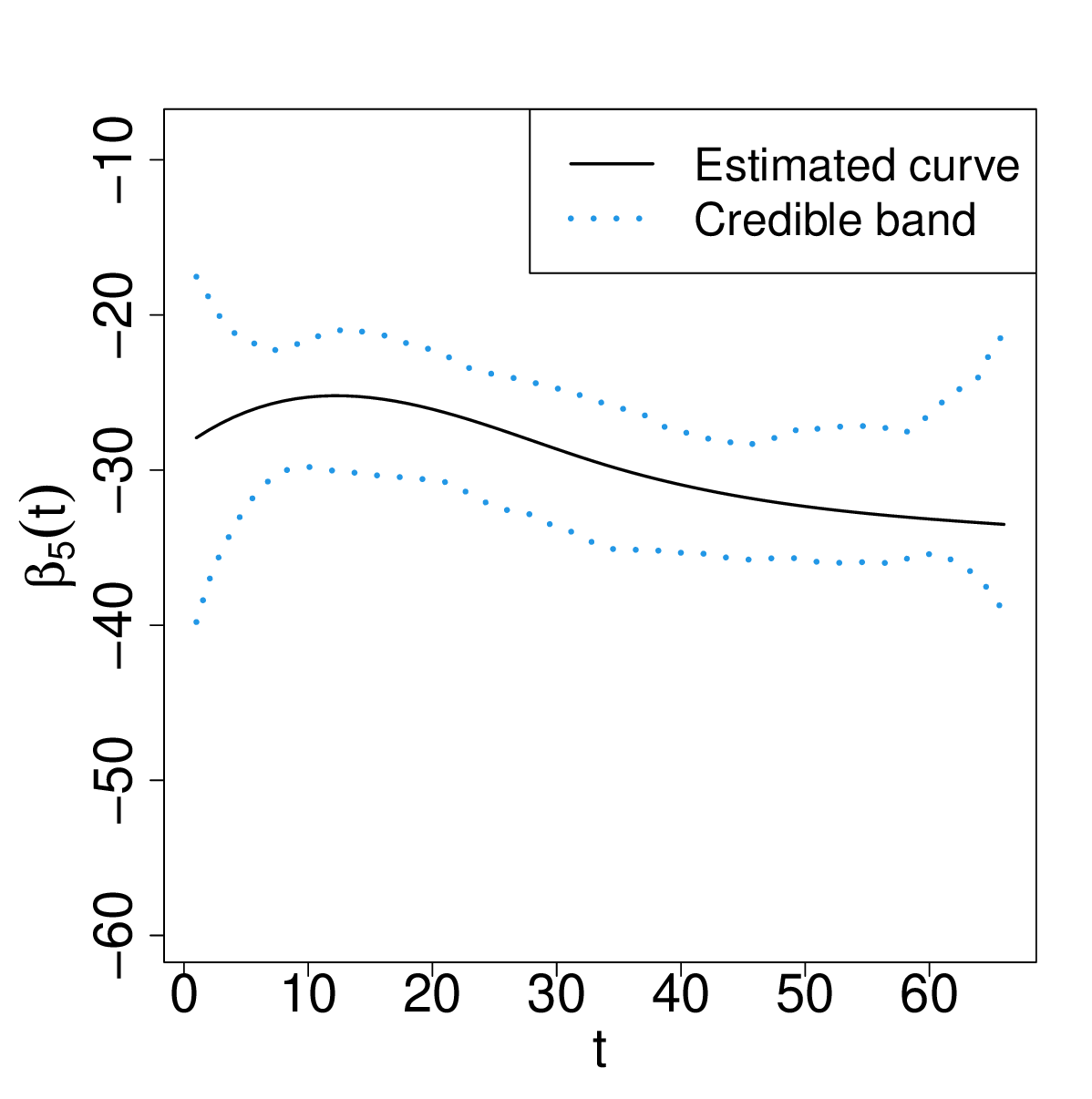}}&
		\subfloat[{\scriptsize Estimated  $\beta_{6}(t)$ (Percentage of population with health insurance).}]{\includegraphics[scale=0.15]{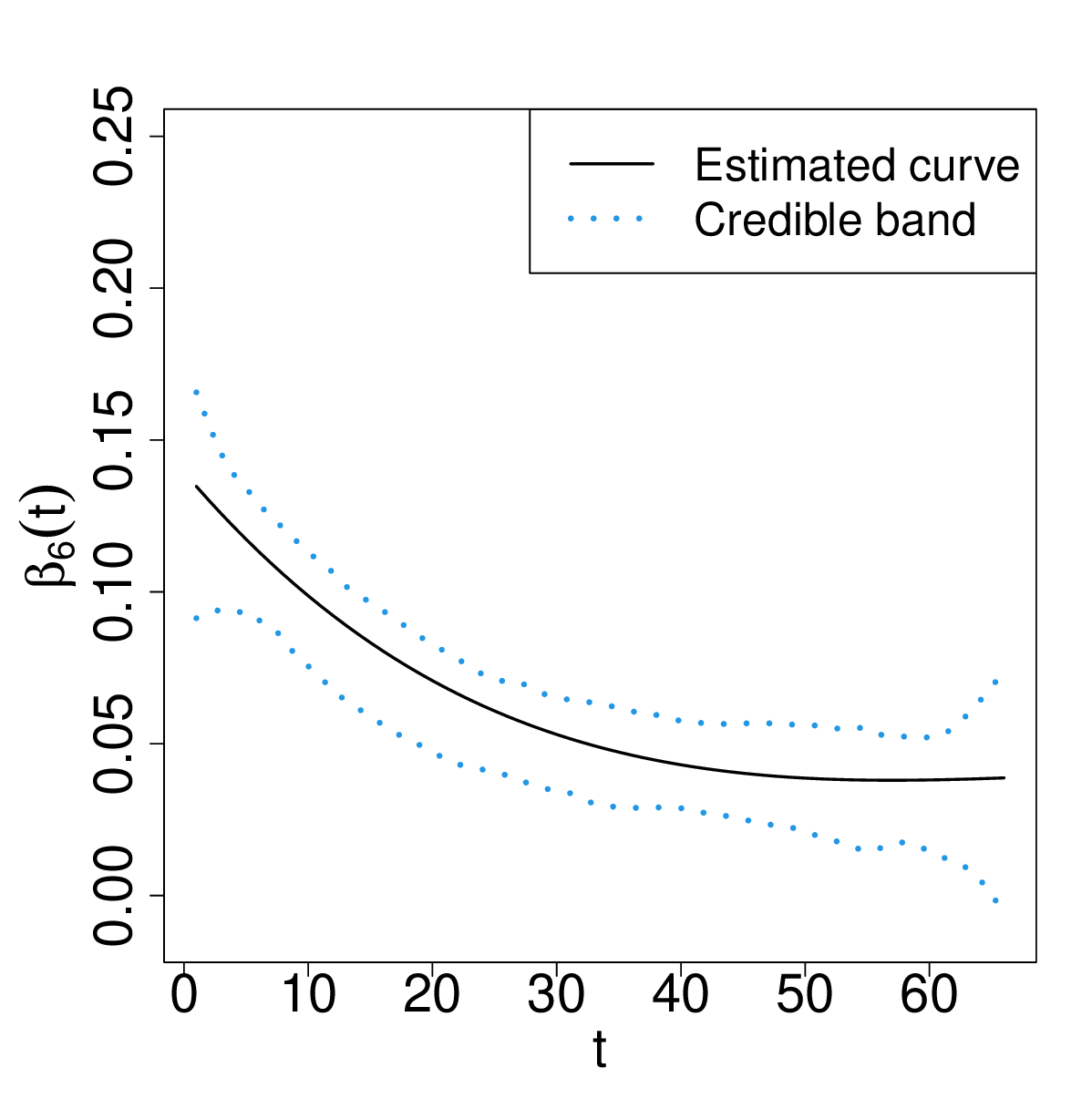}}	&
		\subfloat[{\scriptsize Estimated $\beta_{8}(t)$ (Number of doctors per 10000 inhabitants).}]{\includegraphics[scale=0.15]{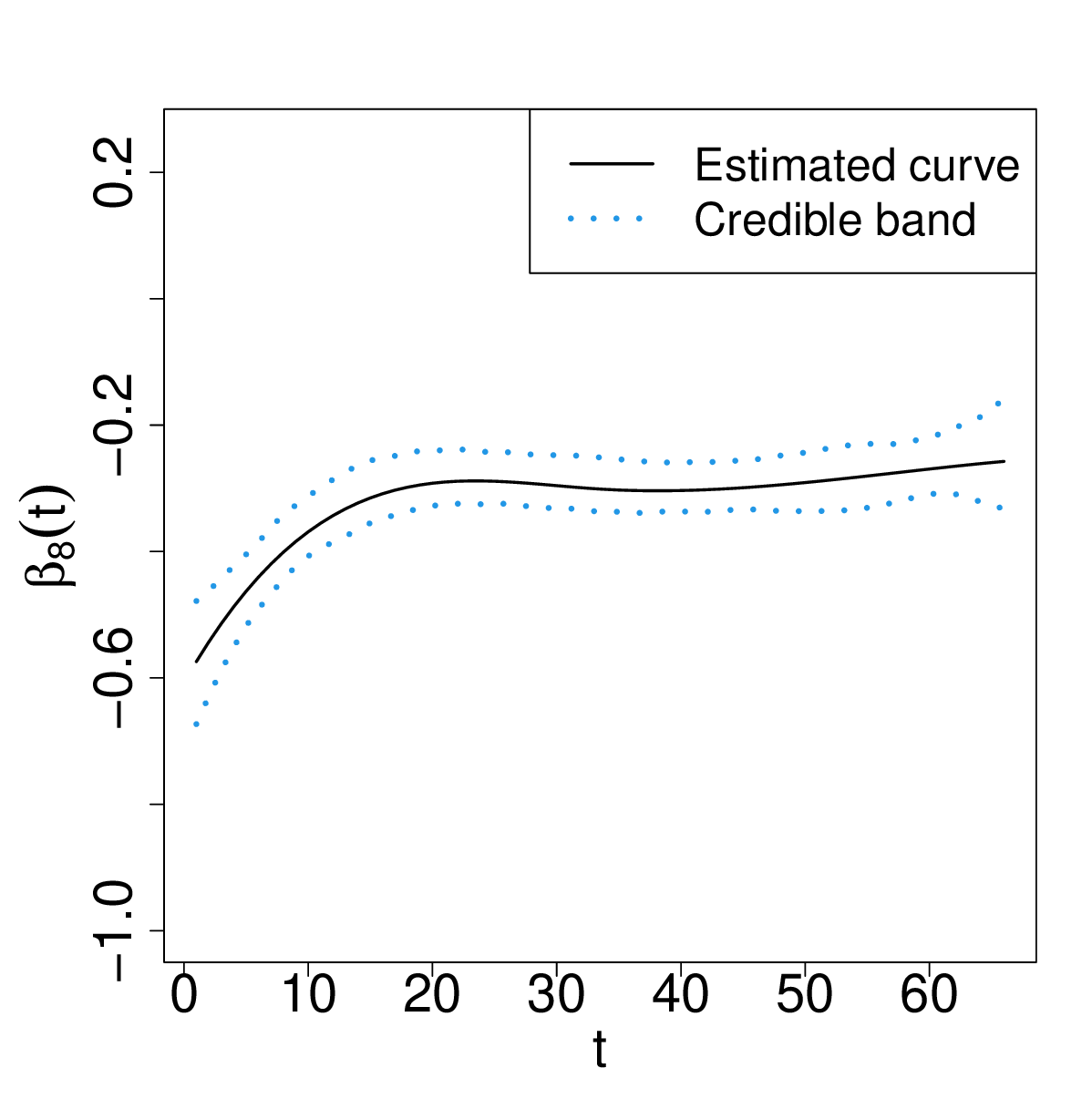}}	&
		\subfloat[{\scriptsize Estimated $\beta_{9}(t)$ (Income).}]{\includegraphics[scale=0.15]{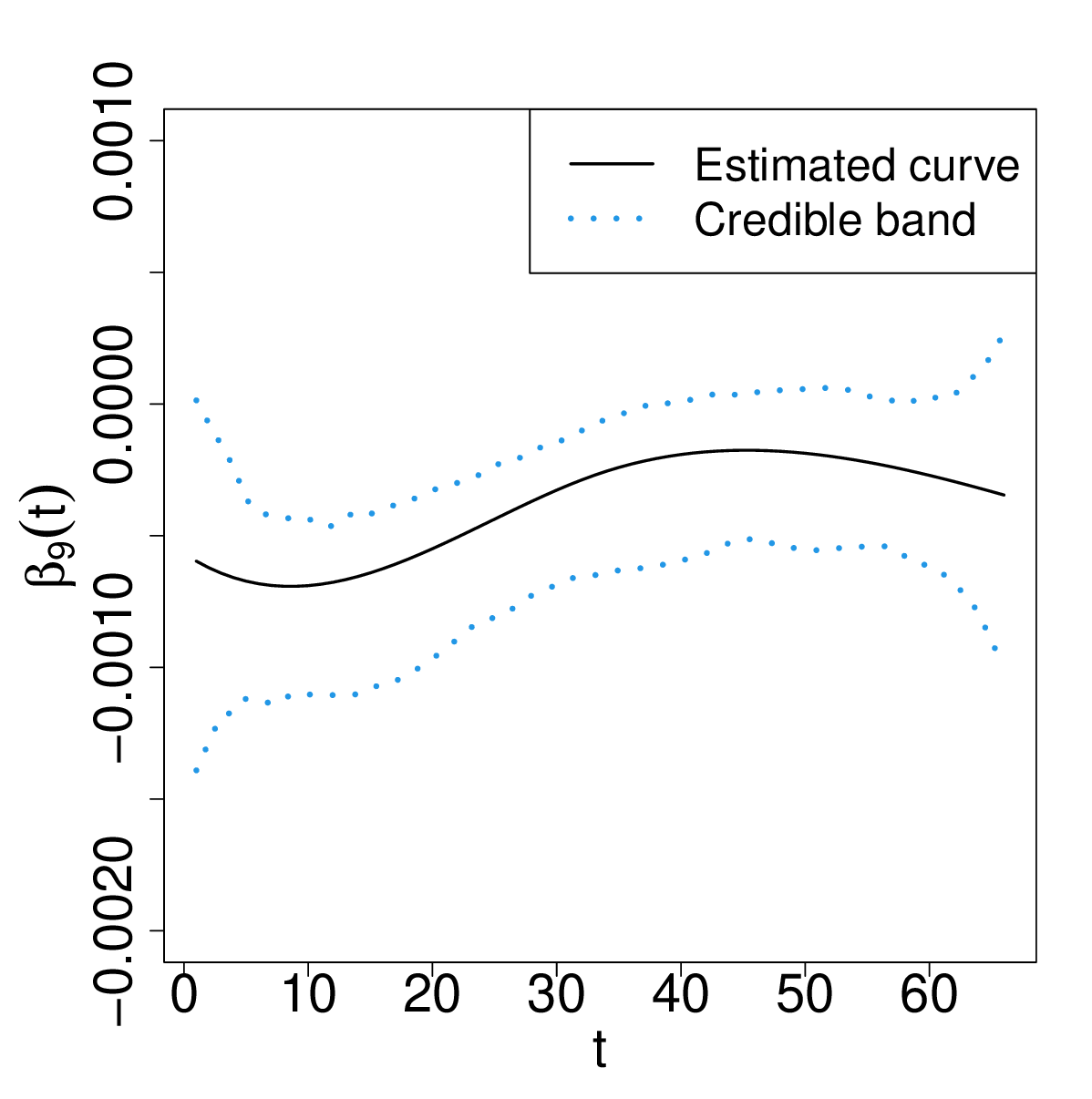}}		
	\end{tabular}
    \caption{Estimated curves for the functional coefficients associated with the covariates that were selected by the proposed model with $K=5$ basis functions.}
	\label{coefs_selecionados_rd}
\end{figure}

Figure \ref{coefs_selecionados_rd} presents the estimated curves and the credible bands for the functional coefficients ($\hat{Z}_{l}\hat{\beta}_{l}(.)$) associated with the covariates that were selected by the proposed model with $K=5$. At the same time, Supplementary Figure 11 shows the estimated curves of the partial functional coefficients ($\hat{\beta}_{l}(.)$) associated with the covariates that were excluded. It is worth remembering that the estimated value for the functional coefficients associated with the covariates that were excluded is equal to zero for every $t$ point. Therefore, Supplementary Figure 11 presents only the estimated curves of the partial functional coefficients. 

Observing the posterior samples of each $Z_{l}$ (not shown here), we notice that there was perfect agreement between the points of the posterior samples and the summary statistics since the respective samples for $\hat{Z}_{l}$’s equal to zero are entirely characterized by zeros. Similarly, the respective samples for $\hat{Z}_{l}$’s equal to one are entirely characterized by ones. This reinforces the model's limited uncertainty about which variables to retain and which to exclude.

The Gini index and QoLLI covariates were the most influential in predicting the response variable, as their respective functional coefficients are larger in magnitude. It is worth remembering that any analysis of the results must take into account that they are conditioned to some initial determining factors, such as the initial set of candidate covariates in the model and the hypothesis of linearity.

\section{Final considerations}
\label{cap_conclusao}

This work focuses on functional data modelling and proposes a model-based approach under the Bayesian paradigm for variable selection in FOSR. The proposed methodology shows low sensitivity to the hyperparameter $\mu$, which is desirable since this component does not control the regularization. On the other hand, $\lambda$ plays the role of a regularization hyperparameter, and naturally, the model has some sensitivity to it. If desired, it is still possible to add a Gamma prior to $\lambda^2$, leading to an improper prior and guaranteeing minor sensitivity of the model to $\lambda$. However, in this case, $\lambda$ would no longer play the role of regularizer and uniqueness problems in the stationary distribution could appear if there is multicollinearity between the covariates.

Through simulations, the proposed model proved superior in correctly selecting the covariates compared with other variable selection methods in FOSR. At the same time, it maintained an optimal level of goodness-of-fit. The frequentist competing methods in the simulation scenario with the highest data dispersion ($\sigma=20$) could not balance selection accuracy well with the goodness of fit. Therefore, they need to sacrifice goodness of fit and increase regularization to achieve better accuracy in selection.

When applying the proposed variable selection model to real data, our proposed methodology could perform variable selection. However, the fit could not explain much of the data variation (value of metric \eqref{eq:r2_PARTE2}, $K = 5$, 44.9$\%$). The not-so-great fit to the data is explained mainly by the small number of curves ($m=27$) used. As the proposed model is consistent, if the same study were carried out at the municipal level (Brazil has about 5568 municipalities, plus the Federal District and the State District of Fernando de Noronha), higher values would certainly be observed for the metric in \eqref{eq:r2_PARTE2}, and naturally smaller values for the GCV criterion. The choice to carry out the study with state data is justified by the high computational cost generated when having $m=5570$.

We conducted a brief simulation study to compare the computational costs of our proposed model with its Bayesian competitor, BGLSS, which was also the most competitive alternative overall. For this comparison, the Gibbs sampler outlined in Panel \ref{GIBBS} was implemented both with the \texttt{nimble} package \citep{NIMBLE} in R and through a naive implementation, which is more pedagogical in nature but computationally less efficient than the nimble-based version. Ten simulated datasets were generated according to Section 3.1 with $m = 10$, $n = 25$, $p = 6$, $K = 10$. Running posterior inference for all ten replications required 29 minutes for the nimble-based implementation of the proposed model, whereas the naive implementation took 14 hours and 40 minutes. In contrast, BGLSS took 2 minutes and 55 seconds. All runs were performed on a computer equipped with an 11th Gen Intel(R) Core(TM) i7-1195G7 @ 2.90GHz processor, using R version 4.4.1.  Given that the nimble-based implementation took 29 minutes for 10 datasets, these findings motivate future work in which inference for the proposed Bayesian hierarchical model is conducted using variational inference, potentially achieving computational performance comparable to, or possibly exceeding, that of BGLSS.

In addition to optimizing and reducing the computational cost by approximating the posterior distribution via variational inference, another possibility for future work is to develop an alternative approach to verify whether the functional coefficients are only null at specific points or regions of $t$ since the model for selecting variables proposed in this article is only able to indicate whether a covariate should be part of the model or not.

\section*{Supplementary information}

The file supplement.pdf contains supplementary figures, tables and additional details regarding variable standardization.

\section*{Acknowledgments}

We acknowledge the significant contributions of our coauthor, Ronaldo Dias, who sadly passed away after the completion of this work. His insight, dedication, and expertise were essential to the development of this research.

This study was financed in part by the Coordenação de Aperfeiçoamento de Pessoal de Nível Superior – Brasil (CAPES) – Finance Code 001, by the Natural Sciences and Engineering Research Council of Canada (NSERC), grant number RGPIN-2019-05915, and Fundação de Amparo à Pesquisa do Estado de São Paulo (FAPESP) grants 2018/04654-9 and 2019/00787-7.

\section*{Statements and Declarations}

\begin{itemize}
\item The authors declare that they have no conflict of interest.
\item The real data sets are available at the Brasil.IO accessible open data portal \citep{BRASILio} and IBGE data portal;
\item R code implementation is available at \url{https://github.com/phtosEST/Bayesian_Variable_Selection_for_FOSR};
\item Author Contributions: \begin{itemize}
    \item Pedro Henrique T. O. Sousa: Conceptualization, Methodology, Code and implementation, Formal analysis, Writing - Original Draft, Writing - Review \& Editing; 
    \item Camila P. E. de Souza: Conceptualization, Methodology, Writing - Review \& Editing;  
    \item Ronaldo Dias: Conceptualization, Methodology, Writing - Review \& Editing.
    \end{itemize}
\item The authors declare that they used generative AI tools (ChatGPT-3.5 and Grammarly Premium) for language improvement.

%P.H.T.O.S. developed the code for model implementation, performed formal analysis, and wrote the original manuscript draft. P.H.T.O.S., C.P.E.S. and R.D. were involved in the development of the conceptual and methodological aspects, as well as reviewed the manuscript.
\end{itemize}



\begin{appendices}
\section{Full conditional distributions}
\label{Ap_full}
\setcounter{equation}{0}
\numberwithin{equation}{section}

Here we present the details of how to obtain the full conditional distributions used in our Gibbs sampler (see panel in Section \ref{gibbs}) for the model given in \eqref{modeloAutoparte2}. Similarly, one can obtain the full conditional distributions for the model in \eqref{modeloparte2}.

First, consider $g_{i}(t_{ij})=\sum_{l=1}^{p}\sum_{k=1}^{K}x_{li}Z_{l}b_{kl}B_{k}(t_{ij})$, so that
\begin{equation}
	y_{ij}=\beta_{0}(t_{ij})+g_{i}(t_{ij})+\epsilon_{ij}=\sum_{k=1}^{K}b_{k0}B_{k}(t_{ij})+g_{i}(t_{ij})+\epsilon_{ij}\text{.}\nonumber%sem_citar
\end{equation}

\noindent We obtain the full conditional distributions of the coefficients $\vec{\bb}$’s as follows: 

 \begin{align*} 	f(\vec{\bb}|\vec{\theta},\vec{\mu},\sigma^2,\vec{\tau^2},\vec{Z},\vec{y})& 	\propto\pi(\sigma^2)\pi(\vec{\tau^2})\pi(\vec{\mu})f(\vec{\theta}|\vec{\mu})f(\vec{\bb}|\sigma^2,\vec{\tau^2})p(\vec{Z}|\vec{\theta})f(\vec{y}|\vec{\bb},\vec{Z},\sigma^2)\nonumber\\
 	&\propto f(\vec{\bb}|\sigma^2,\vec{\tau^2})f(\vec{y}|\vec{\bb},\vec{Z},\sigma^2)\nonumber\\
 	&=\left[\prod_{k=1}^{K}\prod_{l=1}^{p}f(\bb_{kl}|\sigma^2,\tau_{kl}^2)\right]\left[\prod_{i=1}^{m}\prod_{j=1}^{n_{i}}f(y_{ij}|\vec{\bb},\vec{Z},\sigma^2)\right]\nonumber\\
 	&\propto \exp\left\{-\sum_{k=1}^{K}\sum_{l=1}^{p}\frac{\bb_{kl}^{2}}{2\sigma^2\tau_{kl}^2}\right\}
 	\exp\left\{-\frac{\sum_{i=1}^{m}\sum_{j=1}^{n_{i}}(y_{ij}-\beta_{0}(t_{ij})-g_{i}(t_{ij}))^2}{2\sigma^2}\right\}\nonumber\\
 	&= \exp\left\{-\frac{\sum_{k=1}^{K}\sum_{l=1}^{p}\frac{\bb_{kl}^{2}}{\tau_{kl}^2}}{2\sigma^2}\right\}
 	\exp\left\{-\frac{\sum_{i=1}^{m}\sum_{j=1}^{n_{i}}(\tilde{y}_{ij}-g_{i}(t_{ij}))^2}{2\sigma^2}\right\}\text{,}\nonumber%sem_citar
 \end{align*}
 \noindent in which $\tilde{y}_{ij}=y_{ij}-\beta_{0}(t_{ij})$. Remember that $g_{i}(t_{ij})=\sum_{l=1}^{p}\sum_{k=1}^{K}x_{li}Z_{l}\bb_{kl}B_{k}(t_{ij})$. Define 
$\vec{O_{lij}}=(x_{li}Z_{l}B_{1}(t_{ij}),\,x_{li}Z_{l}B_{2}(t_{ij}),\dots,\,x_{li}Z_{l}B_{K}(t_{ij}))^{'}$, a $(K\times1)$ vector, and stack these vectors across covariates to obtain 
$\vec{O_{.ij}}=(\vec{O_{1ij}}^{'},\;\vec{O_{2ij}}^{'},\dots,\vec{O_{pij}}^{'})^{'}$, a vector of dimension $(Kp\times1)$. With this notation, we have $g_{i}(t_{ij})=\vec{\bb}^{'}\vec{O_{.ij}}$. Similarly, let $\vec{\eta^{2}}=(\vec{\eta^{2}_{.1}}^{'},\;\vec{\eta^{2}_{.2}}^{'},\dots,\;\vec{\eta^{2}_{.p}}^{'})^{'}$, in which $\vec{\eta^{2}_{.l}}=(1/\tau^{2}_{1l},1/\tau^{2}_{2l},\dots,1/\tau^{2}_{Kl})^{'}$. Thus, we have

 \begin{align*}
%f(\vec{\bb}\mid\vec{\theta},\vec{\mu},\sigma^2,\vec{\tau^2},\vec Z,\vec y) 
f(\vec{\bb}|{\sigma^2},\vec{\tau^2},\vec{Z},\vec{y})
 &\propto
 \exp\!\left\{
 -\frac{
 \vec{\bb}'\diag(\vec{\eta^2})\vec{\bb}
 +
 \sum_{i=1}^{m}\sum_{j=1}^{n_i}
 (\tilde y_{ij}-\vec{\bb}'\vec O_{.ij})^2
 }{2\sigma^2}
 \right\}
 \\
 &\propto
 \exp\!\left\{
 -\frac{
 \vec{\bb}'
 \Bigl[
 \diag(\vec{\eta^2})
 +\sum_{i=1}^{m}\sum_{j=1}^{n_i}\vec O_{.ij}\vec O_{.ij}'
 \Bigr]
 \vec{\bb}
 -
 2\vec{\bb}'\sum_{i=1}^{m}\sum_{j=1}^{n_i}\vec O_{.ij}\tilde y_{ij}
 }{2\sigma^2}
 \right\}.
 \end{align*}

 In what follows, for notational simplicity, we write 
$f(\vec{\bb}\mid \ldots)$ to denote conditioning on 
$(\sigma^2,\vec{\tau^2},\vec Z,\vec y)$. By the Cholesky decomposition, let $$\vec{Q}=\left[\diag(\vec{\eta^2})+\sum_{i=1}^{m}\sum_{j=1}^{n_{i}}\vec{O_{.ij}}\vec{O_{.ij}}^{'}\right]=\vec{L}\vec{L}^{'}\text{,}$$ a $(Kp\times Kp)$ matrix. Then, 
 \begin{align*} 	%f(\vec{\bb}|\vec{\theta},\vec{\mu},\sigma^2,\tau^2,\vec{Z},\vec{y})
 f(\vec{\bb}|\ldots)&
 	\propto \exp\left\{-\frac{\vec{\bb}^{'}\vec{L}\vec{L}^{'}\vec{\bb}-2\vec{\bb}^{'}\sum_{i=1}^{m}\sum_{j=1}^{n_{i}}\vec{O_{.ij}}\tilde{y}_{ij}}{2\sigma^2}\right\}\nonumber\\
 	&= \exp\left\{-\frac{(\vec{L}^{'}\vec{\bb})^{'}\vec{L}^{'}\vec{\bb}-2\vec{\bb}^{'}\vec{L}\vec{L}^{-1}\sum_{i=1}^{m}\sum_{j=1}^{n_{i}}\vec{O_{.ij}}\tilde{y}_{ij}}{2\sigma^2}\right\}\nonumber\\
 	&= \exp\left\{-\frac{(\vec{L}^{'}\vec{\bb})^{'}\vec{L}^{'}\vec{\bb}-2(\vec{L}^{'}\vec{\bb})^{'}\vec{L}^{-1}\sum_{i=1}^{m}\sum_{j=1}^{n_{i}}\vec{O_{.ij}}\tilde{y}_{ij}}{2\sigma^2}\right\}\nonumber\\
 	&= \exp\left\{-\frac{(\vec{L}^{'}\vec{\bb}-\vec{L}^{-1}\sum_{i=1}^{m}\sum_{j=1}^{n_{i}}\vec{O_{.ij}}\tilde{y}_{ij})^{'}(\vec{L}^{'}\vec{\bb}-\vec{L}^{-1}\sum_{i=1}^{m}\sum_{j=1}^{n_{i}}\vec{O_{.ij}}\tilde{y}_{ij})}{2\sigma^2}\right\}\nonumber\\
 	&= \exp\left\{-\frac{(\vec{L}^{'}\vec{\bb}-\vec{L}^{-1}\sum_{i=1}^{m}\sum_{j=1}^{n_{i}}\vec{O_{.ij}}\tilde{y}_{ij})^{'}\vec{L}^{-1}\vec{L}\vec{L}^{'}(\vec{L}^{'})^{-1}(\vec{L}^{'}\vec{\bb}-\vec{L}^{-1}\sum_{i=1}^{m}\sum_{j=1}^{n_{i}}\vec{O_{.ij}}\tilde{y}_{ij})}{2\sigma^2}\right\}\nonumber\\
 	&= \exp\left\{-\frac{(\vec{\bb}-(\vec{L}\vec{L}^{'})^{-1}\sum_{i=1}^{m}\sum_{j=1}^{n_{i}}\vec{O_{.ij}}\tilde{y}_{ij})^{'}\vec{L}\vec{L}^{'}(\vec{\bb}-(\vec{L}\vec{L}^{'})^{-1}\sum_{i=1}^{m}\sum_{j=1}^{n_{i}}\vec{O_{.ij}}\tilde{y}_{ij})}{2\sigma^2}\right\}\text{.}\nonumber%sem_citar
 \end{align*}
 
 As a result, $f(\vec{\bb}|{\sigma^2},\vec{\tau^2},\vec{Z},\vec{y})$ is proportional to
 \begin{align}
 	\exp\left\{-\frac{(\vec{\bb}-\vec{Q}^{-1}\sum_{i=1}^{m}\sum_{j=1}^{n_{i}}\vec{O_{.ij}}\tilde{y}_{ij})^{'}(\vec{Q}^{-1})^{-1}(\vec{\bb}-\vec{Q}^{-1}\sum_{i=1}^{m}\sum_{j=1}^{n_{i}}\vec{O_{.ij}}\tilde{y}_{ij})}{2\sigma^2}\right\}\text{.}
 	\label{cmpos_beta_PARTE2}
 \end{align}

\noindent So, $f(\vec{\bb}|\sigma^2,\vec{\tau^2},\vec{Z},\vec{y})$ is precisely a multivariate normal distribution with mean $\vec{Q}^{-1}\sum_{i=1}^{m}\sum_{j=1}^{n_{i}}\vec{O_{.ij}}\tilde{y}_{ij}$ and variance $\sigma^2\vec{Q}^{-1}$. The mean of this normal distribution can still be described in a simplified way. Let
\begin{equation}
	\vec{O_{...}}=\begin{pmatrix}
		\vec{O_{.11}}^{'}\\
		\vec{O_{.12}}^{'}\\
		\vdots\\
		\vec{O_{.1n_{1}}}^{'}\\
		\vdots\\
		\vec{O_{.m1}}^{'}\\
		\vec{O_{.m2}}^{'}\\
		\vdots\\
		\vec{O_{.mn_{m}}}^{'}\\
	\end{pmatrix}_{\left(\sum_{i=1}^{m}n_{i}\times Kp\right)}=\begin{pmatrix}
		\vec{O_{111}}^{'}&\vec{O_{211}}^{'}&\dots&\vec{O_{p11}}^{'}\\
		\vec{O_{112}}^{'}&\vec{O_{212}}^{'}&\dots&\vec{O_{p12}}^{'}\\
		\vdots&\vdots&\ddots&\vdots\\
		\vec{O_{11n_{1}}}^{'}&\vec{O_{21n_{1}}}^{'}&\dots&\vec{O_{p1n_{1}}}^{'}\\
		\vdots&\vdots&\ddots&\vdots\\
		
		\vec{O_{1m1}}^{'}&\vec{O_{2m1}}^{'}&\dots&\vec{O_{pm1}}^{'}\\
		\vec{O_{1m2}}^{'}&\vec{O_{2m2}}^{'}&\dots&\vec{O_{pm2}}^{'}\\
		\vdots&\vdots&\ddots&\vdots\\
		\vec{O_{1mn_{m}}}^{'}&\vec{O_{2mn_{m}}}^{'}&\dots&\vec{O_{pmn_{m}}}^{'}
	\end{pmatrix}_{\left(\sum_{i=1}^{m}n_{i}\times Kp\right)}\text{,}\nonumber%sem_citar
\end{equation} so we obtain,
\begin{align*}	\E(\vec{\bb}|\sigma^2,\vec{\tau^2},\vec{Z},\vec{y})=\vec{Q}^{-1}\vec{O_{...}}^{'}\vec{\tilde{y}_{..}}\text{,}
\nonumber%sem_citar\label{mediabeta_PARTE2}
\end{align*}in which $\vec{\tilde{y}_{..}}=(\vec{\tilde{y}_{1.}}^{'},\;\vec{\tilde{y}_{2.}}^{'},\dots,\;\vec{\tilde{y}_{m.}}^{'})^{'}$ and each $\vec{\tilde{y}_{i.}}=(\tilde{y}_{i1},\;\tilde{y}_{i2},\dots,\;\tilde{y}_{in_{i}})^{'}$.

Similarly, we obtain the full conditionals of the $\theta_{l}$’s, taking 
\begin{align*}
	f(\theta_{l}|\vec{\bb},\vec{\theta_{-[l]}},\vec{\mu},\sigma^2,\vec{\tau^2},\vec{Z},\vec{y})\nonumber&
	\propto 	\pi(\sigma^2)\pi(\vec{\tau^2})\pi(\vec{\mu})f(\theta_{l}|\mu_{l})f(\vec{\theta_{-[l]}}|\vec{\mu_{-[l]}})\nonumber\\&\times f(\vec{\bb}|\sigma^2,\vec{\tau^2})p(\vec{Z}|\vec{\theta})p(\vec{y}|\vec{\bb},\vec{Z},\sigma^2)\nonumber\\
	&\propto f(\theta_{l}|\mu_{l}) p(Z_{l}|\theta_{l})\nonumber\\&\propto(\theta_{l})^{\mu_{l}-1}(1-\theta_{l})^{(1-\mu_{l})-1}(\theta_{l})^{Z_{l}}(1-\theta_{l})^{1-Z_{l}}\text{.}\nonumber%sem_citar
\end{align*}
Thus,
\begin{align}
	f(\theta_{l}|\vec{\bb},\vec{\theta_{-[l]}},\vec{\mu},\sigma^2,\vec{\tau^2},\vec{Z},\vec{y})&=f(\theta_{l}|\mu_{l},Z_{l})\nonumber\\&
	\propto(\theta_{l})^{\mu_{l}+Z_{l}-1}(1-\theta_{l})^{(1-\mu_{l})-Z_{l}+1-1}\text{,}
	\label{cmpos_theta_PARTE2}
\end{align}

\noindent which shows that $f(\theta_{l}|\mu_{l},Z_{l})$ is a beta with the first parameter being $\mu_{l}+Z_{l}$ and the second being $2-Z_{l}-\mu_{l}$.

In order to obtain the full conditional distributions of the latent variables, it is necessary to calculate the probability
\begin{gather}
	\prob(Z_{l}=1|\vec{\bb},\vec{\theta},\vec{\mu},\sigma^2,\vec{\tau^2},\vec{Z_{-[l]}},\vec{y})\nonumber\\
	=\frac{\prob(Z_{l}=1|\vec{\theta})\prob(\vec{Z_{-[l]}}=z_{-[l]}|\vec{\theta})f(\vec{y}|\vec{\bb},Z_{l}=1,\vec{Z_{-[l]}}=z_{-[l]},\sigma^2)}{\sum_{z=0}^{1}\prob(Z_{l}=z|\vec{\theta})\prob(\vec{Z_{-[l]}}=z_{-[l]}|\vec{\theta})f(\vec{y}|\vec{\bb},Z_{l}=z,\vec{Z_{-[l]}}=z_{-[l]},\sigma^2)}\text{.}\nonumber%sem_citar
\end{gather} 

\noindent Simplifying this, we obtain
\begin{gather}
	\frac{\theta_{l}f(\vec{y}|\vec{\bb},Z_{l}=1,\vec{Z_{-[l]}}=z_{-[l]},\sigma^2)}{(1-\theta_{l})f(\vec{y}|\vec{\bb},Z_{l}=0,\vec{Z_{-[l]}}=z_{-[l]},\sigma^2)+\theta_{l}f(\vec{y}|\vec{\bb},Z_{l}=1,\vec{Z_{-[l]}}=z_{-[l]},\sigma^2)}\nonumber\\
	=\frac{\theta_{l}}{(1-\theta_{l})\frac{f(\vec{y}|\vec{\bb},Z_{l}=0,\vec{Z_{-[l]}}=z_{-[l]},\sigma^2)}{f(\vec{y}|\vec{\bb},Z_{l}=1,\vec{Z_{-[l]}}=z_{-[l]},\sigma^2)}+\theta_{l}}\text{.}\nonumber%sem_citar
\end{gather}

\noindent Now let $\vec{B_{ij}}=(B_{1}(t_{ij}),\;B_{2}(t_{ij}),\dots,\;B_{K}(t_{ij}))^{'}$ and $g_{i,0}(t_{ij})=\sum_{q\ne l}x_{qi}Z_{q}\vec{\bb_{.q}}^{'}\vec{B_{ij}}$ and $g_{i,1}(t_{ij})=x_{li}\vec{\bb_{.l}}^{'}\vec{B_{ij}}+\sum_{q\ne l}x_{qi}Z_{q}\vec{\bb_{.q}}^{'}\vec{B_{ij}}$, so we have
\begin{align*}
	\frac{f(\vec{y}|\vec{\bb},Z_{l}=0,\vec{Z_{-[l]}}=z_{-[l]},\sigma^2)}{f(\vec{y}|\vec{\bb},Z_{l}=1,\vec{Z_{-[l]}}=z_{-[l]},\sigma^2)}
	&=\exp\Bigg\{\frac{1}{2\sigma^2}\Bigg[\sum_{j=1}^{n_{i}}(y_{ij}-\beta_{0}(t_{ij})-g_{i,1}(t_{ij}))^2\\&+\sum_{s\ne i}\sum_{j=1}^{n_{s}}(y_{sj}-\beta_{0}(t_{sj})-g_{i}(t_{sj}))^2-\sum_{j=1}^{n_{i}}(y_{ij}-\beta_{0}(t_{ij})-g_{i,0}(t_{ij}))^2\\&-\sum_{s\ne i}\sum_{j=1}^{n_{s}}(y_{sj}-\beta_{0}(t_{sj})-g_{i}(t_{sj}))^2\Bigg]\Bigg\}\nonumber\\&
	%=\exp\Bigg\{\frac{1}{2\sigma^2}\Bigg[\sum_{j=1}^{n_{i}}(y_{ij}-\beta_{0}(t_{ij})-g_{i,1}(t_{ij}))^2-\sum_{j=1}^{n_{i}}(y_{ij}-\beta_{0}(t_{ij})-g_{i,0}(t_{ij}))^2\Bigg]\Bigg\}\nonumber\\
	=\exp\Bigg\{\frac{1}{2\sigma^2}\Bigg[\sum_{j=1}^{n_{i}}(\tilde{y}_{ij}-g_{i,1}(t_{ij}))^2-\sum_{j=1}^{n_{i}}(\tilde{y}_{ij}-g_{i,0}(t_{ij}))^2\Bigg]\Bigg\}\text{.}\nonumber%sem_citar
\end{align*}

\noindent Then, $\prob(Z_{l}=1|\vec{\bb},\vec{\theta},\vec{\mu},\sigma^2,\vec{\tau^2},\vec{Z_{-[l]}},\vec{y}) =\prob(Z_{l}=1|\vec{\bb},\theta_{l},{\sigma^2},\vec{Z_{-[l]}},\vec{y})$ is equal to
\begin{align}
%\prob(Z_{l}=1|\vec{\bb},\vec{\theta},\vec{\mu},\sigma^2,\vec{\tau^2},\vec{Z_{-[l]}},\vec{y}) &=\prob(Z_{l}=1|\vec{\bb},\theta_{l},{\sigma^2},\vec{Z_{-[l]}},\vec{y})\nonumber\\
	\frac{\theta_{l}}{(1-\theta_{l})\exp\Bigg\{\frac{1}{2\sigma^2}\Bigg[\sum_{j=1}^{n_{i}}(\tilde{y}_{ij}-g_{i,1}(t_{ij}))^2-\sum_{j=1}^{n_{i}}(\tilde{y}_{ij}-g_{i,0}(t_{ij}))^2\Bigg]\Bigg\}+\theta_{l}}\text{.}
	\label{cmpos_Z_PARTE2}
\end{align}

The construction of the full conditional distribution of $\sigma^{2}$ is also straightforward. Taking into account that

\begin{align*}
f(\sigma^2|\vec{\bb},\vec{\theta},\vec{\mu},\vec{\tau^2},\vec{Z},\vec{y})&\propto 
	\pi(\sigma^2)\pi(\vec{\tau^2})\pi(\vec{\mu})f(\vec{\theta}|\vec{\mu})f(\vec{\bb}|\sigma^2,\vec{\tau^2})p(\vec{Z}|\vec{\theta})f(\vec{y}|\vec{\bb},\vec{Z},\sigma^2)\nonumber\\
	&\propto\pi(\sigma^2)f(\vec{\bb}|\sigma^2,\vec{\tau^2})f(\vec{y}|\vec{\bb},\vec{Z},\sigma^2)\\
    &=\pi(\sigma^2)\left[\prod_{k=1}^{K}\prod_{l=1}^{p}f(\bb_{kl}|\sigma^2,\tau_{kl}^2)\right]\left[\prod_{i=1}^{m}\prod_{j=1}^{n_{i}}f(y_{ij}|\vec{\bb},\vec{Z},\sigma^2)\right]\nonumber\\
	&\propto\left(\frac{1}{\sigma^2}\right)^{\delta_{1}+1}\exp\left\{-\frac{\delta_{2}}{\sigma^2}\right\}\left[\prod_{k=1}^{K}\prod_{l=1}^{p}\left(\frac{1}{\sigma^2\tau_{kl}^2}\right)^{\frac{1}{2}}\exp\left\{-\frac{\bb_{kl}^2}{2\sigma^2\tau_{kl}^2}\right\}\right]\nonumber\\&\times\left[\prod_{i=1}^{m}\prod_{j=1}^{n_{i}}\left(\frac{1}{\sigma^2}\right)^{\frac{1}{2}}\exp\left\{-\frac{(y_{ij}-\beta_{0}(t_{ij})-g_{i}(t_{ij}))^2}{2\sigma^2}\right\}\right]\text{,}\nonumber%sem_citar
\end{align*}
\noindent we obtain
\begin{align}	f(\sigma^2|\vec{\bb},\vec{\theta},\vec{\mu},\vec{\tau^2},\vec{Z},\vec{y})&=f(\sigma^2|\vec{\bb},\vec{\tau^2},\vec{Z},\vec{y})\nonumber\\
	&\propto\left(\frac{1}{\sigma^2}\right)^{\frac{\sum_{i=1}^{m}n_{i}}{2}+\frac{pK}{2}+\delta_{1}+1}\nonumber\\&\times\exp\left\{-\frac{\sum_{i=1}^{m}\sum_{j=1}^{n_{i}}(\tilde{y}_{ij}-g_{i}(t_{ij}))^{2}+\sum_{k=1}^{K}\sum_{l=1}^{p}\frac{\bb_{kl}^2}{\tau_{kl}^2}+2\delta_{2}}{2\sigma^2}\right\}\text{.}
	\label{cmpos_sigma2_PARTE2}
\end{align}

\noindent Thus, the full conditional distribution of $\sigma^{2}$ is an inverse-gamma distribution, with $\left(\sum_{i=1}^{m}n_{i}\right)/2+\left(pK\right)/2+\delta_{1}$ as the shape parameter and $\left(\sum_{i=1}^{m}\sum_{j=1}^{n_{i}}(\tilde{y}_{ij}-g_{i}(t_{ij}))^{2}+\sum_{k=1}^{K}\sum_{l=1}^{p}\frac{\bb_{kl}^2}{\tau_{kl}^2}+2\delta_{2}\right)/2$ as the rate parameter.

Finally, we compute the full conditional for the $\tau_{kl}^{2}$’s. It is known that
\begin{align*}
	f(\tau_{kl}^2|\vec{\bb},\vec{\theta},\vec{\mu},\sigma^2,\vec{Z},\vec{y})&\propto 
	\pi(\sigma^2)\pi(\vec{\tau^2})\pi(\vec{\mu})f(\vec{\theta}|\vec{\mu})f(\vec{\bb}|\sigma^2,\vec{\tau^2})p(\vec{Z}|\vec{\theta})f(\vec{y}|\vec{\bb},\vec{Z},\sigma^2)\nonumber\\
	&\propto\pi(\tau_{kl}^2)f(\bb_{kl}|\sigma^2,\tau_{kl}^2)\nonumber\\&\propto\frac{\lambda^{2}}{2}\exp\left\{-\frac{\lambda^{2}}{2}\tau_{kl}^2\right\}\left(\frac{1}{\sigma^2\tau_{kl}^2}\right)^{\frac{1}{2}}\exp\left\{-\frac{\bb_{kl}^2}{2\sigma^2\tau_{kl}^2}\right\}\text{.}\nonumber%sem_citar
\end{align*}
Therefore,
\begin{align}	f(\tau_{kl}^2|\vec{\bb},\vec{\theta},\vec{\mu},\sigma^2,\vec{Z},\vec{y})&=f(\tau_{kl}^2|\bb_{kl},{\sigma^2})\nonumber\\
	&\propto\left(\frac{1}{\tau_{kl}^2}\right)^{\frac{1}{2}}\exp\left\{-\left(\frac{\frac{\bb_{kl}^{2}}{\sigma^2}}{2\tau_{kl}^2}+\frac{\lambda^{2}}{2}\tau_{kl}^2\right)\right\}\text{.}
	\label{jacob}
\end{align}

Under this structure, it is possible to show that $\eta_{kl}^{2}=1/\tau_{kl}^{2}$, conditional on the other terms of the hierarchical formulation, follows an inverse-normal distribution, with parameters $\lambda^{'}$ and $\mu^{'}$, so that
\begin{align*}
	f(x)=\sqrt{\frac{\lambda^{'}}{2\pi}}x^{-\frac{3}{2}}\exp\left\{-\frac{\lambda^{'}(x-\mu^{'})^{2}}{2(\mu^{'})^{2}x}\right\},\quad x >0\text{.}\nonumber%sem_citar
\end{align*}

\noindent To verify this, we need to rewrite the expression in \eqref{jacob} in terms of $\eta_{kl}^{2}$, including the Jacobian ($-1/(\eta_{kl}^{2})^{2}$). So, the full conditional of $\eta_{kl}^{2}$ is proportional to
\begin{align*}
	(\eta_{kl}^{2})^{-\frac{3}{2}}\exp\left\{-\frac{1}{2}\left(\frac{\bb_{kl}^{2}}{\sigma^2}\eta_{kl}^{2}+\frac{\lambda^{2}}{\eta_{kl}^{2}}\right)\right\}
	&=
		(\eta_{kl}^{2})^{-\frac{3}{2}}\exp\left\{-\frac{\bb_{kl}^{2}\left[(\eta_{kl}^{2})^{2}+\frac{\lambda^{2}\sigma^{2}}{\bb_{kl}^{2}}\right]}{2\sigma^{2}\eta_{kl}^{2}}\right\}\nonumber\\
	&\propto
		(\eta_{kl}^{2})^{-\frac{3}{2}}\exp\left\{-\frac{\bb_{kl}^{2}\left[(\eta_{kl}^{2})^{2}-2\eta_{kl}^{2}\sqrt{\frac{\lambda^{2}\sigma^{2}}{\bb_{kl}^{2}}}+\frac{\lambda^{2}\sigma^{2}}{\bb_{kl}^{2}}\right]}{2\sigma^{2}\eta_{kl}^{2}}\right\}\text{.}\nonumber%sem_citar
\end{align*}

As a result,
\begin{align}
	f(\eta_{kl}^2|\vec{\bb},\vec{\theta},\vec{\mu},\sigma^2,\vec{Z},\vec{y})&=f(\eta_{kl}^2|\bb_{kl},\sigma^2)\nonumber\\&
	\propto
	(\eta_{kl}^{2})^{-\frac{3}{2}}\exp\left\{-\frac{\bb_{kl}^{2}\left(\eta_{kl}^{2}-\sqrt{\frac{\lambda^{2}\sigma^{2}}{\bb_{kl}^{2}}}\right)^{2}}{2\sigma^{2}\eta_{kl}^{2}}\right\}\text{.}
	\label{cmpos_eta2_PARTE2}
\end{align}

\noindent Therefore, $\eta_{kl}^{2}$ follows an inverse-normal distribution with $\mu^{'}=\sqrt{(\lambda^{2}\sigma^{2})/\bb_{kl}^{2}}$ and $\lambda^{'}=\lambda^{2}$. Thus, to simulate $\tau_{kl}^{2}$ in the Gibbs sampler, it is enough to first simulate $\eta_{kl}^{2}$ to then obtain $\tau_{kl}^{2}$ through the transformation $\tau_{kl}^{2}=1/\eta_{kl}^{2}$.

An additional conditional distribution must be derived when considering $\mu_{l}$ as a parameter:
\begin{gather}
	f(\mu_{l}|\vec{\bb},\vec{\theta},\vec{\mu_{-[l]}},\sigma^2,\vec{\tau^2},\vec{Z},\vec{y})\nonumber\\
	\propto 
	\pi(\sigma^2)\pi(\vec{\tau^2})\pi(\mu_{l})\pi(\vec{\mu_{-[l]}})f(\theta_{l}|\mu_{l})f(\vec{\theta_{-[l]}}|\vec{\mu_{-[l]}})f(\vec{\bb}|\sigma^2,\vec{\tau^2})p(\vec{Z}|\vec{\theta})f(\vec{y}|\vec{\bb},\vec{Z},\sigma^2)\text{.}\nonumber%sem_citar
\end{gather}
Thus,
\begin{align}	
		f(\mu_{l}|\vec{\bb},\vec{\theta},\vec{\mu_{-[l]}},\sigma^2,\vec{\tau^2},\vec{Z},\vec{y})&=f(\mu_{l}|\theta_{l})\nonumber\\&
	\propto \pi(\mu_{l})f(\theta_{l}|\mu_{l})\nonumber\\&\propto I_{(0,\psi)}(\mu_{l})\theta_{l}^{\mu_{l}}(1-\theta_{l})^{1-\mu_{l}}\text{.}
	\label{cmpos_mu_PARTE2}
\end{align}Then, $f(\mu_{l}|\theta_{l})$ is a continuous Bernoulli distribution \citep{NEURIPS2019_f82798ec} with parameter $\theta_{l}$ truncated on the interval $(0,\psi)$.

\end{appendices}


\bibliographystyle{apalike}
\bibliography{sn-bibliography}


\clearpage

\end{document}